\documentstyle[preprint,eqsecnum,amsmath,amsfonts,aps,longtable]{revtex}
\allowdisplaybreaks

\newcommand{\soussur}[3]{\underset{#1}{\overset{#2}{#3}}}
\newcommand{\sous}[2]{\underset{#1}{#2}}

\begin{document}
\draft
\title{General relativistic dynamics of compact binaries 
\\at the third post-Newtonian order}
\author{Luc Blanchet}
\address{D\'epartement d'Astrophysique Relativiste et de
Cosmologie, \\ 
Centre National de la Recherche Scientifique (UMR 8629),\\
Observatoire de Paris, 92195 Meudon Cedex, France}

\author{Guillaume Faye}
\address{D\'epartement d'Astrophysique Relativiste et de Cosmologie,\\
Centre National de la Recherche Scientifique (UMR 8629),\\
Observatoire de Paris, 92195 Meudon Cedex, France}

\date{\today}
\maketitle
\widetext
\begin{abstract}
The general relativistic corrections in the equations of motion and
associated energy of a binary system of point-like masses are derived
at the third post-Newtonian (3PN) order. The derivation is based on a
post-Newtonian expansion of the metric in harmonic coordinates at the
3PN approximation. The metric is parametrized by appropriate
non-linear potentials, which are evaluated in the case of two
point-particles using a Lorentzian version of an Hadamard
regularization which has been defined in previous
works. Distributional forms and distributional derivatives constructed
from this regularization are employed systematically. The equations of
motion of the particles are geodesic-like with respect to the
regularized metric. Crucial contributions to the acceleration are
associated with the non-distributivity of the Hadamard regularization
and the violation of the Leibniz rule by the distributional
derivative. The final equations of motion at the 3PN order are
invariant under global Lorentz transformations, and admit a conserved
energy (neglecting the radiation reaction force at the 2.5PN
order). However, they are not fully determined, as they depend on one
arbitrary constant, which reflects probably a physical incompleteness
of the point-mass regularization. The results of this paper should be
useful when comparing theory to the observations of gravitational
waves from binary systems in future detectors VIRGO and LIGO.
\end{abstract}


\narrowtext

\section{Introduction}\label{I}

The present work is a contribution to the problem of the dynamics of
two compact objects at the so-called third post-Newtonian (3PN)
approximation of general relativity. By 3PN we mean the relativistic
corrections in the binary's equations of motion corresponding to the
order $1/c^6$ relatively to the Newtonian acceleration, when the speed
of light $c$ tends to infinity. Why studying the equations of motion
to such a frightful post-Newtonian order? A side reason is the strange
beauty of the post-Newtonian expansion, which becomes quite intricate
at the 3PN order, where it requires some interesting mathematical
methods. The main reason, however, is that inspiralling compact
binaries, namely systems of two neutron stars or black holes (or one
of each) moving on a relativistic orbit prior to their final merger,
should be routinely observed by the gravitational-wave detectors LIGO,
VIRGO and their fellows. Several analysis show that the post-Newtonian
templates required for the detection and parameter extraction of
inspiralling compact binaries should include the relativistic
corrections in the binary's orbital phase at approximately the level
of the 3PN order
\cite{3mn,CFPS93,FCh93,CF94,TNaka94,P95,DIS98}. 

Lorentz and Droste \cite{LD17} were the first to obtain the correct
equations of motion of two non-spinning particles at the 1PN
approximation (see \cite{D83a,D300} for reviews). An important work by
Einstein, Infeld and Hoffmann \cite{EIH,EI40,EI49} showed that the 1PN
acceleration can in fact be deduced from the vacuum gravitational
field outside the masses. This result is interesting because, in their
approach, the bodies are allowed to carry a strong internal
gravity. Unfortunately, the computation of the surface integrals
surrounding the masses is very difficult even at the 1PN order (see
\cite{asada} for a recent derivation of the Einstein-Infeld-Hoffmann
equations).  The same equations were also obtained by Fock and
followers \cite{Fock39,P49,Papa51} for the motion of the centers of
mass of bodies with finite size.  The next approximation, 2PN, has
been tackled by Otha, Okamura, Kimura and Hiida
\cite{O73,O74a,O74b} with a direct post-Newtonian computation of the
Hamiltonian of $N$ point-particles; however the first complete
2-particle case in their framework is only given by Damour and
Sch\"afer \cite{DS85}, and the fully explicit 3-particle case is due
to Sch\"afer \cite{S87}. Up to the 2PN level, the equations of motion
are conservative (existence of ten conserved quantities, including a
conserved energy). The non-conservative effect, which is associated to
the radiation reaction force, arises at the 2.5PN order. The first
correct equations of motion of two masses at the 2.5PN order were
obtained by Damour, Deruelle and collaborators
\cite{BeDD81,DD81a,Dthese,D82} in harmonic coordinates. These 
equations are applicable to systems of strongly self-gravitating
bodies such as neutron stars (see Damour \cite{D83a,D300} for the
proof). Moreover, Kopejkin \cite{Kop85} and Grishchuk and Kopejkin
\cite{GKop86} obtained the same equations in the case of weakly
self-gravitating extended bodies. The corresponding result at 2.5PN
order was also derived by Sch\"afer \cite{S85,S86} using the ADM
Hamiltonian approach. Later, the harmonic-coordinates equations of
motion were re-computed by Blanchet, Faye and Ponsot \cite{BFP98}
following a direct post-Newtonian iteration of the field
equations. Some of the latter derivations \cite{BeDD81,S85,BFP98} opt
for a formal description of the compact objects by
point-particles. There is a nice agreement between all these different
methods at the 2.5PN order. In addition, the complete 2.5PN
gravitational field generated by point-particles in harmonic
coordinates was derived in
\cite{BFP98}. 

At the 3PN order, the equations of motion have been obtained using a
Hamiltonian and formal delta functions by Jaranowski and Sch\"afer
\cite{JaraS98,JaraS99} in the center-of-mass frame, and by Damour, 
Jaranowski and Sch\"afer \cite{DJS00} in an arbitrary frame. These
authors found an irreducible ambiguity linked probably with an
incompleteness in the regularization of the infinite self-field of the
particles. In this paper, following the method initiated in
\cite{BFP98}, we address the problem of the 3PN dynamics of
point-particles in harmonic coordinates. Earlier in \cite{BF00}, our
result has been already discussed and reported in the case of circular
orbits. We find the presence of one (and only one) undetermined
coefficient in the 3PN equations of motion, in agreement with
\cite{JaraS98,JaraS99,DJS00}. Recently, the physical equivalence between our 
result in harmonic coordinates and the result given by the
ADM-Hamiltonian approach has been established \cite{DJS00c,ABF00}.

Another line of research, initiated by Chandrasekhar and collaborators
\cite{CN69,CE70}, consists of working with continuous hydrodynamical
fluids from the start, and derived the metric and
equations of motion of an isolated fluid ball up to the 2.5PN order
\cite{Ehl80,Ker80,Ker80',PapaL81} (the derivation in the case 
of two fluid balls, in the limit of zero size of the bodies, being due
to \cite{Kop85,GKop86}). Our iteration of the gravitational field and
equations of motion in the previous paper
\cite{BFP98} is close to the latter line of work in the sense that it
is based on the reduction of some general expressions of the
post-Newtonian metric, initially valid for continuous fluids, to
point-like particles. The choice of point-particles, adopted here as
well, is motivated by the efficiency of the delta-functions in
performing some complicated non-linear integrations. The price we have
to pay is the necessity of a self-field regularization. We apply
systematically in this paper the regularization of Hadamard, based on
the concept of ``partie finie'' of singular functions and divergent
integrals \cite{Hadamard,Schwartz,Sellier}. This technique is indeed
extensively used in this field \cite{BeDD81,S85,BFP98,JaraS98}. More
precisely, we apply a variant of the Hadamard regularization, together
with a theory of pseudo-functions and distributional derivatives, that
is compatible with the Lorentzian structure of the gravitational
field. All the details about this regularization can be found in
\cite{BFreg,BFregM}. We use notably a specific form of distributional
stress-energy tensor based on ``delta-pseudo-functions'' (with support
limited to the world lines of the particles). In a sense, these
delta-pseudo-functions constitute some mathematically well-defined
versions of the so-called ``good delta functions'' introduced long ago
by Infeld \cite{Infeld} (see also an appendix in the book of Infeld
and Plebanski \cite{InfeldP}).

Thus, we are using a formal regularization method, based on a clear
mathematical framework \cite{BFreg,BFregM}, but that we cannot justify
physically (why should the compact objects be described by such
delta-pseudo-function singularities?). Definitely, our main
justification is that this method permits the derivation of a result
in a consistent and well-defined way (i.e. all the difficult
non-linear integrals at the 3PN order are computed
unambiguously). Furthermore, we shall check that some different
regularization prescriptions yield equations of motion that are
physically the same, in the sense that they differ from each other by
merely a coordinate transformation. Moreover, a justification {\it a
posteriori} is that the end result owns all the physical properties
that we expect the true equations of motion of compact objects to
obey. In particular, there is agreement with the known results at the
previous post-Newtonian orders, we get the correct geodesic limit for
the motion of a test particle in a Schwarzschild background, find that
the 3PN equations of motion stay invariant under global Lorentz
transformations, and obtain a conserved energy at 3PN (neglecting the
radiation reaction). The investigation of the Lagrangian formulation
of the equations is dealt with in a separate work \cite{ABF00}.

Ideally, one should perform, instead of a computation valid only for
point particles (and necessitating a regularization), a complete
calculation in the case of extended bodies, i.e. taking into account
the details of the internal structure of the bodies. By considering
the limit where the radius of the two objects tend to zero, one should
recover the same result as obtained by means of the point-mass
regularization. This would demonstrate the suitability of the
regularization. In fact, this program has been achieved at the 2PN
order by Grishchuk and Kopejkin \cite{Kop85,GKop86}, who proved that
the compactness parameters associated with each object disappear from
the equations of motion, and obtained the same equations as in the
case of point particles. At the 3PN order there is no such a proof
that the method with extended bodies would give the same result as
with point particles.

The main problem is that from the 3PN level one cannot compute the
most difficult of the non-linear integrals in closed form for two
extended fluid bodies of finite radius (though these integrals could
perhaps be obtained as power series valid when the two radius tend to
zero). Presently the only approach which is able to overcome this
problem is the one followed in this paper: namely, to model the source
by delta functions and to use a regularization. The price we have to
pay is the appearance of one physical undetermined coefficient at the
3PN order. As a consequence, this method should be completed
(hopefully in a future work) by the study of the limit relation of the
point-particle result with the physical result valid for extended
bodies in the limit of zero size. This study should probably give the
value of the undetermined parameter left out by the regularization.

The plan of this paper is the following. In Section \ref{II}, we
review some necessary tools concerning the regularization and the
definition of the point-particle model. In Section \ref{III}, we
perform the post-Newtonian iteration of the field equations and write
the 3PN metric in terms of some convenient non-linear
potentials. Section \ref{IV} is devoted to the computation of the
compact-support and quadratically non-linear parts of the
potentials. The most difficult potentials, involving notably some
non-compact cubic non-linearities at 1PN, are obtained in Section
\ref{V}. The so-called Leibniz and non-distributivity contributions to
the equations of motion are derived in Section \ref{VI}. Finally, we
present in Section \ref{VII} the result for the compact binary's 3PN
acceleration (in the case of general orbits) and the associated 3PN
energy.

\section{Hadamard regularization}\label{II}

In this section we present a short account about the regularization of
Hadamard \cite{Hadamard,Schwartz}, the associated generalized or
pseudo-functions, and the choice of stress-energy tensor for
point-particles.  We follow (and refer to) the detailed investigations
in \cite{BFreg,BFregM}.  Consider the class ${\cal F}$ of functions
$F({\bf x})$ which are smooth ($C^\infty$) on ${\mathbb R}^3$ deprived
from two singular points ${\bf y}_1$ and ${\bf y}_2$, around which
they admit a power-like singular expansion of the type

\begin{equation}\label{II1}
\forall n\in {\mathbb N}\;,\quad
F({\bf x})=\sum_{a_0\leq a\leq n} r_1^a \sous{1}{f_a}({\bf
n}_1)+o(r_1^n) \;,
\end{equation}
and similarly for the other point 2. Here $r_1=|{\bf x}-{\bf y}_1|\to
0$, and the coefficients ${}_1f_a$ of the various powers of $r_1$
depend on the unit direction ${\bf n}_1=({\bf x}-{\bf y}_1)/r_1$ of
approach to the singular point. The powers $a$ of $r_1$ are real,
range in discrete steps (i.e. $a\in (a_i)_{i\in {\mathbb N}}$) and are
bounded from below ($a_0\leq a$). The coefficients ${}_1f_a$ (and
${}_2f_a$) for which $a<0$ are referred to as the {\it singular}
coefficients of $F$. If $F$ and $G$ belong to ${\cal F}$ so does the
ordinary pointwise product $FG$, as well as the ordinary gradient
$\partial_iF$. We define the Hadamard ``partie finie'' of $F$ at the
location of the singular point 1 as

\begin{equation}\label{II2}
(F)_{\textstyle {}_1}= \int \frac{d\Omega_1}{ 4\pi} \sous{1}{f_0}({\bf n}_1) \;,
\end{equation}
where $d\Omega_1= d\Omega ({\bf n}_1)$ denotes the solid angle element
centered on ${\bf y}_1$ and of direction ${\bf n}_1$. Furthermore, the
Hadamard partie finie (${\rm Pf}$) of the integral $\int d^3{\bf
x}~F$, which is in general divergent at the two singular points ${\bf
y}_1$ and ${\bf y}_2$ (we assume no divergence at infinity), is
defined by

\begin{eqnarray}\label{II3}
{\rm Pf}_{s_1,s_2}\int d^3{\bf x}~ F &=&~\lim_{s\to
0}~\biggl\{\int_{{\cal D}(s)} d^3{\bf x}~ F\nonumber\\ &&\qquad
+~4\pi\sum_{a+3< 0}{\frac{s^{a+3}}{ a+3}} \left(\frac{F}{
r_1^a}\right)_{\textstyle {}_1}+ 4
\pi \ln\left(\frac{s}{s_1}\right) \left(r_1^3 F\right)_{\textstyle {}_1}
+1\leftrightarrow 2\biggr\}\;.
\end{eqnarray}
The first term integrates over a domain ${\cal D}(s)$ defined as
${\mathbb R}^3$ to which the two spherical balls $r_1\leq s$ and
$r_2\leq s$ of radius $s$ and centered on the singularities are
removed. The other terms, in which the value of a function at 1 takes
the meaning (\ref{II2}), are such that they cancel out the divergent
part of the first term in the limit where $s\to 0$ (the symbol
$1\leftrightarrow 2$ means the same terms but corresponding to the
other point 2). Note that the Hadamard partie finie depends on two
strictly positive constants $s_1$ and $s_2$, associated with the
logarithms in (\ref{II3}). See \cite{BFreg} and Section \ref{V} below
for alternative expressions of the partie-finie integral.

To any $F\in {\cal F}$ we associate a partie finie pseudo-function
${\rm Pf}F$ defined as the linear form on ${\cal F}$ given by the
duality bracket,

\begin{equation}\label{II4}
\forall G\in {\cal F}\;,\quad <{\rm Pf}F,G>={\rm Pf}\int d^3{\bf x}~FG \;.
\end{equation}
The pseudo-function ${\rm Pf}F$, when restricted to the set of smooth
functions with compact support, is a distribution in the sense of
Schwartz
\cite{Schwartz}. The product of pseudo-functions coincides with the ordinary
pointwise product, namely ${\rm Pf}F\!~.\!~{\rm Pf}G={\rm Pf}(FG)$. A
particularly interesting pseudo-function, constructed in \cite{BFreg}
on the basis of the Riesz delta function \cite{Riesz}, is the
delta-pseudo-function ${\rm Pf}\delta_1$, which plays the same role as
the Dirac measure in distribution theory, in the sense that

\begin{equation}\label{II5}
\forall F\in {\cal F}\;,\quad <{\rm Pf}\delta_1,F>={\rm Pf}\int d^3{\bf
x}~\delta_1 F=(F)_{\textstyle {}_1}  \;,
\end{equation}
where $(F)_{\textstyle {}_1}$ is the partie finie of $F$ as defined by
(\ref{II2}). From the product of ${\rm Pf}\delta_1$ with any ${\rm
Pf}F$ we obtain the new pseudo-function ${\rm Pf}(F\delta_1)$ which is
such that

\begin{equation}\label{II5'}
\forall G\in {\cal F}\;,\quad <{\rm Pf}(F\delta_1),G>=(FG)_{\textstyle {}_1} \;.
\end{equation}

Next, the spatial derivative of a pseudo-function of the type ${\rm
Pf}F$, namely $\partial_i({\rm Pf}F)$, is treated as
follows. Essentially, we require in \cite{BFreg} the so-called rule of
integration by parts, namely that we are allowed to freely operate by
parts any duality bracket, with the all-integrated (``surface'') terms
always zero exactly like in the case of non-singular functions. This
requirement is motivated by our will that a computation involving
singular functions be as much as possible the same as a computation
valid for regular functions. Thus,

\begin{equation}\label{II6}
\forall F,G\in {\cal F}\;,\quad <\partial_i({\rm Pf}F),G>=-<\partial_i({\rm
Pf}G),F> \;.
\end{equation}
Furthermore, we assume that when all the singular coefficients of $F$
vanish, the derivative of ${\rm Pf}F$ reduces to the ordinary
derivative, i.e.  $\partial_i({\rm Pf}F)={\rm Pf}(\partial_iF)$. As a
particular case, we see from these assumptions that the integral of a
gradient is always zero: $<\partial_i({\rm Pf}F),1>=0$. Certainly this
should be the case if we want to apply to the case of singular sources
a formula which is defined modulo a total divergence for continuous
sources. We have also at our disposal a distributional time derivative
and the associated partial derivatives with respect to the points 1
and 2 (see Section IX in
\cite{BFreg}). The difference between the distributional derivative and the
ordinary one gives the distributional terms ${\sc D}_i[F]$ present
in the derivative of $F$, 

\begin{equation}\label{II7}
\partial_i ({\rm Pf} F) = {\rm Pf}(\partial_i F) + {\sc D}_i[F]  \;.
\end{equation}
A simple solution of our basic relation (\ref{II6}), denoted ${\sc
D}_i^{\rm part}[F]$ standing for the ``particular'' solution, was
obtained in \cite{BFreg} as the following functional of the singular
coefficients of $F$,

\begin{equation}\label{II8}
{\sc D}_i^{\rm part}[F] = 4\pi~\! {\rm Pf} \Biggl( n_1^i
\biggl[\frac{1}{2}~\!r_1 \sous{\!\!1}{f_{-1}}+\sum_{k\geq 0} \frac{1}{
r_1^k} \sous{\!\!\!\!\!\!1}{f_{-2-k}}\biggl] \delta_1 \Biggr)
+1\leftrightarrow 2\;,
\end{equation}
where we assume for simplicity that the powers $a$ in the expansion of
$F$ are relative integers, $a\in {\mathbb Z}$. (The sum over $k$ is 
always finite.) The distributional term
(\ref{II8}) is of the form ${\rm Pf}(G\delta_1)$ (plus
$1\leftrightarrow 2$). However, the particular solution (\ref{II8})
does not represent the most satisfying derivative operator acting on
pseudo-functions. It is shown in
\cite{BFreg} that one can require also the rule of commutation of successive
derivatives, which is not satisfied in general by (\ref{II8}). Still
we are motivated when asking for the commutation of derivatives that
the properties of our distributional derivative be the closest
possible to those of the ordinary derivative.  The most general
derivative operator satisfying the same properties as (\ref{II8}) and,
in addition, the commutation of derivatives (Schwarz lemma) is given
by
  
\begin{equation}\label{II9}
{\sc D}_i[F] = 4\pi\sum_{l=0}^{+\infty}~\!{\rm
Pf} \Biggl(~\!C_l~\!\biggl[n_1^{iL}\sous{\!\!\!\!1}{{\hat
f}_{-1}^{L}}-n_1^{L}\sous{\!\!\!\!1}{{\hat f}_{-1}^{iL}}\biggr]
r_1\delta_1 +\sum_{k\geq 0} \frac{n_1^{iL}}{r_1^k}
\sous{\!\!\!\!\!\!\!\!1}{{\hat
f}_{-2-k}^{L}}\delta_1\Biggr)+1\leftrightarrow 2\;,
\end{equation}
where we denote by ${}_1{\hat f}_a^L$ the STF-harmonics of the
expansion coefficient ${}_1f_a$, which is such that
${}_1f_a=\sum_{l\geq 0} n_1^L{}_1{\hat f}_a^L$ (see \cite{BFreg} for
details).  A particularity of this derivative is that it depends on an
arbitrary constant $K$ through the $l$-dependent coefficient

\begin{equation}\label{II10}
C_l=(l+1)\biggl[K+\sum_{j=1}^l\frac{1}{ j+1}\biggr]\;.
\end{equation}
Both the derivative operators (\ref{II8}) and (\ref{II9})-(\ref{II10})
represent some generalizations of the Schwartz distributional
derivative
\cite{Schwartz}, that are appropriate to the singular functions 
of the class ${\cal F}$. In was shown in Section VIII in \cite{BFreg}
that the distributional terms associated with the $l$th distributional
derivative, i.e. ${\sc D}_L[F]=\partial_L{\rm Pf}F-{\rm Pf}\partial_L
F$, where $L=i_1i_2\dots i_l$ denotes a multi-index composed of $l$
indices, is given by

\begin{equation}\label{II10'}
{\sc D}_{L}[F]=\sum_{k=1}^l\partial_{i_1\dots i_{k-1}}{\sc
D}_{i_k}[\partial_{i_{k+1}\dots i_l}F]\;.
\end{equation}
Though this is not manifest on this formula, ${\sc D}_L[F]$ in the
case of the ``correct'' derivative (\ref{II9})-(\ref{II10}) is fully
symmetric in the $l$ indices forming $L$.  Note that neither of the
derivatives (\ref{II8}) and (\ref{II9}) satisfy the Leibniz rule for
the derivation of a product. Rather, the investigation in
\cite{BFreg} has suggested that, in order to construct 
a consistent theory (using the ``ordinary'' product for
pseudo-functions), the Leibniz rule should in a sense be weakened, and
replaced by the rule of integration by part (\ref{II6}), which is in
fact nothing but an ``integrated'' version of the Leibniz rule. In
this paper, we shall be careful about taking into account the
violation of the Leibniz rule by the distributional derivative. We
shall also investigate the fate of the constant $K$ appearing in
(\ref{II10}) when deriving the 3PN equations of motion.

The Hadamard regularization $(F)_{\textstyle {}_1}$ is defined by
(\ref{II2}) in a preferred spatial hypersurface $t=$const of a
coordinate system, and consequently is not {\it a priori} compatible
with the global Lorentz invariance of special relativity. If we
restrict the coordinates to satisfy the usual harmonic gauge
conditions, we introduce a preferred Minkowski metric, and thus we can
view the gravitational field as a relativistic Lorentz tensor field in
special relativity, that we certainly want to regularize in a
Lorentz-invariant way. To achieve this we defined in
\cite{BFregM} a new regularization, denoted $[F]_{\textstyle {}_1}$, 
by performing the Hadamard regularization within the spatial
hypersurface which is geometrically orthogonal (in the Minkowskian
sense) to the four-velocity of the particle. In a sense, the
regularization $[F]_{\textstyle {}_1}$ permits us to get rid of the
anisotropic Lorentz contraction due to their motion when defining the
point-masses. The Lorentzian regularization $[F]_{\textstyle {}_1}$
differs from the old one $(F)_{\textstyle {}_1}$ by relativistic
corrections of order $1/c^2$ at least. All the formulas for its
computation are given in
\cite{BFregM} in the form of some infinite expansion series in the
relativistic parameter $1/c^2$. The regularization $[F]_{\textstyle
{}_1}$ plays a crucial role in the present computation, as it will be
seen that the breakdown of the Lorentz invariance due to the old
regularization $(F)_{\textstyle {}_1}$ occurs precisely at the 3PN
order in the equations of motion. Associated with the new
regularization in \cite{BFregM} we can define, exactly like in
(\ref{II5}), a ``Lorentzian'' delta-pseudo-function ${\rm Pf}\Delta_1$
which when applied on any $F$ gives $[F]_{\textstyle {}_1}$. More
generally we have, similarly to (\ref{II5'}),

\begin{equation}\label{II11}
\forall G\in {\cal F}\;,\quad <{\rm Pf}(F\Delta_1),G>
=[FG]_{\textstyle {}_1} \;.
\end{equation}
Notice that as a general rule we are not allowed to replace $F$ in the
pseudo-function ${\rm Pf}(F\Delta_1)$ by its regularized value,
i.e. ${\rm Pf}(F\Delta_1)\not= [F]_{\textstyle {}_1}~{\rm
Pf}\Delta_1$.  This is a consequence of the ``non-distributivity'' of
the Hadamard partie finie with respect to the multiplication, i.e.
$[FG]_{\textstyle {}_1}\not= [F]_{\textstyle {}_1}[G]_{\textstyle
{}_1}$.  In this paper, we shall (heuristically) model the compact
objects by point-particles, and in order to describe those
point-particles we shall use a particular representation of the
stress-energy tensor which has been derived in Section V of
\cite{BFregM} on the basis of an action principle compatible with the
Lorentzian regularization $[F]_{\textstyle {}_1}$. The proposal made
in \cite{BFregM} is that

\begin{equation}\label{II12}
T^{\mu\nu}= m_1 c \frac{v_1^\mu
v_1^\nu}{\sqrt{-[g_{\rho\sigma}]_{\textstyle {}_1} v_1^\rho
v_1^\sigma}} {\rm Pf}\left(\frac{\Delta_1}{\sqrt{-g}}\right) +
1\leftrightarrow 2 \;.
\end{equation}
Most importantly about this expression are the facts that (i)
$[g_{\rho\sigma}]_{\textstyle {}_1}$ within the first factor means the
Lorentzian regularization of the metric in the previous sense, (ii)
the pseudo-function ${\rm Pf}\left(\frac{1}{\sqrt{-g}}\Delta_1\right)$
is of the type ${\rm Pf}(F\Delta_1)$ which is defined by
(\ref{II11}). We denote by $m_1$ the (constant) mass of the particle
1, by ${\bf y}_1(t)$ its trajectory parametrized by the
harmonic-coordinate time $t$, and by ${\bf v}_1(t)=d{\bf y}_1/dt$ the
coordinate velocity [with $v_1^\mu=(c,{\bf v}_1)$]. In the next
section, we look for solutions in the form of post-Newtonian
expansions of the Einstein field equations having the latter
stress-energy tensor as a matter source.
 
\section{The third post-Newtonian metric}\label{III} 

\subsection{The Einstein field equations}

We base our investigation on a system of harmonic coordinates
$x^0=ct$, $(x^i)={\bf x}$, since such coordinates are especially
well-suited to a post-Newtonian (or post-Minkowskian) iteration of the
field equations. We define the gravitational perturbation $h^{\mu\nu}$
associated with the ``gothic metric'' as

\begin{equation}\label{III1}
h^{\mu\nu}=\sqrt{-g} g^{\mu\nu}-\eta^{\mu\nu}\;,
\end{equation} 
with $g^{\mu\nu}$ and $g$ being the inverse and the determinant of the
covariant metric $g_{\mu\nu}$, and where $\eta^{\mu\nu}={\rm
diag}(-1,1,1,1)$ denotes an auxiliary Minkowski metric. Under the
condition of harmonic coordinates,

\begin{equation}\label{III2}
\partial_{\nu} h^{\mu\nu}=0\;,
\end{equation}
the Einstein field equations take the form

\begin{equation}\label{III3}
\Box h^{\mu \nu} = \frac{16 \pi G}{c^4} |g|
T^{\mu\nu}+\Lambda^{\mu\nu} \;,
\end{equation}
where $\Box=\eta^{\mu\nu}\partial_\mu\partial_\nu$ denotes the flat
d'Alembertian operator, where $T^{\mu\nu}$ is the matter stress-energy
tensor defined in our case of point-particle binaries by (\ref{II12}),
and where $\Lambda^{\mu\nu}$ is the gravitational source term.  Using
the integral of the retarded potentials given by

\begin{equation}\label{III3'}
\Box_{\cal R}^{-1}\tau({\bf x},t)=
\int \frac{d^3{\bf x}'}{-4\pi} \frac{\tau({\bf x}',t-|{\bf x}-{\bf x}'|/c)}{
|{\bf x}-{\bf x}'|}\;,
\end{equation}
we can also re-write the solution of the field equations (\ref{III3}),
under a condition of no-incoming radiation, under the form  

\begin{equation}\label{III3''}
h^{\mu \nu} = \Box_{\cal R}^{-1}\bigg[\frac{16 \pi G}{c^4} |g|
T^{\mu\nu}+\Lambda^{\mu\nu} \bigg]\;.
\end{equation}
The gravitational source term $\Lambda^{\mu\nu}$ is related to the
Landau-Lifchitz pseudo-tensor $t^{\mu\nu}_{\rm LL}$ by

\begin{equation}\label{III4}
\Lambda^{\mu\nu}=
\frac{16 \pi G}{c^4} |g| t^{\mu\nu}_{\rm LL}+
\partial_{\rho} h^{\mu\sigma} \partial_{\sigma} h^{\nu\rho}-h^{\rho\sigma}
\partial_{\rho\sigma} h^{\mu\nu} \;,
\end{equation}
and can be expanded as an infinite non-linear series in $h$ and its
first and second space-time derivatives; in this paper we need only
the non-linear terms up to the quartic ($h^4$ or $G^4$) level, {\it
viz}

\begin{equation}\label{III5}
\Lambda^{\mu\nu}=N^{\mu\nu}(h,h)+M^{\mu\nu}(h,h,h)+L^{\mu\nu}(h,h,h,h)+{\cal O}(h^5)\;,
\end{equation}
where the quadratic non-linearity $N^{\mu\nu}$, the cubic one
$M^{\mu\nu}$ and the quartic $L^{\mu\nu}$ are explicitly given by

\begin{mathletters}\label{III6}\begin{eqnarray}
N^{\mu\nu} =
&-&h^{\rho\sigma} \partial_{\rho\sigma} h^{\mu\nu}+\frac{1}{2} \partial^\mu
h_{\rho\sigma} \partial^\nu h^{\rho\sigma}-\frac{1}{4} \partial^\mu h
\partial^\nu h+   \partial_\sigma h^{\mu \rho}
(\partial^\sigma h^\nu_\rho+
\partial_\rho h^{\nu \sigma} )    \nonumber \\ 
&-&2 \partial^{(\mu}
h_{\rho\sigma} \partial^\rho h^{\nu) \sigma}  +
\eta^{\mu\nu} \bigg[ -\frac{1}{4} \partial_\tau h_{\rho\sigma} 
\partial^\tau h^{\rho\sigma} +\frac{1}{8} \partial_\rho h
\partial^\rho h+\frac{1}{2} \partial_\rho h_{\sigma\tau}
\partial^\sigma h^{\rho \tau}  \bigg] \; ,
\\
M^{\mu\nu} = &-&h^{\rho\sigma}
\Big(\partial^\mu h_{\rho\tau} 
\partial^\nu h^\tau_\sigma+\partial_\tau h^\mu_\rho \partial^\tau
h^\nu_\sigma -\partial_\rho h^\mu_\tau \partial_\sigma h^{\nu \tau}\Big)
\nonumber \\ &+& h^{\mu\nu} \bigg[
-\frac{1}{4} \partial_\tau h_{\rho\sigma} \partial^\tau
h^{\rho\sigma}+\frac{1}{8} \partial_\rho h \partial^\rho h+
\frac{1}{2} \partial_\rho h_{\sigma\tau} \partial^\sigma h^{\rho\tau}
\bigg]+ \frac{1}{2} h^{\rho\sigma} \partial^{(\mu} h_{\rho\sigma}
\partial^{\nu)} h \nonumber \\ &+ & 2h^{\rho\sigma} \partial_\tau
h_\rho^{(\mu} \partial^{\nu)} h_\sigma^\tau+h^{\rho(\mu} \bigg[
\partial^{\nu)} h_{\sigma\tau}\partial_\rho h^{\sigma \tau}-2 \partial_\sigma
h_\tau^{\nu)} \partial_\rho h^{\sigma\tau}-\frac{1}{2} \partial^{\nu)}
h \partial_\rho h \bigg] \nonumber \\ &+&  \eta^{\mu\nu} \bigg[
\frac{1}{8} h^{\rho\sigma} \partial_\rho h \partial_\sigma
h-\frac{1}{4} h^{\rho\sigma} \partial_\tau h_{\rho \sigma}
\partial^\tau h -\frac{1}{4} h^{\tau\lambda}
\partial_\tau h_{\rho\sigma} \partial_\lambda h^{\rho\sigma}
\nonumber
\\ &-&\frac{1}{2} h^{\tau\lambda} \partial_\rho h_{\tau\sigma} 
\partial^\sigma h^\rho_\lambda+\frac{1}{2} h^{\tau\lambda}
\partial_\rho h^\sigma_\tau \partial^\rho h_{\lambda\sigma} \bigg] \; ,
\\
L^{\mu\nu} =&-&\frac{1}{2}
h^{\mu\nu} h_{\rho\sigma} 
\partial_\tau h^{\rho\lambda} \partial_\lambda h^{\sigma\tau}-
\frac{1}{4} h^{\mu\nu} h_{\rho\sigma} \partial^\rho h_{\tau\lambda}
\partial^\sigma h^{\tau\lambda}+\frac{1}{8} h^{\mu\nu} h_{\rho\sigma}
\partial^\rho h \partial^\sigma h \nonumber \\ &+& \frac{1}{2} h^{\mu\nu}
h_{\rho\sigma} 
\partial_\tau h^\rho_\lambda \partial^\tau h^{\sigma \lambda}-
\frac{1}{4} h^{\mu\nu} h_{\rho\sigma} \partial_\tau h^{\rho\sigma}
\partial^\tau h+h_{\rho\lambda} h^\lambda_\sigma \partial_\tau
h^{\mu\rho} \partial^\tau h^{\nu\sigma} \nonumber \\ &-& 2 h_{\rho\lambda}
h^\lambda_\sigma \partial_\tau h^{\rho(\mu} \partial^{\nu)}
h^{\sigma\tau}+ h_{\rho\lambda} h^\lambda_\sigma \partial^\mu
h^\rho_\tau \partial^\nu h^{\sigma\tau}-\frac{1}{2} h_{\rho\lambda}
h^\lambda_\sigma \partial^{(\mu} h^{\rho\sigma} \partial^{\nu)} h
\nonumber \\ &-&
h_{\rho\sigma} h_{\tau\lambda} \partial^\tau h^{\mu\rho}
\partial^\lambda h^{\nu\sigma}+\frac{1}{2} h_{\rho\sigma}
h_{\tau\lambda} \partial^\mu h^{\rho\tau} \partial^\nu
h^{\sigma\lambda}-
\frac{1}{4} h_{\rho\sigma} h_{\tau\lambda} \partial^\mu h^{\rho\sigma}
\partial^\nu h^{\tau \lambda} \nonumber \\ &+&2 h_{\sigma\tau} h^{\rho(\mu}
\partial_\lambda h^{\nu)\tau} \partial_\rho h^{\sigma\lambda}-
2 h_{\sigma\tau} h^{\rho(\mu} \partial^{\nu)} h^\sigma_\lambda
\partial_\rho h^{\tau\lambda}+\frac{1}{2} h_{\sigma\tau} h^{\rho(\mu}
\partial^{\nu)} h^{\sigma\tau} \partial_\rho h \nonumber \\ &+& \frac{1}{2}
h_{\sigma\tau} h^{\rho(\mu} \partial^{\nu)} h \partial_\rho
h^{\sigma\tau}+ \frac{1}{2} h^{\mu\rho} h^{\nu\sigma} \partial_\rho
h_{\tau\lambda} \partial_\sigma h^{\tau\lambda} -
\frac{1}{4} h^{\mu\rho} h^{\nu\sigma} \partial_\rho h \partial_\sigma
h \nonumber \\ &+&  \eta^{\mu\nu} \bigg[ \frac{1}{2} h_{\rho\pi}
h^\pi_\sigma \partial_\tau 
h^{\rho\lambda} \partial_\lambda h^{\sigma\tau}-\frac{1}{2}
h_{\rho\pi} h^\pi_\sigma \partial_\lambda h^\rho_\tau \partial^\lambda
h^{\sigma\tau}+\frac{1}{4} h_{\rho\pi} h^\pi_\sigma \partial_\tau
h^{\rho\sigma} \partial^\tau h \nonumber \\ 
&-&\frac{1}{4} h_{\rho\sigma} 
h_{\tau\lambda} \partial_\pi h^{\rho\tau} \partial^\pi
h^{\sigma\lambda}+\frac{1}{8} h_{\rho\sigma} h_{\tau\lambda}
\partial_\pi h^{\rho\sigma} \partial^\pi h^{\tau\lambda}+
\frac{1}{2} h_{\rho\sigma} h_{\tau\lambda} \partial^\rho h^\tau_\pi
\partial^\sigma h^{\lambda \pi} \nonumber \\ &-&\frac{1}{4}
h_{\rho\sigma} 
h_{\tau\lambda} \partial^\rho h^{\tau\lambda} \partial^\sigma h
\bigg] \; . 
\end{eqnarray}\end{mathletters}$\!\!$
All indices are lowered and raised with the Minkowski metric
$\eta_{\mu\nu}$; $h=\eta^{\mu\nu}h_{\mu\nu}$; 
the parenthesis around indices indicate the
symmetrization.

To describe the matter source we find convenient to introduce the
density of mass $\sigma$, of current $\sigma_i$ and of stress
$\sigma_{ij}$ defined by

\begin{mathletters}\label{III7}\begin{eqnarray}
&&\sigma c^2 = T^{00}+T^{ii}\;,\label{III7a}\\
&&\sigma_{i} c = T^{0i}\;, \\
&&\sigma_{ij} = T^{ij} \;,
\end{eqnarray}\end{mathletters}$\!\!$
(where $T^{ii}=\delta_{ij}T^{ij}$).
These definitions are such that $\sigma$, $\sigma_i$ and $\sigma_{ij}$ admit a
finite non-zero limit when $c\to +\infty$ (since $T^{\mu\nu}$ has the
dimension of an energy density). In the case of our model of point-particles
[stress-energy tensor given by (\ref{II12})], we obtain  

\begin{mathletters}\label{III8}\begin{eqnarray}
\sigma ({\bf x},t) &=& {\rm Pf}(\tilde{\mu}_1 \Delta_1) + 1\leftrightarrow 2\;,
\label{III8a}\\ 
\sigma_i ({\bf x},t) &=& {\rm Pf}(\mu_1 v_1^i \Delta_1)+ 1\leftrightarrow 2\;,
\\\sigma_{ij}({\bf x},t) &=& {\rm Pf}(\mu_1 v_1^i v_1^j \Delta_1)+
1\leftrightarrow 2 \;,
\end{eqnarray}\end{mathletters}$\!\!$
where $\Delta_1\equiv \Delta [{\bf x}-{\bf y}_1(t)]$, and where $\mu_1$ and
${\tilde \mu}_1$ represent some effective masses defined by  

\begin{mathletters}\label{III9}
\begin{eqnarray}
\mu_1({\bf x},t) &=& \frac{m_1 c}{\sqrt{-[g_{\rho\sigma}]_{\textstyle {}_1} v_1^\rho
v_1^\sigma}}~\!.~\!\frac{1}{\sqrt{-g({\bf x},t)}} \;,\label{III9a}\\ 
{\tilde\mu}_1({\bf x},t)&=&\mu_1({\bf x},t)\Bigg[1+\frac{{\bf
v}_1^2}{c^2}\Bigg]  \;.
\end{eqnarray}\end{mathletters}$\!\!$
Note that $\mu_1$ and ${\tilde \mu}_1$ depend on time {\it and}
space. Indeed, while the first factor in (\ref{III9a}) is clearly a
mere function of time through the values of the positions and
velocities of the particles at the instant $t$, the second factor
$(-g)^{-1/2}$ is evaluated at the {\it field} point $t,{\bf x}$
instead of the source point $t,{\bf y}_1$. From the non-distributivity
of the Hadamard regularization, one is not allowed to replace
$(-g)^{-1/2}$ by its (regularized) value at the point 1, even though
it is multiplied by a delta-pseudo-function at 1.

\subsection{The 3PN iteration of the metric}

In what follows we sketch the main steps of our iteration of the Einstein
field equations (\ref{III3})-(\ref{III6}) generated by two particles at the
3PN order. For more clarity in the presentation, we reason by induction over
the post-Newtonian order $n$. However, we do not have
proved the validity of this method to any order $n$; simply we applied the
method outlined below to construct the metric at the 3PN order. 

\medskip\noindent
(I) Suppose by induction over $n$ that we have succeeded in obtaining
some approximate post-Newtonian metric coefficients
$h^{\mu\nu}_{[2n-2]}$, as well as the previous coefficients
$h^{\mu\nu}_{[m]}$ for any $m$ such that $2\leq m\leq 2n-2$, which
approach the true metric modulo a small post-Newtonian remainder,

\begin{equation}\label{III10}
h^{\mu\nu}=h^{\mu\nu}_{[2n-2]}+{\cal O}(2n-1)\;,
\end{equation}
with the notation ${\cal O}(2n-1)={\cal O}(1/c^{2n-1})$, here and
elsewhere, for the post-Newtonian error terms. We assume that the
$h^{\mu\nu}_{[m]}$'s are at once some explicit functions of the field
point ${\bf x}$ and functionals of the two trajectories ${\bf
y}_{1}(t)$, ${\bf y}_{2}(t)$ and velocities ${\bf v}_{1}(t)$, ${\bf
v}_{2}(t)$. Since the matter source of the field equations is made of
delta-pseudo-functions, the metric coefficients become singular at the
location of the particles (indeed, this is already true at the
Newtonian order). As a matter of fact, we assume for the present
iteration that,
 
\begin{equation}\label{III11}
\forall m\leq 2n-2\;,\qquad 
h^{\mu\nu}_{[m]} \in {\cal F}\;,
\end{equation}
where ${\cal F}$ is the class of functions considered in \cite{BFreg,BFregM}
and Section \ref{II}. This is not a completely rigorous assumption because of
the presence of logarithms in the expansions around the singularities; but we
shall see that this assumption is justified at the 3PN order where one can
consider these logarithms as mere constants.

\medskip\noindent
(II) Consider for simplicity the combination $\frac{1}{2}(h^{00}+h^{ii})$
only, for which we need the maximal post-Newtonian precision since it is
directly connected to $g_{00}$. The structure of the Einstein field equations
(\ref{III3}), containing notably the gravitational source term
(\ref{III5})-(\ref{III6}), reads as 

\begin{equation}\label{III12}
\Box\left(\frac{h^{00}+h^{ii}}{2}\right)=\frac{8 \pi G}{c^2}|g|\sigma + \sum
h \dots h ~\!\partial h ~\!\partial h \;,
\end{equation}
where $\sigma$ is given by (\ref{III7a}), where the sum runs over
non-linearities and the two partial derivatives $\partial\partial$
have to be distributed among the $h$'s (with double derivatives
allowed in the quadratic term). In order to obtain an equation valid
at the next post-Newtonian order $n$, we replace the approximate
metric (\ref{III10}) into the right-hand side of (\ref{III12}).
Furthermore, we replace the partial derivatives $\partial\partial$ in
(\ref{III12}) by the {\it distributional} derivatives (\ref{II7}) [we
shall discuss the effect of using either the particular derivative
(\ref{II8}) of the more correct one (\ref{II9})]. Using also the
density of particles in the form (\ref{III8a}), we get
 
\begin{eqnarray}\label{III13}
\Box \left(\frac{h^{00}+h^{ii}}{2}\right)&=&\bigg\{\frac{8 \pi G}{c^2}\Big({\rm
Pf}~\!|g|{\tilde \mu}_1\Delta_1+{\rm Pf}~\!|g|{\tilde
\mu}_2\Delta_2\Big)\nonumber\\ 
&+&\!\!\!\!\sum_{m_1, \dots, m_p\leq 2n-2} h_{[m_1]}\dots h_{[m_{p-2}]}\partial
\left({\rm Pf}~\! h_{[m_{p-1}]}\right)\partial \left({\rm
Pf}~\! h_{[m_p]}\right)\bigg\}_{[2n]}\nonumber\\&+&{\cal O}(2n+1)\;, 
\end{eqnarray}
where the $h_{[m_1]}$, $\dots$, $h_{[m_p]}$ (with $2\leq p\leq n$) denote
the metric coefficients known from the previous iterations, and where
as indicated by the label $[2n]$ a truncation up to the
post-Newtonian order $1/c^{2n}$ is understood. At this stage, any
subsequent transformation of the right-hand side must be done using
the rules for handling the pseudo-functions and their derivatives
\cite{BFreg}.

\medskip\noindent
(III) We integrate the latter equation by means of the retarded integral given
by (\ref{III3'}):

\begin{eqnarray}\label{III14}
\frac{h^{00}+h^{ii}}{ 2}&=&\Box_{\cal R}^{-1}\biggl\{\frac{8 \pi
G}{c^2}\Big({\rm Pf}~\!|g|{\tilde \mu}_1\Delta_1+{\rm Pf}~\!|g|{\tilde
\mu}_2\Delta_2\Big)\nonumber\\ 
&+&\!\!\!\!\sum_{m_1, \dots, m_p\leq 2n-2} h_{[m_1]}\dots h_{[m_{p-2}]}\partial
\left({\rm Pf}~\!h_{[m_{p-1}]}\right)\partial \left({\rm
Pf}~\!h_{[m_p]}\right)\biggr\}_{[2n]}\nonumber\\&+&{\cal O}(2n+1) \;.
\end{eqnarray}
This defines the solution to the $n$th order, and so, by recursion, to
any order (in principle).  The partie-finie symbols ${\rm Pf}$ take
care of the divergences of the retarded integral at the locations of
the particles; that is, the retarded integral is considered as a
partie-finie integral in the sense of
\cite{BFreg}. More precisely, the retardations in (\ref{III14}) are expanded
to the $n$PN order and the resulting Poisson-like integrals computed
using the duality brackets in the way specified by Section V in
\cite{BFreg}. Actually, the Poisson-like integrals, which have a
non-compact support, become rapidly divergent at infinity when $n$
increases, and the correct solution we use is not the Poisson-like
integral but is obtained by a matching of the inner metric to the
multipole expansion of the exterior field. So, in fact,

\begin{eqnarray}\label{III15}
\Big(\Box_{\cal R}^{-1}{\rm Pf}F\Big)({\bf x}',t)&=&-\frac{1}{
4\pi}\sum_{k=0}^{2n}\frac{(-)^k}{k!c^{k}} < \left(\frac{\partial}{\partial
t}\right)^k\!\Big[{\rm Pf}F({\bf
x},t)\Big],|{\bf x}-{\bf x}'|^{k-1}>_{\rm match} \nonumber\\&+&{\cal O}(2n+1) \;,
\end{eqnarray}
where the subscript ``match'' refers to the matching process that is described
in Section {IV} in the case of the 3PN order. 
Notice that the time-derivatives $(\partial /\partial t)^k$ resulting from the
Taylor expansion of the retardations are distributional derivatives and
therefore can be put outside the duality bracket (see Section IX in
\cite{BFreg}). Thus, equivalently, 

\begin{eqnarray}\label{III15'}
\Big(\Box_{\cal R}^{-1}{\rm Pf}F\Big)({\bf x}',t)&=&-\frac{1}{
4\pi}\sum_{k=0}^{2n}\frac{(-)^k}{ k!c^{k}}\left(\frac{\partial}{ \partial
t}\right)^k \!\biggl[ < {\rm Pf}F({\bf x},t),|{\bf x}-{\bf
x}'|^{k-1}>_{\rm match}
\biggr]\nonumber\\&+&{\cal O}(2n+1) \;.
\end{eqnarray}

\medskip\noindent
(IV) Once the solution (\ref{III14}) to the $n$th post-Newtonian order is in
hands we perform many simplifications of the expression, following the rules
of application of the distributional derivative. In particular, we find very
useful to use the fact that a double gradient can be re-expressed in terms of
d'Alembertians as 

\begin{equation}\label{III16}
\partial_\mu F\partial^\mu G=\frac{1}{2}\left[\Box \left(FG\right)-F\Box
G-G\Box F\right] \;,
\end{equation}
which implies that the retarded integral reads as

\begin{equation}\label{III17}
\Box_{\cal R}^{-1}[\partial_\mu F\partial^\mu G]
=\frac{1}{2}FG-\frac{1}{2}\Box_{\cal R}^{-1}\left[F\Box G+G\Box F\right]\;.
\end{equation}
The first term is ``all-integrated'', while the second term, in which one can
replace the d'Alembertians by their corresponding sources, brings in
general many interesting cancellations with other terms. Unfortunately, the
formula (\ref{III16}) is valid only in an ordinary sense but not in the
distributional sense, because the distributional derivative does not satisfy,
in general, the Leibniz rule. Thus, in general, 

\begin{equation}\label{III18}
\partial_\mu ({\rm Pf}F)\partial^\mu ({\rm Pf}G)\not=\frac{1}{2}\left[\Box
\left({\rm Pf}FG\right)-F\Box ({\rm Pf}G)-G\Box ({\rm Pf}F)\right] \;.
\end{equation}
Nevertheless, the strategy we have chosen to follow in this paper is to take
advantage of the many simplifications brought about by the latter process, at
the price of introducing some extra terms (named ``Leibniz'') accounting for
the violation of the Leibniz rule. This means that we shall write, similarly
to (\ref{III17}), 

\begin{equation}\label{III19}
\Box_{\cal R}^{-1}[\partial_\mu ({\rm Pf}F)\partial^\mu ({\rm
Pf}G)]=\frac{1}{2}{\rm Pf}FG-\frac{1}{2}\Box_{\cal R}^{-1}\left[F\Box ({\rm
Pf}G)+G\Box ({\rm Pf}F)\right]+\delta_{\rm Leibniz}T \;,
\end{equation}
where the Leibniz term is given by

\begin{equation}\label{III20}
\delta_{\rm Leibniz}T=\Box_{\cal R}^{-1}\left[\partial_\mu ({\rm
Pf}F)\partial^\mu ({\rm Pf}G)-\frac{1}{2}\Box({\rm Pf}FG)+\frac{1}{2}F\Box
({\rm Pf}G)+\frac{1}{2}G\Box ({\rm Pf}F)\right] \;.
\end{equation}
Obviously the Leibniz term depends only on the purely distributional
part of the derivative. See the Appendix \ref{D} for the complete list
of the Leibniz terms.  As it will turn out these terms are not too
difficult to compute, and, of course, arise precisely at the 3PN
order. They give a contribution to the metric and the equations of
motion that we shall be able to check from the requirement of Lorentz
invariance (see Section \ref{VI}).

\subsection{The 3PN non-linear potentials}

The post-Newtonian iteration sketched in the previous subsection is
implemented to the 3PN order. The computation is long but straightforward.
After the simplification process described above we find that the metric is
parametrized by certain non-linear potentials, which do not carry a physical
signification by themselves, but turn out to be useful in the present
computation. The 3PN metric reads as 

\begin{mathletters}\label{III21}\begin{eqnarray}
g_{00} &=& -1+\frac{2}{c^2} V-\frac{2}{c^4} V^2+
\frac{8}{c^6} \left( \hat{X}+V_i V_i+\frac{V^3}{6} \right) \nonumber\\
& & +\frac{32}{c^8} \left( \hat{T}-\frac{1}{2} V \hat{X}+\hat{R}_i
V_i-\frac{1}{2} V V_i V_i-\frac{1}{48} V^4 \right) +{\cal O}(10) \; , \\
g_{0i} &=& -\frac{4}{c^3} V_i -\frac{8}{c^5} \hat{R}_i-\frac{16}{c^7}
\left( \hat{Y}_i+\frac{1}{2} \hat{W}_{ij} V_j+\frac{1}{2} V^2 V_i
\right) + {\cal O}(9) \; , \\
g_{ij} &=& \delta_{ij} \left[ 1+\frac{2}{c^2} V+\frac{2}{c^4} V^2+
\frac{8}{c^6} \left(\hat{X}+V_k V_k+\frac{V^3}{6} \right) \right]+
\frac{4}{c^4} \hat{W}_{ij} \nonumber \\ & &+\frac{16}{c^6} \left( \hat{Z}_{ij}
+\frac{1}{2} V \hat{W}_{ij}-V_i V_j \right)+{\cal O}(8) \; .
\end{eqnarray}\end{mathletters}$\!\!$
We recall our notation for the small post-Newtonian remainders: ${\cal
O}(n)={\cal O}(1/c^n)$. The various post-Newtonian orders are
parametrized by some potentials which are defined by means of the
retarded integral (\ref{III3'}).  At the ``Newtonian'' and 1PN orders
we pose

\begin{mathletters}\label{III22}\begin{eqnarray}
V&=&\Box^{-1}_{\cal R} [-4 \pi G \sigma] \;,\label{III22a}\\
V_i &=& \Box^{-1}_{\cal R} [-4 \pi G \sigma_i] \;,
\end{eqnarray}\end{mathletters}$\!\!$
in which the source densities were defined by (\ref{III7}). Next, at
the 2PN order, we define
 
\begin{mathletters}\label{III23}\begin{eqnarray}
\hat{X}\,\, &=&  \Box^{-1}_{\cal R} [
- 4 \pi G V \sigma_{ii}+\hat{W}_{ij} \partial_{ij} V+2 V_i
\partial_t \partial_i V+ V \partial_t^2 V+\frac{3}{2} (\partial_t
V)^2-2 \partial_i V_j \partial_j V_i]  \; ,\label{III23a}
\\
\hat{R}_i\, &=&  \Box^{-1}_{\cal R} [
-4 \pi G (V \sigma_i-V_i \sigma)-2 \partial_k V \partial_i
V_k-\frac{3}{2} \partial_t V \partial_i V] \; , 
\\
\hat{W}_{ij} &=& \Box^{-1}_{\cal R} [-4 \pi G
(\sigma_{ij}-\delta_{ij} \sigma_{kk})-\partial_i V \partial_j V]  \;.
\end{eqnarray}\end{mathletters}$\!\!$
Finally, at the 3PN order, we have

\begin{mathletters}\label{III24}\begin{eqnarray}
\hat{T}\,\, &=& \Box^{-1}_{\cal R} \biggl[ -4 \pi G \left( \frac{1}{4}
\sigma_{ij} \hat{W}_{ij}+\frac{1}{2} V^2 \sigma_{ii}+\sigma V_i V_i
\right)+\hat{Z}_{ij} \partial_{ij} V+\hat{R}_i \partial_t \partial_i
V -2 \partial_i V_j \partial_j \hat{R}_i  \nonumber \\ & &
  -\partial_i V_j \partial_t 
\hat{W}_{ij} + V V_i \partial_t \partial_i V+2 V_i \partial_j V_i
\partial_j V+\frac{3}{2} V_i \partial_t V \partial_i V+\frac{1}{2} V^2
\partial^2_t V \nonumber \\ & &   +\frac{3}{2} V (\partial_t
V)^2-\frac{1}{2} (\partial_t V_i)^2 \biggr]+\delta_{\rm Leibniz}
\hat{T}\; ,\label{III24a}
\\
\hat{Y}_i \, &=& \Box^{-1}_{\cal R} \biggl[ -4 \pi G \left( 
-\sigma \hat{R}_i-\sigma V V_i+\frac{1}{2} \sigma_k \hat{W}_{ik}+
\frac{1}{2} \sigma_{ik} V_k +\frac{1}{2} \sigma_{kk} V_i \right)
+\hat{W}_{kl} \partial_{kl} V_i-\partial_t \hat{W}_{ik} \partial_k
V  \nonumber \\ & &   + \partial_i \hat{W}_{kl}
\partial_k V_l -\partial_k \hat{W}_{il} 
\partial_l V_k -2 \partial_k V \partial_i \hat{R}_k  -\frac{3}{2} V_k
\partial_i V \partial_k V-\frac{3}{2} V \partial_t V \partial_i V
\nonumber \\ & &  -
2 V \partial_k V \partial_k V_i+ V \partial^2_t V_i+2 V_k \partial_k
\partial_t V_i \biggr] +\delta_{\rm Leibniz} \hat{Y}_i \; , \label{III24b}
\\
\hat{Z}_{ij}\! &=& \Box^{-1}_{\cal R} \biggl[ -4 \pi G V
(\sigma_{ij}-\delta_{ij} \sigma_{kk} )-2 \partial_{(i} V \partial_t
V_{j)}+ \partial_i V_k \partial_j V_k +\partial_k V_i \partial_k V_j-
2 \partial_{(i} V_k \partial_k V_{j)}  \nonumber \\ & &
  -\delta_{ij} \partial_k V_m 
(\partial_k V_m-\partial_m V_k)-\frac{3}{4} \delta_{ij} (\partial_t
V)^2 \biggr] +\delta_{\rm Leibniz} \hat{Z}_{ij} \; .
\end{eqnarray}\end{mathletters}$\!\!$
Note the presence in the 3PN potentials of the Leibniz contributions
described in the previous subsection, which are due to the
simplifications we did to arrive at these relatively simple
expressions (with respect to what could be expected at the high 3PN
order). The Leibniz contributions will be computed in Section
\ref{VI}. Of course, in the case where the matter source is continuous
--- an hydrodynamical fluid for instance ---, the 3PN metric
(\ref{III21}) and all the expressions of non-linear potentials are
valid with simply the Leibniz contributions set to zero.

The potentials (\ref{III22})-(\ref{III24}) are connected by the following
approximate post-Newtonian differential identities (equivalent to the
condition of harmonic coordinates at the 3PN order): 

\begin{mathletters}\label{III25}\begin{eqnarray}
&&\partial_t \left\{ V +\frac{1}{c^2} \left[ \frac{1}{2} \hat{W}_{kk}+ 2
V^2 \right]+\frac{4}{c^4} \left[ \hat{X} +\frac{1}{2}
\hat{Z}_{kk}+\frac{1}{2} V \hat{W}_{kk}+\frac{2}{3} V^3 \right] \right\}
\nonumber \\ 
&&\quad +\partial_i \left\{ V_i +\frac{2}{c^2} \left[\hat{R}_i+V V_i\right]
+\frac{4}{c^4} \left[\hat{Y}_i-\frac{1}{2} \hat{W}_{ij} V_j+\frac{1}{2}
\hat{W}_{kk} V_i+V \hat{R}_i +V^2 V_i\right] \right\}={\cal O}(6) \;,
\\ 
&&\partial_t \left\{ V_i+\frac{2}{c^2} \left[ \hat{R}_i+V V_i
\right] \right\}+\partial_j \left\{ \hat{W}_{ij}-\frac{1}{2}
\hat{W}_{kk} \delta_{ij} +\frac{4}{c^2} \left[\hat{Z}_{ij}-\frac{1}{2}
\hat{Z}_{kk} \delta_{ij}  \right] \right\} = {\cal O}(4) \; . 
\end{eqnarray}\end{mathletters}$\!\!$
We shall check that the (regularized) potentials we compute satisfy
these identities.  They are in turn respectively equivalent to the
equation of continuity at the 2PN order and the equation of motion at
the 1PN order:

\begin{mathletters}\label{III26}\begin{eqnarray}
\partial_t \Biggl[\sigma \Biggl(1+\frac{2\hat{W}_{ii}}{c^4}\Biggr)
\Biggr]&+& \partial_j \Biggl[\sigma_j
\Biggl(1+\frac{2\hat{W}_{ii}}{c^4}\Biggr) \Biggr] \nonumber\\
&=& \frac{1}{c^2}
\left(\partial_t \sigma_{jj} -\sigma \partial_t V\right)-\frac{4}{c^4}
\left(\sigma V_j \partial_j V+\sigma_{jk} \partial_j V_k\right)+ {\cal O}(6)\; , 
\\ 
\partial_t \left[\sigma_i \left(1+\frac{4V}{c^2} \right) \right] &+&
\partial_j \left[\sigma_{ij} \left(1+\frac{4V}{c^2} \right) \right]
\nonumber\\ 
&=&\sigma \partial_i V +\frac{4}{c^2} \left[\sigma
\partial_t V_i+\sigma_j \left(\partial_j V_i -\partial_i V_j\right)
\right]+{\cal O}(4) \; .  
\end{eqnarray}\end{mathletters}$\!\!$

\subsection{Computing the equations of motion}

The equations of motion of the particle 1 are deduced from the covariant
conservation of the stress-energy tensor of the particles, 

\begin{equation}\label{III27}
\nabla_\nu T^{\mu\nu}=0\;,
\end{equation}
where $T^{\mu\nu}$ is given by the definite expression (\ref{II12})
made of the delta-pseudo-functions defined in \cite{BFreg,BFregM}. It
is shown in Section V of \cite{BFregM} that by integrating
(\ref{III27}) over a volume surrounding the particle 1 (and only 1),
i.e. by constructing the duality bracket of (\ref{III27}) with the
characteristic function of that volume, we obtain the equations of
motion of the particle 1 in the form

\begin{equation}\label{III28}
\frac{d}{ dt}\left(\frac{[g_{\lambda\mu}]_{\textstyle {}_1} v_1^\mu}{
\sqrt{-[g_{\rho\sigma}]_{\textstyle {}_1} \frac{v_1^\rho
v_1^\sigma}{ c^2}}}\right)=\frac{1}{2} \frac{[\partial_\lambda
g_{\mu\nu}]_{\textstyle {}_1}v_1^\mu v_1^\nu}{
\sqrt{-[g_{\rho\sigma}]_{\textstyle {}_1} \frac{v_1^\rho v_1^\sigma}{ c^2}}}\;.
\end{equation}
These equations of motion take the same form as the geodesic equations
for a test particle moving on a smooth background, but with the role
of the background metric played by the true metric generated by the
two bodies and regularized according to the Lorentzian prescription
\cite{BFregM}.

In this paper we compute the spatial acceleration of body 1, which
corresponds to the equation with spatial index $\lambda=i$ in
(\ref{III28}); we do not consider the energy which would be given by
the equation with time index $\lambda=0$. Indeed, the energy of the
binary system will be determined directly from the (fully
order-reduced) acceleration. From (\ref{III28}) we can write the
equations into the form

\begin{equation}\label{III29}
\frac{d P_1^i}{ dt} = F_1^i \;,
\end{equation}
where the ``linear momentum density'' $P_1^i$ and ``force density'' $F_1^i$
are given by 

\begin{mathletters}\label{III30}\begin{eqnarray} 
P_1^i&=&\frac{[g_{i\mu}]_{\textstyle {}_1} v_1^\mu}{
\sqrt{-[g_{\rho\sigma}]_{\textstyle {}_1} \frac{v_1^\rho
v_1^\sigma}{ c^2}}}\;,\\
F_1^i&=&\frac{1}{2}\frac{[\partial_i
g_{\mu\nu}]_{\textstyle {}_1}v_1^\mu v_1^\nu}{
\sqrt{-[g_{\rho\sigma}]_{\textstyle {}_1} \frac{v_1^\rho v_1^\sigma}{ c^2}}}\;.
\end{eqnarray}\end{mathletters}$\!\!$ 
The expressions of both $P_1^i$ and $F_1^i$ in terms of the non-linear
potentials follow from insertion of the 3PN metric coefficients
(\ref{III21}).  We obtain some complicated sums of products of
potentials which are regularized at the point 1 following the
prescription $[F]_{\textstyle {}_1}$. Since the computation will turn
out to be quite involved, we decide to adopt the following
``step-by-step'' strategy:

\medskip\noindent
(A) We compute, in Sections \ref{IV} and \ref{V}, all the needed {\it
individual} potentials and their gradients at the point 1 following
the non-Lorentzian regularization $(F)_{\textstyle {}_1}$; for
instance we obtain $(\partial_iV)_{\textstyle {}_1}$ at the 3PN order,
$(V)_{\textstyle {}_1}$ at the 2PN order, $(\partial_i{\hat
X})_{\textstyle {}_1}$ at the 1PN order, $(\partial_i{\hat
T})_{\textstyle {}_1}$ at the Newtonian order, and so on. (Because of
the length of the formulas, and since the results for each of these
individual regularized potentials are only intermediate, we shall not
give them in this paper; see the appendices of \cite{Fthese} for
complete expressions.)

\medskip\noindent
(B) We add up the corrections brought about by the Lorentzian
regularization $[F]_{\textstyle {}_1}$ with respect to
$(F)_{\textstyle {}_1}$. We find, at the end of Section
\ref{V}, that the only effect of the new regularization at the 3PN order, when
computing the values of potentials at 1 (but the new regularization
affects also the corrections due to the non-distributivity), is a
crucial 1PN correction arising from the so-called ``cubic
non-compact'' part of ${\hat X}$; that is, we find $[\partial_i{\hat
X}^{\rm (CNC)}]_{\textstyle {}_1}-(\partial_i{\hat X}^{\rm
(CNC)})_{\textstyle {}_1}\not=0$.

\medskip\noindent
(C) We replace all the individually regularized potentials
$[F]_{\textstyle {}_1}$ and their gradients into the equations of
motion (\ref{III29})-(\ref{III30}) which would be obtained while
supposing that the Hadamard regularization is ``{\it distributive}''
with respect to the multiplication, i.e. supposing incorrectly that we
are allowed to write everywhere $[FG]_{\textstyle
{}_1}=[F]_{\textstyle {}_1}[G]_{\textstyle {}_1}$. In doing this we
obtain what we call the ``distributive'' parts of the linear momentum
and force densities (\ref{III30}), namely $(P^{i}_1)_{\textstyle
{}_{\rm distr}}$ and $(F^{i}_1)_{\textstyle {}_{\rm distr}}$. (Other
types of non-distributivity arising in the potentials themselves are
discussed in Section \ref{IV}.)

\medskip\noindent
(D) Finally, we compute separately, in Section VI, the corrections due
to the non-distributivity, i.e. the differences
$P^{i}_1-(P^{i}_1)_{\textstyle {}_{\rm distr}}$ and
$F^{i}_1-(F^{i}_1)_{\textstyle {}_{\rm distr}}$. Note that these
corrections reflect quantitatively the specific form that we have
adopted for the stress-energy tensor of point-particles
(\ref{II12}). Had we used another stress-energy tensor, for instance
by replacing incorrectly ${\rm Pf}(\frac{1}{\sqrt{-g}}\Delta_1)$ by
$[\frac{1}{\sqrt{-g}}]_{\textstyle {}_1}{\rm Pf}\Delta_1$ inside
(\ref{II12}), we would have obtained a different non-distributivity,
and thereby some different equations of motion. Note also that thanks
to the new regularization $[F]_{\textstyle {}_1}$ the corrections due
to the non-distributivity do not alter the Lorentz invariance of the
equations of motion. At last, we find the 3PN acceleration of body 1
as

\begin{equation}\label{III31}
a_1^i=F^{i}_1-\frac{d}{dt}\Big( P^{i}_1-v_1^i\Big) \;.
\end{equation}

We report now the expressions of the distributive parts of the linear
momentum and force densities as straightforwardly obtained by
substitution of the 3PN metric (\ref{III21}). The expressions of the
correcting terms due to the non-distributivity (i.e. $[FG]_{\textstyle
{}_1}\not=[F]_{\textstyle {}_1}[G]_{\textstyle {}_1}$) are relegated
to Section
\ref{VI}, where it is seen that they contribute only at the 3PN order. 

\begin{mathletters}\label{III32}\begin{eqnarray}
(P^{i}_1)_{\textstyle {}_{\rm distr}} &=& v_1^i +\frac{1}{c^2}
\left(\frac{1}{2} v_1^2v_1^i +3 
[V]_{\textstyle {}_1} v_1^i-4 [V_i]_{\textstyle {}_1} \right)
\nonumber \\ &+& \frac{1}{c^4} \left(\frac{3}{8} v_1^4
v_1^i+\frac{7}{2} [V]_{\textstyle {}_1} v_1^2 v_1^i-4
[V_j]_{\textstyle {}_1} v_1^i v_1^j -2 [V_i]_{\textstyle {}_1}
v_1^2\right.\nonumber
\\ & &  \left.\quad  
+\frac{9}{2} [V]_{\textstyle {}_1}^2 v_1^i-4 [V]_{\textstyle
{}_1} [V_i]_{\textstyle
{}_1} +4 [\hat{W}_{ij}]_{\textstyle {}_1} v_1^j-8
[\hat{R}_i]_{\textstyle {}_1} \right) \nonumber \\ &+ &
\frac{1}{c^6} \left(\frac{5}{16} v_1^6 v_1^i+\frac{33}{8} [V]_{\textstyle {}_1} v_1^4 
v_1^i-\frac{3}{2} [V_i]_{\textstyle {}_1} v_1^4 -6 [V_j]_{\textstyle
{}_1} v_1^i v_1^j v_1^2+ \frac{49}{4} [V]_{\textstyle {}_1}^2 v_1^2
v_1^i \right.\nonumber
\\ & &  \left.\quad +2 [\hat{W}_{ij}]_{\textstyle {}_1} v_1^j  v_1^2+2 
[\hat{W}_{jk}]_{\textstyle {}_1} v_1^i v_1^j v_1^k-10 [V]_{\textstyle
{}_1} [V_i]_{\textstyle {}_1} v_1^2-20 [V]_{\textstyle {}_1}
[V_j]_{\textstyle {}_1} v_1^i v_1^j \right.\nonumber \\ & &
\left.\quad -4 [\hat{R}_i]_{\textstyle {}_1} v_1^2 -8
[\hat{R}_j]_{\textstyle {}_1} v_1^i v_1^j+ \frac{9}{2} [V]_{\textstyle
{}_1}^3 v_1^i+12 [V_j]_{\textstyle {}_1} [V_j]_{\textstyle {}_1}
v_1^i\right.\nonumber \\ & & \left.\quad +12
[\hat{W}_{ij}]_{\textstyle {}_1} [V]_{\textstyle {}_1} v_1^j +12
[\hat{X}]_{\textstyle {}_1} v_1^i +16 [\hat{Z}_{ij}]_{\textstyle {}_1}
v_1^j -10 [V]_{\textstyle {}_1}^2 [V_i]_{\textstyle
{}_1}\right.\nonumber \\ & & \left.\quad -8 [\hat{W}_{ij}]_{\textstyle
{}_1} [V_j]_{\textstyle {}_1} -8 [V]_{\textstyle {}_1}
[\hat{R}_i]_{\textstyle {}_1}-16 [\hat{Y}_i]_{\textstyle {}_1}\right)
+ {\cal O}(8) \;,\label{III32a}
\\
(F^{i}_1)_{\textstyle {}_{\rm distr}} &=& [\partial_i V]_{\textstyle
{}_1}+\frac{1}{c^2}
\left( -[V]_{\textstyle
{}_1}[\partial_iV]_{\textstyle
{}_1}+\frac{3}{2} 
[\partial_i V]_{\textstyle {}_1} v_1^2-4 [\partial_i V_j]_{\textstyle
{}_1} v_1^j \right)\nonumber
\\  
&+&  \frac{1}{c^4} \left(\frac{7}{8} [\partial_i V]_{\textstyle {}_1} v_1^4-2 [\partial_i
V_j]_{\textstyle {}_1} v_1^j v_1^2  + \frac{9}{2} [V]_{\textstyle {}_1} [\partial_i
V]_{\textstyle {}_1} v_1^2\right.\nonumber \\ & &
 \left.\quad   +2 [\partial_i 
\hat{W}_{jk}]_{\textstyle {}_1} v_1^j v_1^k 
-4 [V_j]_{\textstyle {}_1} [\partial_i V]_{\textstyle {}_1} v_1^j-4
[V]_{\textstyle {}_1} [\partial V_j]_{\textstyle {}_1} v_1^j
\right.\nonumber \\ & &
\left.\quad -8 [\partial_i \hat{R}_j]_{\textstyle {}_1}
v_1^j+\frac{1}{2} [V]_{\textstyle {}_1}^2 [\partial_i V]_{\textstyle
{}_1}+8 [V_j]_{\textstyle {}_1} [\partial_i V_j]_{\textstyle {}_1}+4
[\partial_i \hat{X}]_{\textstyle {}_1} \right)
\nonumber \\  &+& \frac{1}{c^6} \left(\frac{11}{16} v_1^6 [\partial_i 
V]_{\textstyle {}_1}-\frac{3}{2} [\partial_i V_j]_{\textstyle {}_1}
v_1^j v_1^4+\frac{49}{8} [V]_{\textstyle {}_1} [\partial_i
V]_{\textstyle {}_1} v_1^4+ [\partial_i \hat{W}_{jk}]_{\textstyle
{}_1} v_1^2 v_1^j v_1^k \right.\nonumber \\ & & \left.\quad -10
[V_j]_{\textstyle {}_1} [\partial_i V]_{\textstyle {}_1} v_1^2
v_1^j-10 [V]_{\textstyle {}_1} [\partial_i V_j]_{\textstyle {}_1}
v_1^2 v_1^j-4 [\partial_i \hat{R}_j]_{\textstyle
{}_1} v_1^2 v_1^j\right.\nonumber \\ & &
\left.\quad +\frac{27}{4} [V]_{\textstyle {}_1}^2 [\partial_i V]_{\textstyle
{}_1} v_1^2
\quad + 12 [V_j]_{\textstyle {}_1} [\partial_i V_j]_{\textstyle {}_1}
v_1^2+6 [\hat{W}_{jk}]_{\textstyle {}_1} [\partial_i V]_{\textstyle
{}_1} v_1^j v_1^k\right.\nonumber \\ & & \left.\quad+6 [V]_{\textstyle
{}_1} [\partial_i \hat{W}_{jk}]_{\textstyle {}_1} v_1^j v_1^k+6
[\partial_i \hat{X}]_{\textstyle {}_1} v_1^2 +8 [\partial_i
\hat{Z}_{jk}]_{\textstyle {}_1} v_1^j v_1^k-20 [V_j]_{\textstyle {}_1}
[V]_{\textstyle {}_1} [\partial_i V]_{\textstyle {}_1} v_1^j
\right.\nonumber \\ & &
\left.\quad -10 [V]_{\textstyle {}_1}^2 [\partial_i V_j]_{\textstyle {}_1} v_1^j 
-8 [V_k]_{\textstyle {}_1} [\partial_i
\hat{W}_{jk}]_{\textstyle {}_1} v_1^j-8
[\hat{W}_{jk}]_{\textstyle {}_1} [\partial_i V_k]_{\textstyle {}_1}
v_1^j \right.\nonumber \\ & &
\left.\quad -8 [\hat{R}_j]_{\textstyle {}_1} [\partial_i
V]_{\textstyle {}_1} v_1^j -8 [V]_{\textstyle {}_1} [\partial_i
\hat{R}_j]_{\textstyle {}_1} v_1^j -16 [\partial_i
\hat{Y}_j]_{\textstyle {}_1} v_1^j -\frac{1}{6} [V]_{\textstyle {}_1}^3
[\partial_i V]_{\textstyle {}_1} \right.\nonumber \\ & & \left.\quad
-4 [V_j]_{\textstyle {}_1} [V_j]_{\textstyle {}_1} [\partial_i
V]_{\textstyle {}_1} +16 [\hat{R}_j]_{\textstyle {}_1} [\partial_i
V_j]_{\textstyle {}_1}+16 [V_j]_{\textstyle {}_1} [\partial_i
\hat{R}_j]_{\textstyle {}_1} -8 [V]_{\textstyle {}_1} [V_j]_{\textstyle {}_1} 
[\partial_i V_j]_{\textstyle {}_1} \right.\nonumber \\ & &
\left.\quad -4 [\hat{X}]_{\textstyle {}_1} [\partial_i V]_{\textstyle {}_1}
-4 [V]_{\textstyle {}_1} [\partial_i
\hat{X}]_{\textstyle {}_1} +16 [\partial_i \hat{T}]_{\textstyle {}_1}  
\right)+{\cal O}(8) \;.\label{III32b}
\end{eqnarray}\end{mathletters}$\!\!$ 
Recall that it is supposed that all the accelerations appearing in the
potentials have been order-reduced by means of the equations of
motion.  Notably, during the reduction of the ``Newtonian'' term
$[\partial_i V]_{\textstyle {}_1}$ in Section \ref{IV}, we shall need
the equations of motion to the 2PN order.  Furthermore, we see from
(\ref{III32a}) that when computing the time-derivative of $P_i$ we
meet an acceleration at 1PN which is thus also to be replaced by the
2PN equations of motion. We recall here that the latter 2PN (or,
rather, 2.5PN) equations in harmonic coordinates are
\cite{DD81a,Dthese,D82,BFP98}

\begin{eqnarray}\label{III32'}
\frac{dv_1^i}{ dt} = &-& \frac{Gm_2}{ r_{12}^2} n_{12}^i 
+ \frac{Gm_2}{ r_{12}^2c^2}\Biggl\{v_{12}^i \left[4(n_{12}v_1) -
3(n_{12}v_2)\right]  \nonumber\\
 &&\quad  +n_{12}^i \left[ -v^2_1 - 2v^2_2 + 4(v_1v_2)
  + \frac{3}{ 2} (n_{12}v_2)^2 + 5\frac{Gm_1}{r_{12}} + 4\frac{Gm_2}{ r_{12}}
\right]\Biggr\}\nonumber \\ 
  &+&\frac{Gm_2}{ r_{12}^2c^4} n_{12}^i \Biggl\{\left[ -2v^4_2 +
  4v^2_2 (v_1v_2) - 2(v_1v_2)^2 + \frac{3}{ 2} v^2_1 (n_{12}v_2)^2
  +\frac{9}{ 2} v^2_2 (n_{12}v_2)^2 \right.\nonumber\\ &&\quad
  \left. -6(v_1v_2) (n_{12}v_2)^2 - \frac{15}{ 8} (n_{12}v_2)^4
  \right] \nonumber\\ &&\quad +\frac{Gm_1}{ r_{12}}\left[ -\frac{15}{
  4} v^2_1 +\frac{5}{ 4} v^2_2 -\frac{5}{ 2} (v_1v_2)
  \right.\nonumber\\ &&\quad \left.+ \frac{39}{ 2} (n_{12}v_1)^2
  -39(n_{12}v_1)(n_{12}v_2)+\frac{17}{ 2}(n_{12}v_2)^2
  \right]\nonumber \\ &&\quad +\frac{Gm_2}{r_{12}}\left[ 4 v^2_2 -
  8(v_1v_2)+ 2(n_{12}v_1)^2 - 4(n_{12}v_1)(n_{12}v_2) -
  6(n_{12}v_2)^2\right]\nonumber \\ &&\quad
  +\frac{G^2}{r_{12}^2}\left[ -\frac{57}{ 4}m^2_1 - 9m^2_2 -
  \frac{69}{ 2} m_1m_2 \right]\Biggr\}\nonumber\\ &+&
  \frac{Gm_2}{r_{12}^2c^4}v_{12}^i \Biggl\{ v^2_1(n_{12}v_2) +
  4v^2_2(n_{12}v_1) -5v^2_2(n_{12}v_2) -4(v_1v_2)(n_{12}v_1)\nonumber
  \\ &&\quad + 4(v_1v_2)(n_{12}v_2) -6(n_{12}v_1)(n_{12}v_2)^2 +
  \frac{9}{ 2} (n_{12}v_2)^3 \nonumber \\ &&\quad +\frac{Gm_1}{
  r_{12}} \left[ -\frac{63}{ 4}(n_{12}v_1) + \frac{55}{ 4}
  (n_{12}v_2)\right] +\frac{Gm_2}{
  r_{12}}\left[-2(n_{12}v_1)-2(n_{12}v_2)\right] \nonumber\\
  &+&\frac{4G^2m_1m_2}{ 5c^5r_{12}^3}\Biggr\{ n_{12}^i (n_{12}v_{12})
\left[-6\frac{Gm_1}{ r_{12}}+\frac{52}{ 3}\frac{Gm_2}{  
r_{12}}+3v_{12}^2\right]\nonumber\\
&&\quad +v_{12}^i \left[2\frac{Gm_1}{ r_{12}}-8\frac{Gm_2}{ r_{12}}-
v_{12}^2\right]\Biggl\}+{\cal O}(6) \ . 
\end{eqnarray}
Unavoidably, because of the proliferation of possible terms, the
equations of motion at the next 3PN order are even much longer [see
(\ref{VII1}) below].

\section{Compact support and quadratic potentials} \label{IV}

All the potentials that enter the linear momentum $(P^i_1)_{\textstyle
{}_{\rm distr}}$ and the force density $(F^i_1)_{\textstyle {}_{\rm
distr}}$ are computed at the point 1 by means of the Lorentzian
regularization $[F]_{\textstyle {}_1}$. However, we shall first
determine their Hadamard partie finie in the usual sense
$(F)_{\textstyle {}_1}$, i.e. by approaching the singularity in the
spatial slice $t={\rm const}$. The difference between the two
regularization processes does not affect any compact or quadratic
potentials.

\subsection{Iterative computation of compact support potentials}\label{IVA}

In this paragraph, we are interested in the compact terms involved in
the equation of motion [see (\ref{III32a}) and
(\ref{III32b})]. According to our previous remark, it is sufficient to
evaluate them with the classical Hadamard prescription. We need
$(\partial_i V)_{\textstyle {}_1}$ up to 3PN order; $(V)_{\textstyle
{}_1}$, $(V_i)_{\textstyle {}_1}$, and $(\partial_i V_j)_{\textstyle
{}_1}$ at 2PN; $(\hat{W}^{\rm (C)}_{ij})_{\textstyle {}_1}$,
$(\partial_i \hat{W}^{\rm (C)}_{jk})_{\textstyle {}_1}$,
$(\hat{R}^{\rm (C)}_i)_{\textstyle {}_1}$, $(\partial_i \hat{R}^{\rm
(C)}_j)_{\textstyle {}_1}$ and $(\partial_i \hat{X}^{\rm
(C)})_{\textstyle {}_1}$ at 1PN.  The remaining contributions are
Newtonian: $(\hat{X}^{\rm (C)})_{\textstyle {}_1}$, $(\partial_i
\hat{T}^{\rm (C)})_{\textstyle {}_1}$, $(\hat{Y}_i^{\rm
(C)})_{\textstyle {}_1}$ and $(\partial_i \hat{Y}_j^{\rm
(C)})_{\textstyle {}_1}$. We follow the same classification and
nomenclature concerning the various parts of potentials --- compact,
non-compact, etc. --- as in Section II of \cite{BFP98}. The compact
(C) potentials are generated by sources with (spatially compact)
support limited to the particles; for instance, $V^{\rm (C)}=V$, and,
from (\ref{III24a}),

\begin{equation}\label{TC}
\hat{T}^{\rm (C)}\,\, = \Box^{-1}_{\cal R} \biggl[ -4 \pi G \left( \frac{1}{4}
\sigma_{ij} \hat{W}_{ij}+\frac{1}{2} V^2 \sigma_{ii}+\sigma V_i V_i
\right)\biggr]\;.
\end{equation}
Thus, by definition, the source $S({\bf x},t)$ of each compact
potential $P^{\rm (C)}$ is made of Dirac pseudo-functions, multiplied
by some functions of the class ${\cal F}$: $$ S({\bf x},t)= {\rm Pf}
(F \Delta_1) + {\rm Pf}(G \Delta_2) \; ,$$ with $F$, $G \in {\cal
F}$. As a result, it is in general possible to find an explicit
expression of $P^{\rm (C)}$ over the whole space (for any ${\bf
x}$). Besides, the expansion under the integration symbol of the
retardation of $S({\bf x}',t-|{\rm x}-{\rm x}'|/c)$ as $c$ goes to
infinity is perfectly licit, because the integrand has a compact
support:

\begin{eqnarray} \label{eq:PC} P^{\rm (C)} &=&-\frac{1}{4\pi}
\sum_{n=0}^{+\infty} \frac{(-)^n}{n! c^n} \partial_t^n \int \! d^3{\bf x}'
|{\bf x} - {\bf x}'|^{n-1} S({\bf x}',t) \nonumber \\ &=&
-\frac{1}{4\pi}\sum_{n=0}^{+\infty} \frac{(-)^{n}}{n! c^n}
\partial_t^n \big([F |{\bf x}-{\bf x}'|^{n-1}]_{\textstyle {}_1} + [G
|{\bf x}-{\bf x}'|^{n-1}]_{\textstyle {}_2}\big) \; .
\end{eqnarray}
The sources $S({\bf x},t)$ are supposed to be known at the current
order. This implies to proceed iteratively as explained in
\cite{BFP98}. The reader is referred to this paper for more
details. In short, we start from the $V$ and $V_i$ potentials, whose
sources do not depend on any other ones at the lowest order. Indeed,
we have $\Box V=-4\pi G {\rm Pf} (\tilde{\mu}_1
\Delta_1)+1\leftrightarrow 2$ (and similarly for $V_i$), where
$\tilde{\mu}_1=m_1+{\cal O}(2)$, as it follows from insertion of the
``Newtonian'' metric into the definition (\ref{III9}) of the effective
mass. Hence,
\begin{eqnarray} \label{eq:VN}
V &=& G~\int \! \frac{d^3{\bf x}'}{|{\bf x}-{\bf x}'|}~
\left\{ {\rm Pf} \left[\tilde{\mu}_1
\Delta_1\right] -
\frac{1}{c} \partial_t {\rm Pf}  \left[|{\bf x}-{\bf x}'|
\tilde{\mu}_1 \Delta_1\right] \right\} +1
\leftrightarrow 2 +{\cal O}(2) \nonumber \\ 
&=& \frac{G m_1}{r_1}-\frac{G}{c}\partial_t m_1+1 \leftrightarrow 2+{\cal O}(2)
= \frac{G m_1}{r_1}+\frac{G m_2}{r_2}+{\cal O}(2) \; ,
\end{eqnarray} 
To obtain the regularized metric (at the location of the first body,
say), we need the partie finie of the potential $V$ at point 1,
$(V)_{\textstyle {}_1}$.  Since we use Hadamard regularization, it is
simply given by the value of its non-singular part when ${\bf x}={\bf
y}_1$. Here, we find $(V)_{\textstyle {}_1}=G m_2/r_{12}+{\cal O}(2)$,
with the notation $r_{12}=|{\bf y}_1-{\bf y}_2|$.

The computation of more complicated compact terms necessitates the
knowledge of the effective masses $\mu_1$ and $\tilde{\mu}_1$ beyond
the Newtonian approximation. By substituting to $g_{\mu\nu}$ the
explicit 3PN expression (\ref{III21}) for the metric in the equations
(\ref{III9}), we get the general forms of both effective masses.  As
an example, $\tilde{\mu}_1$ at 2PN order reads

\begin{eqnarray}
\frac{\tilde{\mu}_1}{m_1} &=& 1+\frac{1}{c^2}
\left[-2V+[V]_{\textstyle {}_1}+\frac{3}{2} v_1^2 \right]\nonumber\\&+& 
\frac{1}{c^4} \left[-2 \hat{W}_{ii}+2V^2-2V [V]_{\textstyle {}_1}
+\frac{3}{2}[V]_{\textstyle {}_1}^2-[V^2]_{\textstyle {}_1} -3V
v_1^2+\frac{7}{2}[V]_{\textstyle {}_1} v_1^2-4 [V_i]_{\textstyle {}_1}
v_1^i+\frac{7}{8} v_1^4\right]
\nonumber\\&+&{\cal O}(6)\;,
\label{mu1tildfluide} 
\end{eqnarray}
where we are careful at distinguishing the potentials computed at the
field point ${\bf x}$ from those computed at the source point ${\bf
y}_1$, and where we take into account the non-distributivity of the
regularization ($\mu_1$ differs only by some numerical coefficients).
Thus, as emphasized in Section \ref{III}, $\mu_1$ and $\tilde{\mu}_1$
are functions of time \emph{and} space. Replacing them by the
regularized quantities $(\mu_1)_{\textstyle {}_1}$,
$(\tilde{\mu}_1)_{\textstyle {}_1}$ (and $1
\leftrightarrow 2$) is definitely forbidden because, on one side, 
the partie finie is not distributive, and, on the other side, the
usual Hadamard regularization does not coincide with the Lorentzian
one. However, this replacement does not modify any compact potentials,
with the notable exception of the 3PN contributions in $V$ (see
below). It is thus convenient to pose: $$V_{\rm distr} = \Box_{\cal
R}^{-1}
\left[-4\pi G (\tilde{\mu}_1)_{\textstyle {}_1} {\rm Pf} \delta_1
-4\pi G (\tilde{\mu}_2)_{\textstyle {}_2} {\rm Pf} \delta_2
\right] \; ,$$ 
and to calculate $V_{\rm distr}$ and $V-V_{\rm distr}$ separately.  In
the other compact sources, we shall employ $(\mu_1)_{\textstyle
{}_1}$, $(\tilde{\mu}_1)_{\textstyle {}_1}$, etc. instead of $\mu_1$
and $\tilde{\mu}_1$ for practical calculations at the 3PN
approximation.  Furthermore, in all the compact terms, the action of
the Lorentzian delta-pseudo-functions ${\rm Pf} \Delta_1$ and ${\rm
Pf} \Delta_2$, remarkably, reduces to the one of ${\rm Pf} \delta_1$
and ${\rm Pf} \delta_2$.  From what precedes, it becomes obvious that,
after the evaluation of $(\mu_1)_{\textstyle {}_1}$ or
$(\tilde{\mu}_1)_{\textstyle {}_1}$ at a given post-Newtonian order
$n$, we can determine all the potentials to the precision
$1/c^{2n}$. As all the terms involving the retarded potentials in
$\tilde{\mu}_1$ appear already with a factor $1/c^2$ at least, we are
then in a position to compute the right-hand-side of the equation
(\ref{mu1tildfluide}). The process is initiated by the computation of
the Newtonian value of $V$ as presented above. Most of the quantities
needed to get $(\tilde{\mu}_1)_{\textstyle {}_1}$ at the 3PN order are
obtained in
\cite{BFP98}. 
Finally, the regularized value of $\tilde{\mu}_1$ at point 1 is

\begin{align}
\frac{(\tilde{\mu}_1)_{\textstyle {}_1}}{m_1} &= 1+\frac{1}{c^2} \left[-
\frac{G m_2}{r_{12}}+\frac{3}{2} v_1^2 \right] +\frac{1}{c^4} \left[\frac{G
m_2}{r_{12}} \left(\frac{1}{2} v_1^2-4 (v_1v_2)+2 v_2^2+\frac{1}{2}
(n_{12}v_2)^2-\frac{1}{2} \frac{G m_1}{r_{12}}
\right. \right. \nonumber\\ &
\left. \left. \qquad \qquad \qquad \qquad \qquad \qquad
\qquad  \; +\frac{3}{2} \frac{G 
m_2}{r_{12}}\right) + \frac{7}{8} v_1^4 \right]+
\frac{8 G^2 m_1 m_2}{3 c^5 r_{12}^2} \left(
(n_{12}v_1)-(n_{12}v_2)\right) \nonumber\\ & +
\frac{1}{c^6} \left[ \frac{G^2 m_1 m_2}{r_{12}^2} \left(-\frac{3}{2}
\frac{G m_1}{r_{12}}-\frac{39}{4} \frac{G m_2}{r_{12}}+\frac{35}{8}
v_1^2-\frac{41}{4} (v_1v_2)+\frac{41}{8} v_2^2-\frac{9}{8}
(n_{12}v_1)^2 \right. \right. \nonumber\\ &  \left. \qquad \qquad
\qquad \qquad +\frac{25}{4}
(n_{12}v_1)(n_{12}v_2)-\frac{41}{8} 
(n_{12}v_2)^2 \right) \nonumber\\ & + \frac{G^2
m_2^2}{r_{12}^2} \left(\frac{3}{2} 
v_2^2-(n_{12}v_2)^2-3 (v_1 v_2)-(n_{12}v_1)(n_{12}v_2)+\frac{1}{2}
(n_{12}v_1)^2-\frac{3}{2} \frac{G m_2}{r_{12}}+\frac{15}{4}
v_1^2\right)\nonumber\\ &  +\frac{G m_2}{r_{12}} \left(2 (v_1v_2)^2+5
v_1^2 v_2^2-10 
(v_1v_2) v_1^2+\frac{33}{8} v_1^4+2 v_2^4-\frac{1}{2} (n_{12}v_2)^2
v_2^2-\frac{3}{8} (n_{12}v_2)^4 \right. \nonumber\\ & \left. \left. \qquad
\qquad -4 v_2^2 (v_1v_2)+2 (n_{12}v_2)^2 
(v_1v_2) -\frac{1}{4} v_1^2 (n_{12}v_2)^2 \right)+\frac{11}{16} v_1^6
\right]+{\cal O}(7) \; .
\end{align}
In our notation, two vectors ${\bf v}_1$, ${\bf v}_2$ between brackets
represent the scalar product: $(v_1 v_2)=v_1^i v_2^i$; $v_1^2=v_1^i
v_1^i$. We recall that it is important to keep the grouping of factors
imposed by the regularization in products of potentials.  For
instance: $(V \hat{W}_{ij})_{\textstyle {}_1} \neq (V)_{\textstyle
{}_1} (\hat{W}_{ij})_{\textstyle {}_1}$.

Among the compact potentials, the 3PN value of $V$ is certainly the
most difficult one to obtain, since the other quantities require only
lower orders in powers of $1/c$.  We shall focus on $V_{\rm distr}$ to
illustrate the method we have followed. The difference $V-V_{\rm
dist}$ will be handled in the next subsection. We begin with
specializing the general formula for $V$ to the case of $V_{\rm
distr}$:

\begin{eqnarray} V_{\rm distr} &=& \frac{G (\tilde{\mu}_1)_{\textstyle
{}_1}}{r_1}-\frac{G}{c} \partial_t (\tilde{\mu}_1)_{\textstyle {}_1}
+\frac{G}{2c^2} \partial^2_t \left[ (\tilde{\mu}_1)_{\textstyle {}_1} r_1
\right]- \frac{G}{6c^3} \partial_t^3 \left[(\tilde{\mu}_1)_{\textstyle {}_1}
r_1^2 \right]+ \frac{G}{24c^4} \partial_t^4 \left[(\tilde{\mu}_1)_{\textstyle
{}_1} r_1^3 \right]
\nonumber \\ & & - 
\frac{G}{120c^5} \partial_t^5 \left[(\tilde{\mu}_1)_{\textstyle {}_1} r_1^4
\right]+ \frac{G}{720c^6} \partial_t^6 \left[ (\tilde{\mu}_1)_{\textstyle
{}_1} r_1^5 \right]+ 1 \leftrightarrow 2 +{\cal O}(7)\;
. \label{expV3PN}
\end{eqnarray} 
Since the Schwarzschild mass $m_1$ is constant, $\partial_t
\tilde{\mu}_1/c$ is of order ${\cal O}(3)$ and does not contribute at
the 1PN level. For convenience, we shall introduce some special
notation for the terms that occur at this approximation; we pose: $$U
= \frac{G (\tilde{\mu}_1)_{\textstyle {}_1}}{r_1} + \frac{G
(\tilde{\mu}_2)_{\textstyle {}_2}}{r_2} \qquad {\rm and} \qquad X = G
(\tilde{\mu}_1)_{\textstyle {}_1} r_1 + G (\tilde{\mu}_2)_{\textstyle
{}_2} r_2\;.$$ Actually the potentials are to be considered as
pseudo-functions and it is understood that there is a symbol ${\rm
Pf}$ in front of them. Notably, the time derivatives appearing in
(\ref{expV3PN}) are distributional.  The regularized effective mass
$(\tilde{\mu}_1)_{\textstyle {}_1}$ as well as the distance to the
first body $r_1$ depend on time through the trajectories ${\bf
y}_{1,2}(t)$ and velocities ${\bf v}_{1,2}(t)$.  We explicit the
time-differentiations and obtain, at 1PN,

\begin{eqnarray} \label{eq:V1PN} 
V_{\rm distr} &=& U+\frac{1}{2c^2} \partial_t^2 X +{\cal O}(3) \\ &=&
\frac{G m_1}{r_1} \left[1+\frac{1}{c^2}\left( -\frac{G
m_2}{r_{12}}+\frac{3}{2} v_1^2 \right) \right]+ \frac{G m_1}{2c^2}
\left(-a_1^i \partial_i r_1+v_1^i v_1^j \partial^2_{ij} r_1 \right)+1
\leftrightarrow 2 +{\cal O}(3) \; .\nonumber
\end{eqnarray}
The accelerations are order-reduced by means of the equations of
motion at previous orders. Notably, for computing the 1PN term
$\frac{G}{2c^2}\partial_t^2 {\rm Pf} [(\tilde{\mu}_1)_{\textstyle
{}_1} r_1]$ at relative order 3PN, we need the 2PN acceleration given
by (\ref{III32'}).  Once we have got $V_{\rm distr}$ all over the
space, the last stage consists of regularizing it, as well as its
gradient, at ${\bf x}={\bf y}_1$ using the Hadamard partie finie. Now,
$V_{\rm distr}$ can be divided into two parts, $V_{{\rm distr}~\!{\bf
r}_1}$ and $V_{{\rm distr}~\!{\bf r}_2}$ corresponding to the sources
$-4\pi G (\tilde{\mu}_1)_{\textstyle {}_1} {\rm Pf} \delta_1$ and
$-4\pi G (\tilde{\mu}_2)_{\textstyle {}_2} {\rm Pf} \delta_2$
respectively. The first part, $V_{{\rm distr}~\!{\bf r}_1}$, depends
on ${\bf x}$ through ${\bf r}_1$ only, and contains many terms that
are either singular or vanish when ${\bf x} \to {\bf y}_1$, giving no
contribution to the partie finie; on the opposite, the smooth terms
with odd $1/c$-power factors in (\ref{expV3PN}) generally
contribute. The part $V_{{\rm distr}~\!{\bf r}_2}$ does not
necessitate any regularization since it is already regular in the
neighbourhood of ${\bf x}={\bf y}_1$.

The remaining potentials are determined in the same way. However, we
have to apply properly the formalism developed in \cite{BFreg}. In
particular:

\medskip \noindent (1)
The regularized value of some potential $P^{\rm (C)}$ is the partie
finie of $P^{\rm (C)}$ computed initially outside the singularity.  In
the case where $P^{\rm (C)}$ is the Poisson integral of a compact
source ${\rm Pf}( F \delta_1)$, with $F \in {\cal F}$, we must take
care \cite{BFreg} that

\begin{equation}\label{formule}
P^{\rm (C)}_1=\left({\rm Pf} \int \frac{d^3{\bf x}'}{-4 \pi}
\frac{1}{|{\bf x}-{\bf x}'|}~ F \delta_1 \right)_{\textstyle {}_1} \neq  
{\rm Pf} \int \! \frac{d^3{\bf x}}{-4 \pi}~\frac{1}{r_1} F \delta_1
\; . 
\end{equation}
[We generally do not write the dependence of the integrand on the
integration variable, as it is evident from the context; thus,
$F\delta_1$ is computed at point ${\bf x}'$ in the intermediate
expression of (\ref{formule}) and at point ${\bf x}$ in the last one.]

\medskip \noindent
(2) If $F$ is not regular at point 1, we generally have ${\rm Pf} (F
\delta_1) \neq (F)_{\textstyle {}_1} {\rm Pf} \delta_1$, even when both
members act, in the sense of pseudo-functions, on smooth test
functions. This distinction is crucial, for instance, in the
determination of $\hat{T}^{\rm (C)}$ at Newtonian order.  Indeed, one
of its contributions [first term in (\ref{TC})], denoted by
$\hat{T}^{\rm (C1)}$, reads as:
\begin{eqnarray*} \hat{T}^{\rm (C1) } &=& \frac{1}{4}G
~{\rm Pf} \int \! 
\frac{d^3{\bf x}'}{|{\bf x}-{\bf x}'|}~\sigma_{ij}   
\hat{W}_{ij}+{\cal O}(1)
\\ & =& \frac{1}{4} G m_1 v_1^i v_1^j
\left(\frac{\hat{W}_{ij}}{|{\bf x}-{\bf x}'|} 
\right)_{\textstyle {}_1} + 1 \leftrightarrow 2+{\cal O}(1) \; , 
\end{eqnarray*}
which is different from $ G m_1 v_1^i v_1^j (\hat{W}_{ij})_{\textstyle
{}_1}/4r_1+1 \leftrightarrow 2+{\cal O}(1) \; .$ Had we used the
latter expression instead of $\hat{T}^{\rm (C1)}$, we would have
obtained a different potential $\hat{T}_{\rm distr}$; this would have
been correct if the partie finie operation had been ``distributive''
(see Section \ref{I}), but we have actually $$\hat{T}-\hat{T}_{\rm
distr} = \frac{G^3 m_1^3}{r_1^3} \left[-\frac{1}{240} v_1^2 +
\frac{1}{80} (n_1 v_1)^2\right]+ 1 \leftrightarrow 2\; . $$ Notice
that the latter expression is not Galilean-invariant by itself, and
therefore will be checked later when verifying that the final
equations of motion stay invariant under Lorentz transformations.

\subsection{Non-distributivity in the potential $V$}

We call non-distributivity in the potential $V$ that contribution
which arises because the coefficient of the delta-pseudo-function
${\rm Pf}\Delta_1$ in the matter stress-energy tensor (\ref{II12}) is
a function not only on time but also on space through the factor
$1/\sqrt{-g}$. It will turn out that this contribution is purely of
order 3PN. A related contribution, due to the non-distributivity in
${\hat T}$, has just been computed in the previous subsection. The
potential $V$ is generated by the source density $\sigma ({\bf x},t) =
\tilde{\mu}_1 \Delta [{\bf x}-{\bf y}_1(t)] + 1\leftrightarrow 2$,
where ${\tilde \mu}_1$ is a function of space-time given explicitly by

\begin{equation}
{\tilde \mu}_1({\bf x},t)= \frac{m_1c\left(1+\frac{{\bf v}_1^2}{ c^2}\right)}{
\sqrt{-[g_{\rho\sigma}]_{\textstyle {}_1} v_1^\rho
v_1^\sigma}}~\!.~\!\frac{1}{ \sqrt{-g({\bf x},t)}}\;.
\end{equation}
The first factor is a function of time, and the second one depends on
both time {\it and} space (non-distributivity). The potential $V$ is
given by the retarded integral (\ref{III22a}), whose retardations we
expand up to any post-Newtonian order:

\begin{equation}
V({\bf x},t)=G \sum_{n=0}^{+\infty} \frac{(-)^n}{ n! c^n} \partial_t^n
\!\int d^3{\bf x}' ~|{\bf x}-{\bf x}'|^{n-1} \sigma({\bf x}',t)\;.
\end{equation}
(Actually we shall see that an expansion to the 1PN order is
sufficient for our purpose.) We insert into that expression the source
density $\sigma$, use the definition of the delta-pseudo-function
${\rm Pf}(F\Delta_1)$ given by (\ref{II11}), and arrive at

\begin{equation}
V({\bf x},t)=G \sum_{n=0}^{+\infty} \frac{(-)^n}{ n! c^n} \partial_t^n
\biggl[{\tilde \mu}_1({\bf x}',t) |{\bf x}-{\bf
x}'|^{n-1}\biggr]_{\textstyle {}_1}+1\leftrightarrow 2 \;.        
\end{equation}
Here, the square brackets refer to the Lorentzian Hadamard
regularization when ${\bf x}'\to {\bf y}_1$. Using a multipolar expansion, we
obtain immediately these brackets as 

\begin{equation}\label{mult}
\biggl[{\tilde \mu}_1({\bf x}',t) |{\bf x}-{\bf
x}'|^{n-1}\biggr]_{\textstyle {}_1}=\sum_{l=0}^{+\infty}\frac{(-)^l}{
l!}\partial_L\left(r_1^{n-1}\right)\left[{r'}_1^l {n'}_1^L {\tilde
\mu}'_1\right]_{\textstyle {}_1} \;,
\end{equation}
where ${\tilde \mu}'_1 \equiv {\tilde \mu}_1({\bf x},t)$. If ${\tilde
\mu}_1$ were a function of time only, then we see that all the
multipolar contributions in the right-hand side of (\ref{mult}) but
the scalar $l=0$ one would be zero, because of the factor ${r'}_1^l$
with $l\geq 1$ (this is clear with the old regularization, and easily
checked to be true with the Lorentzian regularization as well). We
defined $V_{\rm distr}$ as being $V$ but computed with the function of
time $[{\tilde \mu}_1]_{\textstyle {}_1}$ instead of the true ${\tilde
\mu}_1({\bf x},t)$. This $V_{\rm distr}$ is exactly the one which has
been computed in the subsection \ref{IVA}. [It can be checked that up
to the 3PN order $[{\tilde \mu}_1]_{\textstyle {}_1}=({\tilde
\mu}_1)_{\textstyle {}_1}$.] Therefore, by the previous argument,
$V_{\rm distr}$ is produced entirely by the scalar part $l=0$ in the
latter multipolar expansion, so that its complementary to the true
potential $V$ reads as

\begin{equation}\label{VVdistr}
V-V_{\rm distr}=G \sum_{n=0}^{+\infty} \frac{(-)^n}{ n! c^n} \frac{\partial^n}{
\partial t^n}\Biggl\{ \sum_{l=1}^{+\infty}\frac{(-)^l}{
l!}\partial_L\left(r_1^{n-1}\right)\left[{r'}_1^l {n'}_1^L {\tilde
\mu}'_1\right]_{\textstyle {}_1}\Biggr\}+1\leftrightarrow 2\;,         
\end{equation}
where the sum over $l$ starts with $l=1$. Thus, the problem reduces to
the computation of each regularization terms $[{r'}_1^l {n'}_1^L
{\tilde
\mu}'_1]_{\textstyle {}_1}$. Obviously, at a given post-Newtonian order, these
terms will all become zero for $l$ larger than a certain value. We
find that, up to the 3PN order, all the regularizations are zero
starting at $l=3$, namely $[{r'}_l {n'}_1^{L} {\tilde
\mu}'_1]_{\textstyle {}_1}={\cal O}(8)$ for any $l\geq 3$, while the
non-zero values for $l=1,2$ are given by

\begin{mathletters}\begin{eqnarray}
&&\left[{r'}_1 {n'}_1^i {\tilde \mu}'_1\right]_{\textstyle
{}_1}=3\frac{G^3m_1^3m_2}{ c^6r_{12}^2}n_{12}^i+{\cal O}(8)\;,\\
&&\left[{r'}_2 {n'}_1^{ij} {\tilde \mu}'_1\right]_{\textstyle
{}_1}=\frac{G^2m_1^3}{ c^4}\delta^{ij}- 3\frac{G^3m_1^3m_2}{
c^6r_{12}}\delta^{ij}+\frac{3}{ 2}\frac{G^2m_1^3}{
c^6}v_1^2\delta^{ij}-\frac{G^2m_1^3}{ c^6}v_1^iv_1^j+{\cal O}(8)\;.
\end{eqnarray}\end{mathletters}$\!\!$
Replacing these results back into (\ref{VVdistr}), and using the fact that
$\delta^{ij}\partial_{ij}\frac{1}{r_1}=0$, leads to the intermediate form 

\begin{eqnarray}
V-V_{\rm distr}&=&-3 \frac{G^4m_1^3m_2}{
c^6r_{12}^2}n_{12}^i\partial_i\left(\frac{1}{ r_1}\right)-\frac{G^3m_1^3}{
2c^6}v_1^iv_1^j\partial_{ij}\left(\frac{1}{ r_1}\right)+\frac{G^3m_1^3}{
2c^6}\frac{\partial^2}{ \partial t^2}\left(\frac{1}{ r_1}\right)\nonumber\\ 
&+&{\cal O}(8)+1\leftrightarrow 2\;.
\end{eqnarray}
As we see, the non-distributivity of the potential $V$ is a 3PN
effect.  Expanding the time derivative in the last term we find that
the dependence on the velocity $v_1^i$ cancels out, which is normal
because a velocity-dependent term would violate the Lorentz
invariance, in contradiction with our use of the Lorentzian
regularization $[F]_{\textstyle {}_1}$. The final expression is
simple:

\begin{equation}
V-V_{\rm distr}=-\frac{5}{ 2}\frac{G^4m_1^3m_2}{c^6r_{12}^2}n_{12}^i
\partial_i\left(\frac{1}{ r_1}\right)+{\cal O}(8)+1\leftrightarrow 2 \;.
\end{equation}
The contribution of the non-distributivity in the acceleration of 1 is
given by the gradient at 1 as

\begin{equation}
[\partial_iV]_{\textstyle {}_1}-[\partial_iV_{\rm distr}]_{\textstyle
{}_1}=5\frac{G^4m_1m_2^3}{ 
c^6r_{12}^5}n_{12}^i+{\cal O}(8) \;.
\end{equation}

\subsection{Computation of quadratic potentials $(\partial V \partial V)$}

By definition, the quadratic potentials are those whose sources are
made of products of two compact factors, like $V$, $V_i$, $W_{ij}^{\rm
(C)}$, etc. (or their derivatives, in most of the time).  A typical
source term for them is of the type ``$\partial V \partial V$'', hence
their denomination; for instance

\begin{equation}\label{Wijc}
\hat{W}_{ij}^{(\partial
V \partial V)} = \Box^{-1}_{\cal R} [-\partial_i V \partial_j V]\;.  
\end{equation}
But the quadratic source terms may also involve other quantities of
the same structure, as it is the case for $\partial_t
\hat{W}_{ik}^{\rm (C)} \partial_k V $ appearing in the source of the
potential $\hat{Y}_i^{(\partial V \partial V)}$ [{\it cf}
(\ref{III24b})].

\subsubsection{Matching to the external field} \label{par:matching}

The retardation of the compact potentials defining the metric of an
isolated fluid can be expanded in powers of $1/c$ only in the ``near
zone'' $D_{\rm near}$ of the source, at a distance much smaller than
the typical wave length of the emitted radiation. The question then is
how to incorporate in the post-Newtonian metric the no-incoming
radiation conditions at past null infinity. We achieve this by
performing a matching between the post-Newtonian expansion of the
metric, adequate in the near zone, and its multipole expansion, valid
in the region $D_{\rm ext}$ exterior to the compact support of the
source. Recall that for slowly moving sources, one can always choose
$D_{\rm near}$ and $D_{\rm ext}$ in such a way that their intersection
is not empty: $D_{\rm near} \cap D_{\rm ext} \neq
\emptyset$ (see e.g. \cite{Fock}). The field $h^{\mu\nu}$ admits 
a multipole-type expansion ${\cal M}(h^{\mu\nu})$, in the sense of
\cite{BD86}, at every spatial point ${\bf x} \in D_{\rm ext}$.  As a
matter of fact, it is shown in
\cite{B98mult} that the multipole expansion of the exterior field (a
vacuum solution of the field equations) that {\it matches}, according
to the theory of matched asymptotic expansions, to the post-Newtonian
expansion in the interior of the source, is given by

\begin{equation} {\cal M}(h^{\mu\nu})={\rm FP}_{B \to 0}
\Box_{\cal R}^{-1} 
\left[\left(\frac{r}{r_0}\right)^B {\cal M}(\Lambda^{\mu\nu})
\right]-\frac{4G}{c^4} \sum_{l=0}^{+\infty} \frac{(-)^l}{l!} \partial_L
\left\{\frac{1}{r} {\cal H}^{\mu\nu}_L(t-r/c)\right\}
\label{eq:matching} \end{equation}
(with $L$ a multi-index of order $l \in {\mathbb N}$). The multipole
moments ${\cal H}^{\mu\nu}_L$ entering the right-hand-side read as
$${\cal H}^{\mu\nu}_L(u)={\rm FP}_{B \to 0} \int \!  d^3{\bf y}~
\left(\frac{|{\bf y}|}{r_0} \right)^B y_L
\overline{\tau}^{\mu\nu}({\bf y},u) \; , $$
where $\overline{\tau}^{\mu\nu}$ represents the (formal) {\it
post-Newtonian} expansion of the complete source term $\tau^{\mu\nu}=
|g| T^{\mu\nu} + \frac{c^4}{16\pi G} \Lambda^{\mu\nu}$ of the field
equations (\ref{III3}). These expressions are defined by analytic
continuation in $B$, and the symbol ${\rm FP}_{B \to 0}$ denotes the
finite part when $B$ goes to zero of the Laurent expansion of the
analytic continuation (we refer to \cite{BD86,B98mult} for more
details about this finite part).

Let us show how we find the ``matched'' solution of the equation $\Box
P=S$ at the relative 1PN order (this is all we shall need in this
paper). We neglect all higher-order post-Newtonian contributions in
the source term $S$, and look for the solution of

\begin{equation}\label{eq}
\Box P=S^{1{\rm PN}}+{\cal O}(3)\;.
\end{equation}
Since the formula (\ref{eq:matching}) results from the properties of
the d'Alembertian operator (and is not specific to the field variable
$h^{\mu\nu}$), we can use it with the replacements of ${\cal
M}(\Lambda)$ by ${\cal M}(S^{1{\rm PN}})$ and of $\overline{\tau}$ by
$\frac{c^4}{16\pi G}S^{1{\rm PN}}$. Thus, the multipole expansion of
the solution must satisfy

\begin{eqnarray} {\cal M}(P) &=& {\rm FP}_{B \to 
0} \Box^{-1}_{\cal R} 
\left[\left(\frac{r}{r_0}\right)^B {\cal M}(S^{1{\rm PN}})
\right]-\frac{1}{4 \pi} \sum_{l=0}^{+\infty}
\frac{(-)^l}{l!} \partial_L 
\left\{\frac{1}{r} {\cal P}_L(t-r/c)\right\} \nonumber\\
&+&{\cal O}(3)\; , \label{eq:matching1PN} 
\end{eqnarray}
$${\rm with} \qquad \qquad
{\cal P}_L[S^{1{\rm PN}}](u)={\rm FP}_{B \to 0} \int \!
d^3{\bf y}~  
\left(\frac{|{\bf y}|}{r_0} \right)^B y_L~
S^{1{\rm PN}}({\bf y},u) \; . \qquad \qquad 
\qquad \qquad \; \qquad \;  \mbox{}$$
The partie-finie retarded integral of the multipole source ${\cal
M}(S^{1{\rm PN}})$ has to be handled with care. It is not licit to
develop when $c\to +\infty$ the integrand under the integration symbol
because the source is not compact supported. The correct formula was
shown in \cite{B93} to be:

\begin{eqnarray} & & {\rm FP}_{B\to 0} \Box_{\cal R}^{-1}
\left[\left(\frac{r}{r_0}\right)^B {\cal M}\left(S^{1{\rm PN}}\right)
\right]\qquad \qquad \qquad \qquad \qquad \qquad \qquad \qquad \qquad
\qquad \qquad \mbox{} \nonumber \\ & & \mbox{} \qquad \qquad
={\rm FP}_{B\to 0} \sum_{k=0}^{+\infty} \frac{1}{(2k)!}
\left(\frac{\partial}{c \partial t} \right)^{2k}
\int \! \frac{d^3{\bf x}'}{-4\pi} ~|{\bf x}-{\bf x}'|^{2k-1}
\left( \frac{r'}{r_0}\right)^B
{\cal M}\left(S^{1{\rm PN}}\right)
\nonumber \\  & & \mbox{} \qquad \quad \quad  
-\frac{1}{4\pi}\sum_{l \ge 0} \frac{(-)^l}{l!}\hat{\partial}_L   
\left\{
\frac{{\cal R}_L(t-r/c)-{\cal R}_L(t+r/c)}{2r}
\right\} \;.\label{eq:retarddvpt} 
\end{eqnarray}
The hat on the partial derivatives $\hat{\partial}_L$ indicates that
the trace has been removed, i.e. $\hat{\partial}_L={\rm STF}
(\partial_L)$. The ${\cal R}_L$ functions parametrize the general
solution of d'Alembert equations that are smooth near the origin:
``antisymmetric'' solution as given by the last term in
(\ref{eq:retarddvpt}). We have, more precisely,

\begin{eqnarray}  {\cal R}_L(u)&=&{\rm FP}_{B\to 0}  \int \!
d^3{\bf y}~ \hat{y}^L 
\left(\frac{|{\bf y}|}{r_0}\right)^B  T_{l}({\bf y},u) \; , \label{eq:RL} \\ 
 {\rm with} : \qquad T_{l}({\bf y},u)&=&(-)^{l+1}
\frac{(2l+1)!!}{2^l l!} 
\int_1^{+\infty}\! dz ~(z^2-1)^l {\cal M}
\left(S^{1{\rm PN}}\right)({\bf y},u-z|{\bf y}|/c) \;. 
\nonumber \end{eqnarray} Here, $\hat{y}^L$ denotes the symmetric
trace-free tensor associated with $y^{i_1} \dots y^{i_l}$, for $l \in
{\mathbb N}$. With (\ref{eq:matching1PN}) and (\ref{eq:retarddvpt}) we
can write the multipole expansion ${\cal M}(P)$ at the 1PN order as

\begin{eqnarray} {\cal M}(P) &=& {\rm FP}_{B\to 0} \int \! 
\frac{d^3{\bf x}'}{-4\pi}~\frac{1}{|{\bf x}-{\bf x}'|}
\left( \frac{r'}{r_0}\right)^B
{\cal M}\left(S^{1{\rm PN}}\right) -\frac{1}{4 \pi}
\sum_{l=0}^{+\infty}
\frac{(-)^l}{l!} \partial_L 
\left(\frac{1}{r}\right) {\cal P}_L(t)\nonumber\\
&+&\frac{1}{2c^2}\left(\frac{\partial}{\partial t} \right)^{2}\left\{
{\rm FP}_{B\to 0}\int \! \frac{d^3{\bf x}'}{-4\pi} ~|{\bf x}-{\bf x}'|
\left( \frac{r'}{r_0}\right)^B
{\cal M}\left(S^{1{\rm PN}}\right)-\frac{1}{4 \pi} \sum_{l=0}^{+\infty}
\frac{(-)^l}{l!} \partial_L 
\left({r}\right) {\cal P}_L(t)\right\}\nonumber\\
&+&\frac{1}{4\pi c}\left[{\dot {\cal R}}(t)+{\dot {\cal P}}(t)\right]+{\cal O}(3)
\end{eqnarray}
where the last term, of order $1/c$, is a simple function of time made
of the functions ${\cal R}(t)$ and ${\cal P}(t)$ defined as being
${\cal R}_L(t)$ and ${\cal P}_L(t)$ with $l=0$ (the dot indicates the
time-derivative).  Now, it can be shown that the latter
multipole expansion can be re-written under the new form

$${\cal M}(P) = {\cal M}(P^{\rm (I)}) + \frac{1}{4\pi c}\left[{\dot
{\cal R}}(t)+{\dot {\cal
P}}(t)\right]+\frac{1}{2c^2}\left(\frac{\partial}{\partial t}
\right)^{2}\left[{\cal M}(P^{\rm (II)})\right]+{\cal O}(3)\;,$$ or,
equivalently (indeed the second term is a mere function of time, and
the multipole expansion obviously commutes with the time derivative),

\begin{equation}\label{Pmatch} {\cal M}(P) = {\cal M}\Bigg(P^{\rm (I)}
+ \frac{1}{4\pi c}\left[{\dot {\cal R}}(t)+{\dot {\cal
P}}(t)\right]+\frac{1}{2c^2}\left(\frac{\partial}{\partial t}
\right)^{2}\left[P^{\rm (II)}\right]\Bigg)+{\cal O}(3)\;.
\end{equation}
In these equations, $P^{\rm (I)}$ and $P^{\rm (II)}$ denote the {\it
matched} solutions of the following Poisson equations

\begin{mathletters}\label{PIII}\begin{eqnarray}
\Delta P^{\rm (I)} &=& S^{1{\rm PN}} \;,  \label{PI}\\
\Delta P^{\rm (II)} &=&  2 P^{\rm (I)}\; . \label{PII}
\end{eqnarray}\end{mathletters}$\!\!$ 
Therefore, we have reduced the problem of finding the matched solution
of the d'Alembertian equation (\ref{eq}) to that of solving and
matching the two successive {\it Poisson} equations (\ref{PI}) and
(\ref{PII}). Now, from the equation (\ref{Pmatch}), it is evident that
the correct matched solution of (\ref{eq}) reads in terms of the
matched solutions of (\ref{PIII}) as
 
\begin{equation}\label{P}
P = P^{\rm (I)} + \frac{1}{4\pi c}\left[{\dot {\cal R}}(t)+{\dot {\cal
P}}(t)\right] +\frac{1}{2c^2}\left(\frac{\partial}{\partial t}
\right)^{2}\left[P^{\rm (II)}\right]+{\cal O}(3)\;.
\end{equation}
To recall the meaning of this solution we shall often denote it as
$P=P_{\rm match}$ below; similarly for $P=P_{\rm match}^{\rm (I)}$
(for instance $g_{\rm match}$ computed below) and $P=P_{\rm
match}^{\rm (II)}$ (e.g. $f_{\rm match}$). Actually, we shall find
that the function ${\cal R}(t)$ appearing in the $1/c$ term of our
solution (\ref{P}) is in fact always either zero or cancelled out by a
spatial gradient in the case of the applications made in the present
paper. Thus, it will not be considered in this paper, whereas the
function ${\cal P}(t)$ plays a role and is given by

\begin{equation}\label{Pt}
{\cal P}(t)={\rm Pf} \left\{ {\rm FP}_{B
\to 0} \int \!
d^3{\bf x}~\left(\frac{r}{r_0} \right)^B
S^{1{\rm PN}}({\bf x},t)\right\}
\end{equation}
(of course $S^{1{\rm PN}}$ there could be replaced with this
approximation by $S^{0.5{\rm PN}}$). See (\ref{example}) below for an
example of computation of this function.

In practice, in order to find the matched solution of a Poisson
equation, $P^{\rm (I)}$ for instance, we proceed as follows. Suppose
that we know a particular solution of the equation, say $P_{\rm
part}^{\rm (I)}$. Then the correct solution is necessarily of the type
$P^{\rm (I)}=P_{\rm part}^{\rm (I)}+h_{\rm hom}^{\rm (I)}$, where
$h_{\rm hom}^{\rm (I)}$ denotes an homogeneous solution of the Laplace
equation (harmonic function): $\Delta h_{\rm hom}^{\rm (I)}=0$, which
is moreover {\it regular} at the location of the source points. Note
that its multipole expansion coincides with itself, ${\cal M}(h_{\rm
hom}^{\rm (I)})=h_{\rm hom}^{\rm (I)}$. Now, the latter homogeneous
solution is determined by the matching equation as

\begin{eqnarray}\label{eq:matchingN}
h_{\rm hom}^{\rm (I)} &=& {\rm FP}_{B\to 0} \int \! \frac{d^3{\bf
x}'}{-4\pi} ~\frac{1}{|{\bf x}-{\bf x}'|}
\left( \frac{r'}{r_0}\right)^B
{\cal M}\left(S^{1{\rm PN}}\right) - \frac{1}{4 \pi} \sum_{l=0}^{+\infty}
\frac{(-)^l}{l!} \partial_L 
\left(\frac{1}{r}\right) {\cal P}_L(t)\nonumber\\
&-&{\cal M}(P_{\rm part}^{\rm (I)})\;.
\end{eqnarray}
It is not evident on that expression that the right-hand side is an
harmonic function; but it really is, as can be verified explicitly in
practice. We compute the multipole expansion of the source term
$S^{1{\rm PN}}$ as well as of our particular solution $P_{\rm
part}^{\rm (I)}$. In our case this means computing the formal
expansions of $S^{1{\rm PN}}$ and $P_{\rm part}^{\rm (I)}$ when $r$
tends to infinity or equivalently when the two source points ${\bf
y}_{1,2}$ tend to zero. The computation is greatly simplified if one
considers the dimensionality of the source. Suppose for instance that
$[S^{1{\rm PN}}]=({\rm length})^d$ which means $[h_{\rm hom}^{\rm
(I)}]=({\rm length})^{d+2}$. Then, using the fact that this function
is harmonic, its structure is necessarily of the type $h_{\rm
hom}^{\rm (I)}\sim \sum {\hat x}_L y_1^{L_1}y_2^{L_2}$ with
$l+l_1+l_2=d+2$. This shows that in order to obtain $h_{\rm hom}^{\rm
(I)}$ completely it is sufficient to develop the right side of
(\ref{eq:matchingN}) when ${\bf y}_{1,2}\to 0$ up to the order
$d+2$ included (i.e. to control all the terms $y_1^{L_1}y_2^{L_2}$ in
the expansions which have $l_1+l_2\leq d+2$). All the higher-order
terms, having $l_1+l_2\geq d+3$ in the right side of (\ref{eq:matchingN})
{\it must} manage to give zero. The same method is used to compute the
homogeneous solution $h_{\rm hom}^{\rm (II)}$ contained in $P^{\rm
(II)}$. We shall implement this method in practice below.

\subsubsection{Structure of the quadratic sources}

In the context of the present paper, we will not need to compute the
quadratic sources beyond the 1PN order. As a consequence, we will deal
with only a few kinds of elementary sources. By equation
(\ref{eq:V1PN}), we already know the structure of $V$ at the 1PN
approximation. The other compact retarded potentials have a very
similar form. After expansion of the retardation of any of them, say
$P^{\rm (C)}$, we get:

\begin{eqnarray} P^{\rm (C)}&=&{\rm Pf} \int \frac{d^3{\bf x}'}{-4 \pi}
\frac{1}{|{\bf x}-{\bf x}'|}~ F \delta_1-\frac{1}{c} 
\partial_t {\rm Pf} \int 
\frac{d^3{\bf x}}{-4 \pi}~ F \delta_1 \nonumber \\ &+& \frac{1}{2c^2}
\partial_t^2 {\rm Pf} \int 
\frac{d^3{\bf x}'}{-4 \pi} |{\bf x}-{\bf x}'| ~F \delta_1+1
\leftrightarrow 2 +{\cal O}(3) \; , \label{eq:dvptPC} \end{eqnarray} 
with $F \in {\cal F}$. The first contribution has been calculated in
\cite{BFreg} [see equation (6.18) there]. What is interesting for us is that the
result writes as a sum of space derivative of $1/r_1$ (or $1/r_2$),
i.e.  $\partial_L (1/r_1)$ (with the convention that $L$ designates a
multi-index of length $l$). Similarly, it is easy to convince oneself
that the third contribution is composed of terms $\partial_L r_1$ (or
$\partial_L r_2$). Moreover, the action of time derivatives in front
of the integral leaves the latter structure unchanged, in accordance
with formulas such as $\partial_t \partial_L r_1=-v_1^i
\partial_i \partial_L r_1$. 
The second contribution in (\ref{eq:dvptPC}) is a mere constant with
respect to ${\bf x}$. In fact, as they appear in the quadratic
sources, the compact potentials are preceded by some space or time
derivatives. Now, these derivations have to be performed in the sense
of pseudo-functions \cite{BFreg}. From these considerations, we are
now in a position to tell what is the precise structure of the sources
of quadratic potentials. They read as a sum of what we shall call
elementary terms. As we are interested here in their spatial behaviour
only, we shall omit purely time dependent factors, though they are
normally included. Newtonian elementary terms are themselves products
of two pseudo-function derivatives of contributions coming from the
first integral in the generic expression (\ref{eq:dvptPC}):
$\partial_J {\rm Pf} \partial_K (1/r_1) \times \partial_L {\rm Pf}
\partial_M (1/r_1)$, or $\partial_J {\rm Pf} \partial_K (1/r_1) \times
\partial_L {\rm Pf} \partial_M (1/r_2)$ (and similarly with $1\leftrightarrow 2$), 
where $J$, $K$, $L$, $M$ are
multi-indices of respective length $j$, $k$, $l$, $m$. In the same manner, the
1PN terms result from products of pseudo-function derivatives of 
Newtonian and post-Newtonian integrals as the first and the third ones in
(\ref{eq:dvptPC}): $\partial_J {\rm Pf} \partial_K 
(1/r_1) \times \partial_L {\rm Pf} \partial_M r_1$, or  
$\partial_J {\rm Pf} \partial_K (1/r_1) \times
\partial_L {\rm Pf} \partial_M r_2$, and $1\leftrightarrow 2$. 
As for the 0.5PN terms, they are simply the pseudo-function
derivatives of $\partial_L (1/r_1)$ or $\partial_L (1/r_2)$ (times a
mere function of time).  It is also natural to distinguish between the
``self'' elementary terms on one side, which depend on one body only,
e.g. $\partial_i {\rm Pf} (1/r_1)
\times \partial_j {\rm Pf} (1/r_1)$, and always admit pre-factors
$\mu_1^2$, $\tilde{\mu}_1^2$, $\tilde{\mu}_1 \mu_1$, and the
``interaction'' terms on the other side, involving both objects,
e.g. $\partial_i {\rm Pf} (1/r_1) \times
\partial_j {\rm Pf} (1/r_2)$. The 0.5PN terms $\partial_L (1/r_{1,2})$ 
are considered separately. 
 
To be more explicit, we shall provide as an example the 1PN source of
$\hat{W}_{ij}^{(\partial V \partial V)}$ defined by (\ref{Wijc}):

\begin{eqnarray} \label{eq:WijdVdVsource}
\Box \hat{W}_{ij}^{(\partial V \partial V)}
= &-&G^2 \tilde{\mu}_1^2 \partial_i {\rm Pf} \frac{1}{r_1}~
\partial_j {\rm Pf} \frac{1}{r_1} +\frac{G^2 m_1^2}{c^2} \left( 
a_1^k \partial_{(i} {\rm Pf} \frac{1}{r_1} ~\partial_{j)k} {\rm Pf}
r_1- v_1^k v_1^l \partial_{(i} {\rm Pf} \frac{1}{r_1}~
\partial_{j)kl} {\rm Pf} r_1  \right) \nonumber \\ &-& 
G^2 \tilde{\mu}_1 \tilde{\mu}_2  \partial_i {\rm Pf} \frac{1}{r_1}~
\partial_j {\rm Pf} \frac{1}{r_2} \\ 
&+& \frac{G^2 m_1 m_2}{c^2} \left( 
a_1^k \partial_{(i} {\rm Pf} \frac{1}{r_2}~ \partial_{j)k}
{\rm Pf} r_1 - v_1^k v_1^l \partial_{(i}
{\rm Pf} \frac{1}{r_2}~ 
\partial_{j)kl} {\rm Pf} r_1  \right)+1 \leftrightarrow 2 +{\cal O}(3) \; .
\nonumber \end{eqnarray}
Here, we have used the fact that ${\rm Pf}\partial_i r_1=\partial_i {\rm Pf} r_1$
and ${\rm Pf}\partial_{ij} r_1=\partial_{ij} {\rm Pf} r_1$. 

The sum of the retarded integral of the elementary terms then gives us
the complete quadratic potentials after expansion in $1/c$ and
matching. Therefore, these potentials are generated by the sources
through some partie-finie integrals, which can be regarded as the
result of the action of the elementary terms, considered as
pseudo-functions, on smooth quantities in the field point. By
inspection, it can be shown that the distributional part of the self
terms never contributes to the previous integrals, whereas the
partie-finie derivatives applied to the interaction terms coincide
with those of the Schwartz distribution theory.

\subsubsection{Integration of the elementary sources}

We now come to the solving of $\Box P=S$ at the 1PN order for each of
the elementary terms composing the quadratic sources.  We proceed
following the method we exposed at the end of paragraph
\ref{par:matching}. For this purpose, we first need to find a particular
solution of the following Poisson equations:

\begin{mathletters}  \label{eq:poisson}
\begin{eqnarray} \Delta P^{\rm (I)}_{\rm part}
&=& \partial_L {\rm Pf} \frac{1}{r_1}~ \partial_K {\rm Pf}
\frac{1}{r_1} \;, \qquad \Delta P^{\rm (I)}_{\rm part} =
\partial_L {\rm Pf} r_1 ~\partial_K {\rm Pf}
\frac{1}{r_1} \;, \label{eq:poissona} \\
{\rm and} \qquad \Delta P^{\rm (I)}_{\rm part} &=&
\partial_L {\rm Pf} \frac{1}{r_1}~ \partial_K {\rm Pf}
\frac{1}{r_2} \;, \qquad 
\Delta P^{\rm (I)}_{\rm part} =
\partial_L {\rm Pf} r_1 ~\partial_K {\rm Pf}
\frac{1}{r_2} \; , \label{eq:poissonb}
 \end{eqnarray}
\end{mathletters}$\!\!$
with $L=i_1 \dots i_l$ and $K=j_1 \dots j_k$.  From $P^{\rm (I)}_{\rm
part}$, we deduce the matched value $P^{\rm (I)}$ by computing $h_{\rm
hom}^{\rm (I)}$ according to the relation (\ref{eq:matchingN}) adapted
for each elementary terms.

The equation (\ref{eq:poissona}) involves only the vector variable
${\bf r}_1$, so that it is simple enough to be integrable in a
systematic way. To put the sources into a more suitable form, we start
by applying the derivative operator that enters the self terms in the
sense of functions, since the purely distributional part of the
derivative does not contribute. The result is an adequate power of
$r_1$ times a finite sum of partial terms $\delta^{i_1 i_2} \dots
\delta^{i_{2k-1} i_{2k}} n_1^{i_{2k+1}} \dots n_1^{i_l}$,
we shall denote more compactly as $\delta^{2K} n_1^{L-2K}$. The
solving of the Poisson equations rests then on the well-known
identities (easily checked by direct calculation):
\begin{eqnarray}
  r_1^a \hat{n}_1^L  &=&
\Delta \left\{ \frac{r_1^{a+2} 
\hat{n}_1^L}{(a-l+2)(a+l+3)}\right\} \quad {\rm for}~ a \in {\mathbb C}
\setminus \{l-2,-l-3\} \;,\nonumber  \\ 
 r_1^{l-2} \hat{n}_1^L  &=&
\Delta \left\{ \frac{1}{2l+1} 
\left[ \ln \left( \frac{r_1}{r_{1\,0}} \right)-\frac{1}{2l+1} \right]
r_1^l \hat{n}_1^L \right\} \;,\nonumber \\ r_1^{-l-3} \hat{n}_1^L &=&
\Delta \left\{ -\frac{1}{2l+1} 
\left[ \ln \left(\frac{r_1}{r_{1\,0}} \right)+\frac{1}{2l+1} \right]
\frac{\hat{n}_1^L}{r_1^{l+1}} \right\}  \;,
\label{eq:Matthieu} \end{eqnarray} 
where $\hat{n}_1^L$ is the trace-free part of $n_1^L$, and $r_{1\,0}$
a strictly positive constant. The quantities between braces are
particular solutions $P_{\rm part}^{\rm (I)}$ of $\Delta P^{\rm
(I)}=r_1^a \hat{n}_1^L$, and we must in general add to them some
harmonic functions to be evaluated by matching to the external
field. Equations (\ref{eq:poissonb}) are \emph{a priori} the most
difficult ones, because of the mixing of the sources 1 and 2. As a
matter of fact, determining $P^{\rm (I)}_{\rm part}$ amounts to
solving:

\begin{equation} \Delta g=\frac{1}{r_1 r_2} \qquad \qquad \Delta
{f}^{12}=\frac{r_1}{r_2} \;, \label{eq:gf12} \end{equation} in the
sense of distributions, on account of the fact that, for instance,

$$\partial_{L} {\rm Pf} \frac{1}{r_1}~ \partial_{K} {\rm Pf}
\frac{1}{r_2} = (-)^{k+l} \partial_{1L} \partial_{2K} {\rm Pf}
\frac{1}{r_1 r_2} \;  ,$$ 
where $\partial_{1L}$ and $\partial_{2K}$ denote the partial
derivatives with respect to ${\bf y}_1$ and ${\bf y}_2$ [the same
transformation applies to $\partial_{L} {\rm Pf} r_1 \times
\partial_{K} {\rm Pf} (1/r_2)$].  As a consequence, ${}_L g_K \equiv
\partial_{1L} \partial_{2K} g$ and ${}_L {f}^{12}_K \equiv
\partial_{1L} \partial_{2K} {f}^{12}$ clearly verify:
 
\begin{equation} \label{eq:poissongf12} \Delta \left[ (-)^{k+l} 
~{}_L g_K\right] = 
\partial_L {\rm Pf} \frac{1}{r_1} \partial_K {\rm Pf}
\frac{1}{r_2}\qquad {\rm and} \qquad 
\Delta \left[ (-)^{k+l} ~{}_L {f}^{12}_K \right]=
\partial_L {\rm Pf} {r_1} \partial_K {\rm Pf}
\frac{1}{r_2}  \; . \end{equation}
Note that the derivatives above should be understood as mere
(Schwartz) distributional derivatives.  Luckily, particular solutions
of equations (\ref{eq:gf12}) in the whole space can be exhibited
\cite{BDI95,Fock,DI91a}. We may take:

\begin{mathletters}\label{gf12}\begin{eqnarray}
g&=& \ln S  \; , \qquad \qquad {\rm with}~~
S=r_1+r_2+r_{12} \; ,\\ 
{f}^{12}&=& -\frac{1}{3} r_1 r_{12} (n_1n_{12})
\left(g-\frac{1}{3}\right)+\frac{1}{6} (r_2 r_{12}+r_1 
r_2-r_1 r_{12}) \; ,
\end{eqnarray}\end{mathletters}$\!\!$
where $(n_1n_{12})$ denotes the scalar product of Euclidean vectors.
The function $g$ is symmetric in its three variables ${\bf x}$, ${\bf
y}_1$ and ${\bf y}_2$, so that $$\Delta_1 g=\frac{1}{r_1 r_{12}}
\qquad {\rm and} \qquad
\Delta_2 g=\frac{1}{r_2 r_{12}} \; ,$$
in the sense of distributions. For a more complete list of useful
formulas, see \cite{BFP98}. We have also the two identities:
$$\Delta_1 {f}^{12}=2g \qquad {\rm and}
\qquad  \Delta_2 {f}^{12} =\frac{r_{12}}{r_2} \; .$$
Their proof is straightforward and call for some simple relations
permitting to express the scalar products $(n_1 n_2)$, $(n_1 n_{12})$
and $(n_2 n_{12})$ by means of some fractions involving $r_1$, $r_2$
and $r_{12}$. These relations, given by (5.14) in \cite{BFP98}, are
very convenient in most of our computations. The function $f^{12}$ is
obtained by exchanging ${\bf x}$ and ${\bf y}_1$ in the function $f$
which was introduced in the appendix B of \cite{BDI95},

\begin{equation}\label{f}
{f}= \frac{1}{3} r_1 r_{2} (n_1n_{2})
\left(g-\frac{1}{3}\right)+\frac{1}{6} (r_1 r_{12}+r_2 
r_{12}-r_1 r_{2}) \; ,
\end{equation}
and which satisfies, in the sense of distributions, the equation
$$\Delta f=2g\;.$$

Once Poisson equations with 1PN source are integrated, it remains to
find the homogeneous solutions to be added to get the full matched
solution. Most of the self terms are already correct, namely those
that go to zero when $r \to +\infty$. The other ones are determined
from the interaction terms by taking the limit ${\bf y}_2 \to {\bf
y}_1$, which happens to be always possible. The matching formula
(\ref{eq:matchingN}) provides the function $h_{\rm hom}^{\rm (I)}$
associated to $P_{\rm part}^{\rm (I)}=g$. The computation is very easy
because the dimension of the source is $[1/(r_1r_2)]=({\rm
length})^{-2}$ (i.e. $d=-2$ in the notation of the end of paragraph
\ref{par:matching}), therefore one needs to control only the constant
term in the right side of (\ref{eq:matchingN}) when ${\bf y}_{1,2}\to
0$. We arrive at $h_{\rm hom}^{\rm (I)}=-\ln (2r_0) -1$ hence the
correct $g_{\rm match}$ solution of $\Delta g_{\rm match}=1/r_1r_2$:

$$ g_{\rm match}= \ln \left( \frac{S}{2r_0}\right)-1 \; ,$$ where
$r_0$ is the positive constant occurring in (\ref{eq:matching}).
Similarly, but with a little more work because the dimensionality of
the source is now $d=0$ so we must expand to second order in ${\bf
y}_{1,2}$, we obtain the matched value corresponding to ${f}^{12}$ as

$$f^{12}_{\rm match}= -\frac{1}{3} r_1 r_{12} (n_1n_{12})
\left(g_{\rm match}-\frac{1}{3}\right)+\frac{1}{6} (r_2 r_{12}+r_1 
r_2-r_1 r_{12})-
\frac{1}{6} r (ny_1)-\frac{1}{6} (y_1 y_2)+\frac{1}{2} r (ny_2) $$
(where $(ny_1)$ for instance denotes the scalar product of ${\bf
n}={\bf x}/r$ with ${\bf y}_1$). As a consequence, the potentials
$P^{\rm (I)}$ satisfying (\ref{eq:poissonb}) are respectively
$(-)^{l+k} {}_L g_{{\rm match}~ K}$ and $(-)^{l+k} {}_L f^{12}_{{\rm
match}~K}$.  With this result in hands, we are able to deduce very
simply all the self terms that do not match properly yet. We shall
content ourselves with examining how this works on an example. Let us
suppose we want to solve:

\begin{equation} \label{eq:selfpoisson} 
\Delta P^{\rm (I)}=r_1 \partial_{ij} {\rm Pf}\frac{1}{r_1} \; .
\end{equation}
We make the correspondence with the equation $\Delta f^{12}_{{\rm
match}~ij}=r_1 \partial_{ij} {\rm Pf} (1/r_2)$, whose source coincides
with $r_1 \partial_{ij} {\rm Pf} (1/r_1)$ for ${\bf y}_1={\bf y}_2$
(we recall our notation $f^{12}_{{\rm
match}~ij}=\partial_{2ij}f^{12}_{\rm match}$). The distributional part
of the derivative yields a compact supported contribution to
${f}^{12}_{{\rm match} ~ij}$ given by

$$<{\rm Pf}~r_1 \left(\frac{-4 \pi}{3}\right) \delta^{ij} \delta_2
~,~\frac{1}{-4\pi}\frac{1}{|{\bf x}-{\bf x}'|}>= \frac{r_{12}}{3
r_2}\delta^{ij} \; ,$$ which is zero in the limit ${\bf y}_2\to {\bf
y}_1$, while the ordinary part yields a Poisson integral which is
well-defined and easily evaluated for ${\bf y}_1={\bf y}_2$. We
conclude that the value of $f^{12}_{{\rm match}~ij}$ when ${\bf
y}_2\to {\bf y}_1$,

$$P^{\rm (I)}=(f^{12}_{{\rm match}~ij})_{{\bf y}_2\to {\bf y}_1}=\frac{1}{6}
\delta^{ij} -\frac{1}{2} n_1^i n_1^j \;,$$  
\emph{is} precisely 
the matched solution of the Poisson equation (\ref{eq:selfpoisson}).

Let us complete now the program presented at the end of the paragraph
\ref{par:matching}. Because of the presence of time derivatives at the 1PN order,
we restore in the elementary terms all the coefficients depending only
on time, either through the trajectories ${\bf y}_{1,2}$ or the
velocities ${\bf v}_{1,2}$ [we shall generically call $\alpha (t)$
this time-dependent coefficient; for instance $\alpha=(v_1v_2)$].  It
is worth noting that the potentials assimilated to $P^{\rm (II)}$ in
(\ref{PII}) are needed only at the Newtonian order, as they come with
a $1/c^2$ factor. Consequently, the sources of the $P^{\rm (II)}$'s
are simply (two times) the matched solutions of the Poisson equations
$\Delta P^{\rm (I)}=S^{{\rm N}}+{\cal O}(1)$, where $S^{{\rm N}}$ are
the Newtonian-type sources $\alpha(t)~\! \partial_L {\rm Pf}
(1/r_{1,2})\times \partial_K {\rm Pf} (1/r_{1,2})$; so, all we have to
solve is:

\begin{mathletters} \label{eq:bipoisson}
\begin{eqnarray}
\Delta P^{\rm (II)}&=& 2 \alpha(t)~\! r_1^p \delta^{2K}
n_1^{L-2K} \; , \label{eq:bipoissona} \\
\hbox{or}\quad\Delta P^{\rm (II)} &=& 2 \alpha(t)~\! {}_L g_{{\rm match}~K}\; ,
\label{eq:bipoissonb}
\end{eqnarray}
\end{mathletters}$\!\!$
with $p \in {\mathbb Z}$, and $L$, $K$ some multi-indices. The
elementary self potentials obeying (\ref{eq:bipoissona}) are evaluated
by application of the identities (\ref{eq:Matthieu}), before matching
the full term (i.e. included the $\alpha$ coefficients) to the
external field. The latter stage will be dropped here, because, on one
hand, the general procedure has been explained before, and, on the
other hand, powerful methods permitting to deal with the trickiest
integrals one could encounter here will be expounded in Section
\ref{V}. Let us look next at the second equation (\ref{eq:bipoissonb}). To get
$P^{\rm (II)}$ from some particular $P_{\rm part}^{\rm (II)}$ one must
perform the complete matching including all the time-dependent factors
$\alpha (t)$. Here, for simplicity's sake, we give the result
in the case where the (Newtonian) source is $1/(r_1 r_2)$ hence
$P^{\rm (I)}=g_{\rm match}$ as we have seen before. Then, we have to
find the matched solution of

$$ \Delta {P}^{\rm (II)}= 2 g_{\rm match} \; .$$ A particular solution
$P_{\rm part}^{\rm (II)}$ of this equation is easily obtained with the
help of the function $f$ defined by (\ref{f}) [indeed, the Laplacian
of $P_{\rm part}^{\rm (II)}-f$ is a mere constant]. The corresponding
homogeneous solution $h_{\rm hom}^{\rm (II)}$ is computed using the
same equation as (\ref{eq:matchingN}) but using the source $2P^{\rm
(I)}\equiv 2g_{\rm match}$. The result, that we naturally call $f_{\rm
match}\equiv {P}^{\rm (II)}$, reads as

\begin{eqnarray}
 f_{\rm match} &=& \frac{1}{3} r_1 r_2 (n_1n_{2}) \left(g_{\rm match}
-\frac{1}{3}\right)- 
\frac{1}{6} (r_1 r_{12}+r_2 r_{12}-r_1 r_2) \nonumber \\ &-& \frac{1}{6} r
(ny_1)-\frac{1}{6} r 
(ny_2)+\frac{1}{2} (y_1y_2) \; .
\label{eq:fmatch} 
\end{eqnarray}
Notice that in the case of the source $1/(r_1 r_2)$ the only ``odd''
contribution $1/c$ in the formula (\ref{P}) is that given by the
function ${\cal P}(t)$ defined by (\ref{Pt}); the contribution due to
${\cal R}(t)$ is of higher order in this case. We readily find

\begin{equation}\label{example}
{\cal P}(t)={\rm FP}_{B
\to 0} \int \!
d^3{\bf x}~\left(\frac{r}{r_0} \right)^B
\frac{1}{r_1r_2}= -2\pi ~\!r_{12}
\end{equation}
(no need of the symbol ${\rm Pf}$). This calculation is also done in
equation (5.8) of \cite{BFP98}.  In Section \ref{V}, we shall see more
generally how such integrals can be obtained. Thus, our definitions of
$g_{\rm match}$ and $f_{\rm match}$ are such that

\begin{equation}
\left(\Box^{-1}_{\cal R}\frac{1}{r_1r_2}\right)_{\rm match}
=g_{\rm match}-\frac{1}{2c}{\dot r}_{12}+\frac{1}{2c^2} 
\partial_t^2 f_{\rm match}+{\cal O}(3)\;.
\end{equation}

Finally, we have all the material to integrate the individual
post-Newtonian terms in such a way that the inner metric matches to
the external field at the 3PN order. Let us remark however that, in
fact, the work we have done on the matching is, as seen {\it a
posteriori}, unnecessary. Indeed, summing up all the contributions in
the potentials, we find that, had we made use of some ``un-matched''
elementary functions, e.g. $g$ and ${f}^{12}$ defined by (\ref{gf12}),
to compute the interaction terms instead of the corresponding matched
quantities, and had we deduced jointly the corresponding self terms
from the limit ${\bf y}_2\to {\bf y}_1$, we would have arrived at the
{\it same} potentials up to the 3PN order. This means that the new
contributions brought about by the matching to the external field
actually cancel out in the final 3PN equations of motion. In
particular, the constant $r_0$ which enters into the matched
quantities $g_{\rm match}$, ${f}^{12}_{\rm match}$ and ${f}_{\rm
match}$ disappears from the final result. Though we have verified
this, we stick to our presentation and use systematically all the
matched functions determined previously.

To end this section, we shall achieve the example of the potential
$\hat{W}_{ij}^{(\partial V \partial V)}$ defined by (\ref{Wijc}). We
indeed already know its source from (\ref{eq:WijdVdVsource}). We split
the potential itself into:

\begin{equation} \label{eq:Wijsplit}
\hat{W}_{ij}^{(\partial V
\partial V)}=-U_{ij}-\frac{1}{c^2} K_{ij}+\frac{1}{c}
L_{ij}-\frac{1}{2c^2} \partial_t^2 X_{ij}  \; . \end{equation} 
The first two contributions are respectively the matched solutions of the
Poisson equations
$$\Delta U_{ij} =\partial_i U \partial_j U \qquad {\rm and} \qquad
\Delta K_{ij} =\partial_{(i} U \partial_{j)} \partial_t^2 X \; ,$$
which come from $\partial_i U \partial_j U +\frac{1}{c^2}\partial_{(i}
U \partial_{j)} \partial_t^2 X+{\cal O}(3)=$ $\partial_i V \partial_j
V$. Recall that the potentials $U$, $X$ and so on have to be viewed as
pseudo-functions (for instance $U = {\rm Pf} G
(\tilde{\mu}_1)_{\textstyle {}_1}/r_1 +1\leftrightarrow 2$), so the
derivatives entering the source terms are distributional derivatives.
The self terms can be determined with the help of the relations
(\ref{eq:Matthieu}) and matching. To get the interaction part, we
change the spatial derivatives to partial derivatives with respect to
the source points ${\bf y}_{1,2}$, and next, we make the replacement
$1/(r_1 r_2) \to g_{\rm match}$, $r_1/r_2 \to f^{12}_{\rm match}$, and
$r_2/r_1 \to f^{21}_{\rm match}$. The ``odd'' term $L_{ij}$ is a pure
function of time given by

$$L_{ij}=\partial_t \int \frac{d^3{\bf x}}{-4\pi}~ \partial_i U
\partial_j U \; ,$$ which is already known from equation (5.9) in
\cite{BFP98}; it can also be computed with the methods of Section
\ref{V}. The contribution $X_{ij}$ is the matched solution of the
double-Poisson equation $$\Delta^2 X_{ij}=2 \partial_i U \partial_j U
\; .$$ whose source is to be considered at the Newtonian order
only. The iterative application of (\ref{eq:Matthieu}) plus matching
yields the self terms; for interaction terms, we replace $\partial_L
(1/r_1) \partial_K (1/r_2)$ by $(-)^{l+k}{}_L~\!f_{{\rm match} \;
K}$. The results are

\begin{mathletters} \label{eq:WijdVdV}
\begin{eqnarray}
U_{ij} &=&
\frac{G^2 \tilde{\mu}_1^2}{8} \left(\partial^2_{ij} \ln
r_1+\frac{\delta^{ij}}{r_1^2} \right)  +
 G^2 \tilde{\mu}_1 \tilde{\mu}_2 ~{}_i g_{{\rm match}\; j}\;, \\ 
K_{ij} &=&
G^2 \tilde{\mu}_1^2 \left[ -\frac{a_1^{(i}}{4}  \partial_{j)}\ln r_1 +\
\frac{a_1^k}{8} \delta^{ij}  \partial_k \ln r_1
-\frac{a_1^k}{48}  \partial_{ijk} \left(r_1^2 \ln r_1 \right)
+\frac{v_1^2 \delta^{ij}}{16 r_1^2}
+\frac{v_1^i v_1^i}{8 r_1^2}  -\frac{v_1^k v_1^l}{16} \delta^{ij}
\partial^2_{kl} \ln r_1 
\right. \nonumber 
\\ & &\left. \qquad +\frac{v_1^2}{16} \partial_{ij} \ln r_1
+\frac{v_1^k v_1^l}{96}  \partial_{ijkl} \left(r_1^2 \ln r_1 \right)
\right] \nonumber \\ &+&
 G^2 \tilde{\mu}_1 \tilde{\mu}_2  \left[a_1^k ~{}_{k(i}
f^{12}_{{\rm match} \; j)}+ v_1^k v_1^l ~{}_{kl(i}
f^{12}_{{\rm match}\; j)} \right] + 1 \leftrightarrow 2  \; ,\label{Kij}
\\ 
L_{ij}&=&
G^2 \partial_t \left[\tilde{\mu}_1^2 {\rm Pf} \! \int \!
\frac{d^3{\bf x}}{-4\pi} 
\partial_i \frac{1}{r_1} \partial_j \frac{1}{r_1} \right]+
G^2 \partial_t \left[ \tilde{\mu}_1 \tilde{\mu}_2 \partial_{1i} \partial_{2j
}{\rm Pf} \! \int \!
\frac{d^3{\bf x}}{-4\pi} \frac{1}{r_1 r_2}\right]+1 \leftrightarrow 2 +{\cal O}(2)
\nonumber \\ &=& G^2 \tilde{\mu}_1 \tilde{\mu}_2 \partial_t \partial_{1i}
\partial_{2j} \frac{r_{12}}{2}+ 1 \leftrightarrow 2 +{\cal O}(2) \; ,
\\
X_{ij} &=&
\frac{G^2\tilde{\mu}_1^2}{4}  \left[\frac{1}{6} \partial_{ij}
\left(r_1^2 \ln r_1 \right) + \delta^{ij} \ln r_1
 \right]+G^2 \tilde{\mu}_1 \tilde{\mu}_2
~{}_i f_{{\rm match} \; j}+ 1 \leftrightarrow 2+{\rm const} \; .
\end{eqnarray}
\end{mathletters}$\!\!$
In the last equation we do not write for simplicity a constant
(associated with a function of type ${\cal R}$) which is cancelled out
by the time derivative $\partial_t^2$ in front of that term.  The self
terms have been written in the form of some (ordinary) space
derivatives in order to prepare the computation of the cubic sources.

\section{Cubic potentials}\label{V}

\subsection{methodological scheme}

For methodological reasons, it is convenient to express all the cubic
sources in a similar way, with the help of the same set of elementary
integrals.  The so-called ``cubic-non-compact'' term
\begin{equation} \label{eq:DXcNCCgen}
\hat{X}^{\rm (CNC)}=\Box^{-1}_{\cal R} ~\left\{\hat{W}_{ij}^{(\partial
V \partial V)} \partial_{ij} V \right\} \;,
\end{equation}
which is part of the ${\hat X}$-potential [see (\ref{III23a})], is a
good example to understand the successive transformation operations we
perform in practice. Furthermore, this cubic-non-compact term is the
only one we need to compute at the relative 1PN order; all the other
ones, which enter into ${\hat T}$ and ${\hat Y}_i$, are merely
Newtonian. So the practical computation of (\ref{eq:DXcNCCgen}) is the
most difficult one we face at the 3PN approximation. Recall that
$\hat{W}_{ij}^{(\partial V \partial V)}$ was defined by (\ref{Wijc}).
We start from the expression of the source of $\hat{X}^{\rm (CNC)}$
obtained by insertion of (\ref{eq:V1PN}) and (\ref{eq:Wijsplit}) into
(\ref{eq:DXcNCCgen}). We get:

\begin{eqnarray}
\Box \hat{X}^{\rm (CNC)} = 
&-&U_{ij} \partial_{ij} U+\frac{1}{c} L_{ij} \partial_{ij} U-
\frac{1}{c^2} K_{ij} \partial_{ij} U \nonumber \\ &-&
\frac{1}{2c^2} \left[ \partial^2_t X_{ij} \partial_{ij} U+
U_{ij}  \partial_{ij} \partial^2_t X \right]+{\cal O}(3) \; , \label{eq:DXcNCC}
\end{eqnarray}
using the notation introduced in (\ref{eq:Wijsplit}). In the right
side, the potentials are seen as pseudo-functions (involving a ${\rm
Pf}$) and the derivatives are distributional. After carrying on the
expansion of retardations up to the 1PN approximation, we find:
\begin{eqnarray}
\hat{X}^{\rm (CNC)} &=& \int \frac{d^3{\bf x}'}{4 \pi} 
\frac{1}{|{\bf x}-{\bf x}'|} ~U_{ij} \partial_{ij} U-
\frac{1}{c} \int \frac{d^3{\bf x}'}{4 \pi} 
\frac{1}{|{\bf x}-{\bf x}'|} ~L_{ij} \partial_{ij} U \nonumber
\\ &-&
\frac{1}{c} \partial_t \int \frac{d^3{\bf x}'}{4 \pi} ~U_{ij}
\partial_{ij} U 
+ \frac{1}{c^2} \int \frac{d^3{\bf x}'}{4 \pi}
\frac{1}{|{\bf x}-{\bf x}'|} ~K_{ij} \partial_{ij} U \nonumber
\\ &+&
\frac{1}{2c^2} \int \frac{d^3{\bf x}'}{4 \pi}
\frac{1}{|{\bf x}-{\bf x}'|} ~\partial^2_t X_{ij} \partial_{ij}
U  +
\frac{1}{2c^2} \int \frac{d^3{\bf x}'}{4 \pi}
\frac{1}{|{\bf x} - {\bf x}'|} ~U_{ij}
\partial_{ij} \partial^2_t X \nonumber
\\ &+&
\frac{1}{2c^2} \partial^2_t \int \frac{d^3{\bf x}'}{4 \pi} 
|{\bf x}-{\bf x}'| ~U_{ij} \partial_{ij} U +
\frac{1}{c^2} \partial_t \int \frac{d^3{\bf x}'}{4 \pi} ~
 L_{ij} 
\partial_{ij} U+{\cal O}(3) \; . \label{eq:XcNCC}
\end{eqnarray}
We have checked explicitly that the sum of the integrals occuring in
this formula yields an integral convergent at infinity when
considering the regularized value of the gradient
$(\partial_i\hat{X}^{\rm (CNC)})_{\textstyle {}_1}$ which is the only
thing required; thus we do not need to introduce a finite part at
infinity (but of course the regularization ${\rm Pf}$ is needed to
cure the point-particle singularities). The next step consists of
replacing the potentials $U$, $X$, $L_{ij}$, $K_{ij}$ and $X_{ij}$
given by (\ref{eq:WijdVdV}) above by their values at the field point
${\bf x} \neq {\bf y}_1$ and ${\bf x}
\neq {\bf y}_2$. The spatial and time derivations appearing in 
each of the integrals of (\ref{eq:XcNCC}) are to be understood in the
sense of pseudo-functions (see Section
\ref{II}). Consider, as an example, the term $$ \int \frac{d^3{\bf
x}'}{-4 \pi}
\frac{1}{|{\bf x}-{\bf x}'|} ~K_{ij} \partial_{ij} U \; . $$
Remind that $K_{ij}$ is given by (\ref{Kij}). Let us multiply (\ref{Kij})
by $\partial_{ij}U=\partial_{ij} {\rm Pf} (G \tilde{\mu}_1/r_1+G
\tilde{\mu}_2/r_2)$ and develop the product. The result is made of a sum of
terms of the type 
$(1/r_1^2)\times\partial_{ij} {\rm Pf} (1/r_1)$, $\partial_k \ln r_1
\times\partial_{ij} {\rm Pf} (1/r_1)$, $\partial_{ijkl} (r_2^2 \ln r_2) 
\times\partial_{ij} {\rm Pf} (1/r_1)$, $_{ikl}(f_{\rm match}^{12})_j
\times\partial_{ij} {\rm Pf} (1/r_1)$, etc. Some of them are functions of
${\bf r}_1$ only; we call them ``self'' terms, whereas those
depending on both ${\bf r}_1$ and ${\bf r}_2$ are called 
``interaction'' terms.

\subsection{Self terms} \label{par:selfterms}

We agree on considering only the self terms (i) that are proportional
to $m_1^3$ rather than $m_2^3$ (they are the same modulo the
replacement $1 \rightarrow 2$), (ii) that do contribute to the 1PN
order at most. We leave aside the terms that are generated by
$L_{ij}$, since their structure is especially simple and they are
evaluated at the end of the section. By explicitly writing down all
the sources, as done previously for $K_{ij} \partial_{ij} U$, we can
draw the complete list of intervening terms. There are three types of
terms: the $V\partial V \partial V$-type concerns one kind of term
only, i.e.  $1/r_1\times\partial_i {\rm Pf} (1/r_1)\times\partial_j
{\rm Pf} (1/r_1)$; the so-called ${\cal Y}$-type refers to $\partial_L
{\rm Pf} r_1^p\times\partial_K {\rm Pf} r_1^q$ terms, where $p$ and
$q$ are positive or negative integers; the ${\cal N}$-type terms come
as $\partial_L {\rm Pf} \partial_M (r_1^p \ln r_1)\times\partial_J
{\rm Pf} \partial_K r_1^q$ (the terms ${\cal Y}$ and ${\cal N}$ are
named after some integrals introduced below). There may exist
contracted indices among the set of multi-indices. In particular, some
terms involve a factor $\partial \Delta {\rm Pf} (1/r_1)$ and are thus
purely compact supported. In fact all the terms can be split into
compact and non-compact parts. The latter part is an ordinary function
that we are able to calculate explicitly. The former is determined
from the results of Sections VI-VIII in \cite{BFreg} and depends on
the pseudo-function derivative we use. We shall refer to it as the
self partie-finie-derivative contribution to the potentials.  If we
take the term $\partial_{ijkl} (r_1^2 \ln r_1)\times\partial_{ij} {\rm
Pf}(1/r_1)$ for instance, it reads:

$$ \partial_{ijkl} (r_1^2 \ln r_1) ~\!\partial_{ij} {\rm Pf}
\frac{1}{r_1}  =
\partial_{ijkl} (r_1^2 \ln r_1) ~\!\partial_{ij} \frac{1}{r_1}+
\partial_{ijkl} (r_1^2 \ln r_1) ~\!{\sc D}_{ij}\!\left[\frac{1}{r_1}
\right]\;. $$
In the case of the ``particular'' derivative defined by (\ref{II8}),
we have ${\sc D}_{ij}^{\rm part} [1/r_1]=2 \pi {\rm Pf}(\delta^{ij}-5
n_1^i n_1^j)
\delta_1$; so that

\begin{equation}
\partial_{ijkl} (r_1^2 \ln r_1) ~\!\partial_{ij} {\rm Pf}
\frac{1}{r_1}  = 12 \frac{n_1^kn_1^l-\delta^{kl}}{r_1^5}+
8 \pi {\rm Pf}\frac{-3n_1^kn_1^l+4 \delta^{kl}}{r_1^2} 
\delta_1 \label{eq:exemple_ln}
\end{equation} 
(like in the first term of the right side we sometimes do not write
the ${\rm Pf}$ when there is no possible confusion).  In most of this
section, we shall use the particular derivative ${\sc D}_{ij}^{\rm
part} [F]$ given by (\ref{II8}) instead of the more ``correct''
derivative ${\sc D}_{ij}[F]$ defined by (\ref{II9})-(\ref{II10}); in
Section \ref{VI} we shall discuss the effect on the final 3PN
equations of motion of using the derivative ${\sc D}_{ij}[F]$.  In
order to obtain the self cubic potentials, all we have to do now is to
apply the operator \mbox{$\int \!
\frac{d^3{\bf x'}}{-4\pi}~  |{\bf x}-{\bf x}'|^{-1} \; $} to the
various sources we are focusing on, and \mbox{$ \partial_t^2\int \!
\frac{d^3{\bf x'}}{-4\pi}~ |{\bf x}-{\bf x}'|  \; $} to
the Newtonian source of $\hat{X}^{\rm (CNC)}$. As a matter of fact,
the resulting integrals can be viewed as partie finie pseudo-functions
like (\ref{eq:exemple_ln}) acting on $1/(-4\pi |{\bf x}-{\bf x}'|)$ or
$|{\bf x}-{\bf x}'|/(-4\pi)$; both quantities are smooth at point 1,
so the pseudo-functions associated with the non-compact part reduce to
Schwartz distributions in that case (but, in order to construct the
pseudo-functions themselves, we used the generalized distributions of
\cite{BFreg}). Each integral is indeed a sum of terms of the form
$r_1^p \delta^{2K} n_1^{L-2K}$, where $p$ belongs to ${\mathbb
Z}~\!$. It is convenient to write them as sums of pseudo-function (or,
equivalently here, distributional) derivatives of quantities without
indices (``scalars''), times some possible Kronecker symbols.  We have
for example:

\begin{eqnarray}
\frac{n_1^i n_1^j}{r_1^5} &=& \frac{1}{15} \partial_{ij}
\frac{1}{r_1^3}+\frac{1}{5} \frac{\delta^{ij}}{r_1^5} \nonumber \\
&=& \frac{1}{15} \partial_{ij} {\rm Pf}
\frac{1}{r_1^3}+\frac{1}{5} \frac{\delta^{ij}}{r_1^5}+
\frac{4\pi}{15} (8 n_1^i n_1^j-\delta^{ij})
\frac{\delta_1}{r_1^2}\;
,\label{eq:toscalar} 
\end{eqnarray}
and other similar formulas for $n_1^i/r_1^4$, 
$n_1^i n_1^j n_1^k/r_1^6$, etc. 

As an illustration of our handling of the sources, here are the
effects of these transformations on $\partial_{ijkl} (r_1^2 \ln r_1)
\partial_{ij} {\rm Pf} (1/r_1)$. Starting from equation
(\ref{eq:exemple_ln}) and using (\ref{eq:toscalar}), we find

\begin{eqnarray*}
& & \int \frac{d^3{\bf x}'}{-4 \pi} 
 \frac{1}{|{\bf x}-{\bf x}'|} ~\partial^{'}_{ijkl} (r_1^{'2} \ln r'_1) 
~\partial^{'}_{ij} {\rm Pf} \frac{1}{r'_1}  \nonumber \\ 
& & \qquad =\int \frac{d^3{\bf x}'}{-4 \pi} 
\frac{1}{|{\bf x}-{\bf x}'|}~ \left[ \frac{4}{5} \partial^{'}_{kl} {\rm Pf}
\frac{1}{r_1^{'3}}-\frac{48}{5} {\rm Pf}\frac{\delta^{kl}}{r_1^{'5}}
+\frac{8\pi}{5}{\rm Pf}
\frac{n_1^{'k} n_1^{'l}-18\delta^{kl}}{r_1^{'2}}\delta'_1
\right] \;,\end{eqnarray*}
where the first term is generated by the specific derivative
(\ref{II8}). In this term the derivative can be changed to a partial
derivative with respect to the point 1, and since we employ a
pseudo-function derivative, we are allowed to permute integration and
derivation symbols. This yields
\begin{eqnarray*}
&&{\rm Pf} \int \frac{d^3{\bf x}'}{-4 \pi} 
 \frac{1}{|{\bf x}-{\bf x}'|} ~\partial^{'}_{ijkl} (r_1^{'2} \ln r'_1) 
~\partial^{'}_{ij} {\rm Pf} \frac{1}{r'_1} \\
&&\qquad\qquad =\frac{4}{5} \partial_{1kl} {\rm Pf} \int 
\frac{d^3{\bf x}'}{-4 \pi} 
\frac{1}{|{\bf x}-{\bf x}'|}~ \frac{1}{r_1^{'3}}-
\frac{48}{5} \delta^{kl} {\rm Pf} \int \frac{d^3{\bf x}'}{-4 \pi} 
\frac{1}{|{\bf x}-{\bf x}'|} \frac{1}{r_1^{'5}}\\
&&\qquad\qquad +\left(-\frac{2}{5}\frac{n_1^{'k}
n_1^{'l}}{r_1^{'2}|{\bf x}-{\bf
x}'|}+\frac{36}{5}\frac{\delta^{kl}}{r_1^{'2}|{\bf x}-{\bf
x}'|}\right)_1\; .
\end{eqnarray*}
The first two terms are left in this form for the time being. On the
other hand the last term is computed following the procedure explained
by the equations (6.17)-(6.18) in \cite{BFreg}; see also
(\ref{eq:Fdelta2}) below.  By implementing the previous procedure for
all the self terms entering $K_{ij} \partial_{ij} U$, we finally
arrive at:
\begin{eqnarray}
& &\left( \int \frac{d^3{\bf x}'}{-4 \pi} 
\frac{1}{|{\bf x}-{\bf x}'|}~ K_{ij} \partial_{ij} U\right)_{\rm self} =
G^3 {m}_1^3 \left\{
-\frac{a_1^i}{6} \partial_{1i} {\rm Pf} \int \frac{d^3{\bf x}'}{-4 \pi}
\frac{1}{|{\bf x}-{\bf x}'|}~ \frac{1}{r_1^{'3}} \right.\nonumber \\
& &\qquad \qquad \qquad +
\frac{v_1^i v_1^j}{30} \partial_{1ij} {\rm Pf} \int
\frac{d^3{\bf x}'}{-4 \pi} 
\frac{1}{|{\bf x}-{\bf x}'|}~ \frac{1}{r_1^{'3}}-
\frac{2v_1^2}{5} {\rm Pf} \int
\frac{d^3{\bf x}'}{-4 \pi} 
\frac{1}{|{\bf x}-{\bf x}'|}~ \frac{1}{r_1^{'5}} \nonumber \\
 & &\qquad \qquad \qquad - \left.
\frac{(n_1a_1)}{12r_1^2}+\frac{7(n_1v_1)^2}{150r_1^3}-
\frac{7 v_1^2}{450r_1^3}\right\}+1\leftrightarrow 2+{\cal O}(1) 
\; . \label{eq:KU}
\end{eqnarray}
We follow the same way to treat the self parts of the other cubic
potentials of interest here: 
$\hat{T}^{\rm (CNC)}$, 
$\hat{Y}_i^{\rm (CNC)}$, or the remaining of $\hat{X}^{\rm (CNC)}$.

We find that there are definite contributions, coming at the 3PN
order, due specifically to the pseudo-function derivative introduced
in \cite{BFreg}. Indeed, the distributional part of the derivative
gives some well-defined non-zero contributions, while for instance the
Schwartz derivative yields some terms which are ill-defined in this
case. These contributions of the pseudo-function derivative actually
take part in the values of $\hat{T}^{\rm (CNC)}$ and $\hat{X}^{\rm
(CNC)}$ only. Denoting them by $\delta_{\rm self} \hat{T}$ and
$\delta_{\rm self} \hat{X}$ in the case of the {\it particular}
derivative (\ref{II8}) we find

\begin{mathletters}\label{self}\begin{eqnarray} 
\delta_{\rm self} \hat{T}& = & \frac{7}{12}\frac{G^4 m_1^3m_2}
{r_1^2r_{12}^2}(n_1 n_{12})
+ 1 \leftrightarrow 2 \\
\delta_{\rm self} \hat{X}& = & \frac{G^3 m_1^3}{c^2r_1^3} 
\left(-\frac{17}{72} \frac{Gm_2}{r_{12}^2} r_1(n_1 n_{12}) 
+ \frac{1}{40} (n_1 v_1)^2 - \frac{1}{120}
v_1^2 \right)+ 1 \leftrightarrow 2
\end{eqnarray}\end{mathletters}$\!\!$

\subsection{Interaction terms} \label{par:interaction}

We consider exclusively the interaction terms (i) that are
proportional to $m_1^2 m_2$ rather than $m_1 m_2^2$, (ii) that
contribute at relative 1PN order, leaving aside those which are
generated by $L_{ij}$. Depending on whether they come from ``simple''
or ``composite'' cubic parts as shown respectively below, the
elementary terms composing the sources read schematically

\begin{mathletters} \label{eq:intterms}
\begin{eqnarray} 
& & \mbox{} \qquad \partial {\rm Pf} [F({\bf r}_1)] ~\partial {\rm
Pf} [G({\bf r}_1)] ~\partial {\rm Pf} [H({\bf r}_2)] 
\label{eq:inttermsa} \\
&{\rm and}&  \mbox{} \qquad \partial {\rm Pf} [F({\bf r}_1)]
~\partial_1 \partial_2 {\rm Pf} 
[G({\bf r}_1,{\bf r}_2)]= 
\partial_2 {\rm Pf} \{\partial {\rm Pf} [F({\bf r}_1)]
~\partial_1 {\rm Pf} [G({\bf r}_1,{\bf r}_2)]\} \; ;
\label{eq:inttermsb} 
\end{eqnarray}
\end{mathletters}$\!\!$
the functions $G$, $H$ belong to ${\cal F}$, and it is also the case
of $F$ in general. However, there exist some composite terms for which
$F=\ln r_1$, but this is not a trouble since $\partial F$ is still in
${\cal F}$. In the cases needed in this problem, $G$ is always one of
the four functions: $g_{\rm match}$, $f^{12}_{\rm match}$,
$f^{21}_{\rm match}$, or $f_{\rm match}$. Then, $G$ is ``regular
enough'' so that ${\rm Pf}
\partial G$ and  $\partial {\rm Pf} G$ coincide in any cases, and 
further simplifications of the sources do not seem to be possible at
this level. All we need, thus, is to transform the simple cubic
contributions (\ref{eq:inttermsa}) similarly to the self terms (see
paragraph \ref{par:selfterms}). A typical example of elementary source
we have to handle is $1/(r_1 r_2)\times \partial_L (1/r_1)$, where $L$
represents at most two non-contracted indices. We can check that this
term can be computed to Newtonian order only, hence it is given simply
by a Poisson integral. In the language of pseudo-functions, this means
that we have to evaluate:

$$<\frac{1}{r'_1 r'_2} ~\partial^{'}_L {\rm Pf} \frac{1}{r'_1},
\frac{1}{-4\pi |{\bf x}-{\bf x}'|}>=
<\frac{1}{r'_1} ~\partial^{'}_L {\rm Pf} \frac{1}{r'_1},
\frac{1}{-4\pi |{\bf x}-{\bf x}'| r'_2}> \; .$$
The compact part of the dual bracket, which is associated with the
distributional part of the derivative, when acting on $1/(-4\pi|{\bf
x}-{\bf x}'|)$, i.e.  $$<\frac{1}{r'_1} {\sc
D}_{L}\left[\frac{1}{r'_1}\right],
\frac{1}{-4\pi |{\bf x}-{\bf x}'| r'_2}> $$
($l$=1 or 2), leads \emph{a priori} to a non-zero result. Here, ${\sc
D}_{L}$ denotes the distributional part of the multi-derivative,
obtained in Section VIII of \cite{BFreg} and recalled by the equation
(\ref{II10'}) above. However, the left side of the bracket, which is
homogeneous to the $(-l-2)$th power of a length, is necessarily of the
type $r_1^{1-l} {\rm Pf} \delta_1$, times some dimensionless angular
function whose multipolarity differs from $l$ by an even integer
(because of the index structure of the operator ${\sc D}_L$). Now, the
previous compact part is equal to the angular integral of the
$r_1^{l-1}$ Taylor coefficient of $1/(-4\pi |{\bf x}-{\bf x}'|)$,
times the angular dependence of ${\sc D}_L[1/r_1]$. The integrand then
appears as a sum of terms whose multipolarity differs from
$l+(l-1)=2l-1$ by an even integer and so, is always odd; thereby the
angular integral gives zero. By similar arguments, we can prove that
the other compact sources associated with the distributional
derivatives will never contribute to the Poisson integrals
constituting the potentials we are considering here. Actually, it is
possible to put together all the various kinds of simple-type [see
(\ref{eq:inttermsa})] cubic terms into a unique one, which is
$n_1^L/r_1^{l+2} \times
\partial {\rm Pf} (1/r_2)$. We express at last the first factor 
(if $l \neq 0$) as a sum of derivatives of ``scalars'' 
thanks to identities such
as (\ref{eq:toscalar}). Since the pseudo-function derivatives will
give here the same results as the Schwartz distributional ones, and by
virtue of $\partial_L {\rm Pf} (1/r_1^2)|_{\cal D}={\rm Pf} \partial_L
(1/r_1^2)|_{\cal D}$ [where ${\cal D}$ is the set of smooth functions
with compact support], the last transformation can be done in the
sense of functions. Note however that the multiple derivatives of
$1/r_2$ are indeed distributional and play an important role in the
sources.

To sum up what precedes, all the interaction terms have the
general structure:

\begin{equation}  \int \frac{d^3{\bf x}'}{-4\pi}
|{\bf x}-{\bf x'}|^p~ 
\partial_{2L} {\rm Pf}[\partial^{'}_J {\rm Pf} F({\bf r}'_1) 
~\partial_{1K}{\rm Pf} G({\bf r}'_1,{\bf r}'_2)]
\label{eq:interaspect} \end{equation}
($p=-1$ or 1, $F$ and $G$ functions of ${\cal F}$). After commuting
the integral and the derivative $\partial_{2L}$, which is always licit
for integrals converging at infinity (since $\partial_{2L}$ is
followed by a ${\rm Pf}$), the general cubic term
(\ref{eq:interaspect}) becomes:

$$ \partial_{2L} \int 
\frac{d^3{\bf x'}}{-4\pi} |{\bf x}-{\bf x'}|^p ~
 \partial^{'}_J {\rm Pf} F({\bf r}'_1) ~\partial_{1K} {\rm Pf} G({\bf
 r}'_1,{\bf r}'_2) \; . $$ In the case where $G=G({\bf r}_2)$ (and
 $k=0$), this rearranges as
\begin{eqnarray} & &\partial_{2L} \int 
\frac{d^3{\bf x'}}{-4\pi} |{\bf x}-{\bf x'}|^p ~
 (-)^j \partial_{1J} [{\rm Pf} F({\bf r}'_1) ~
 {\rm Pf} G({\bf r}'_2)] \nonumber \\ & &\qquad =
 (-)^j \partial_{1J} \partial_{2L} {\rm Pf} \int
\frac{d^3{\bf x'}}{-4\pi} |{\bf x}-{\bf x'}|^p ~
 F({\bf r}'_1) G({\bf r}'_2) \; . \end{eqnarray} 
We shall end with implementing concretely our treatment of the source
on two typical terms:

\begin{mathletters}\label{ex}\begin{eqnarray}
&& \int  \frac{d^3{\bf x'}}{-4\pi}
\frac{1}{|{\bf x}-{\bf x'}|} ~
\frac{1}{r'_1} ~\partial'_i {\rm Pf} \frac{1}{r'_1} 
~\partial'_j {\rm Pf} \frac{1}{r'_2} 
= \frac{1}{2} \partial_{1i} \partial_{2j} {\rm Pf} \int
\frac{d^3{\bf x'}}{-4\pi} \frac{1}{|{\bf x}-{\bf x'}|} ~
\frac{1}{r_1^{'2} r'_2} \; ,\\
&& \int  \frac{d^3{\bf x'}}{-4\pi}
\frac{1}{|{\bf x}-{\bf x'}|} ~
{}_{ki} f^{12}_{{\rm match}\; j} ~\partial^{'}_{ij} {\rm Pf}
\frac{1}{r'_1}  
  =  \partial_{2j} \int  \frac{d^3{\bf x'}}{-4\pi}
\frac{1}{|{\bf x}-{\bf x'}|} ~
{}_{ki} f^{12}_{\rm match} ~\partial^{'}_{ij} {\rm Pf} \frac{1}{r'_1}
\; ,
\end{eqnarray}\end{mathletters}$\!\!$
and by providing the complete interaction component corresponding to
the Poisson integral of $K_{ij} \partial_{ij} U$, which completes the
self part obtained previously in (\ref{eq:KU}). The compact support
terms have been explicitly determined, while the other ones are left
un-evaluated for the moment:

\begin{eqnarray} & &
\left( \int  \frac{d^3{\bf x'}}{-4\pi}
\frac{1}{|{\bf x}-{\bf x'}|} ~
K_{ij} \partial_{ij} U\right)_{\rm int} =G^3 {m}_1^2
{m}_2 \left\{
\frac{a_1^i}{4} \partial_{2i} D~\! 
{\rm Pf} \int \frac{d^3{\bf x'}}{-4\pi}
\frac{1}{|{\bf x}-{\bf x'}|}~ \frac{\ln r'_1}{r'_2} \right.
\nonumber \\ & & \qquad \qquad \quad +
\frac{a_1^i}{48} \partial_{1i} D^2~\! 
{\rm Pf} \int  \frac{d^3{\bf x'}}{-4\pi}
\frac{1}{|{\bf x}-{\bf x'}|}~ \frac{r_1^{'2} \ln r'_1}{r'_2}+
a_1^j \partial_{2i}  {\rm Pf} \int  \frac{d^3{\bf x'}}{-4\pi}
\frac{1}{|{\bf x}-{\bf x'}|}~ {}_{jk} f^{12}_{\rm match}~
\partial^{'}_{ik} 
\frac{1}{r'_1} \nonumber \\ & & \qquad \qquad \quad 
+  a_2^j \partial_{2ij} 
 {\rm Pf} \int  \frac{d^3{\bf x'}}{-4\pi}
\frac{1}{|{\bf x}-{\bf x'}|}~ {}_k f^{21}_{\rm match}~
\partial^{'}_{ik} 
\frac{1}{r'_1} - \frac{(n_{12} a_1)}{8r_{12} r_2}
 \nonumber \\ & & \qquad \qquad \quad +
\frac{v_1^i v_1^j}{8} \partial_{2ij}  
 {\rm Pf} \int  \frac{d^3{\bf x'}}{-4\pi}
\frac{1}{|{\bf x}-{\bf x'}|}~ \frac{1}{r_1^{'2} r'_2}+
\frac{v_1^2}{16} D^2~\! {\rm Pf} \int  \frac{d^3{\bf x'}}{-4\pi}
\frac{1}{|{\bf x}-{\bf x'}|}~ \frac{\ln r'_1}{r'_2}
 \nonumber \\ & & \qquad \qquad \quad +
\frac{v_1^i v_1^j}{96} \partial_{1ij} D^2 
{\rm Pf} \int  \frac{d^3{\bf x'}}{-4\pi}
\frac{1}{|{\bf x}-{\bf x'}|}~ \frac{r_1^{'2} \ln r'_1}{r'_2}
\nonumber \\ & & \qquad \qquad \quad +
v_1^j v_1^k \partial_{2i} {\rm Pf} \int  \frac{d^3{\bf x'}}{-4\pi}
\frac{1}{|{\bf x}-{\bf x'}|}~ {}_{jkl} f^{12}_{\rm match}
~\partial^{'}_{il} 
\frac{1}{r'_1} \nonumber \\ & & \left. \qquad \qquad \quad +
v_2^j v_2^k \partial_{2ijk} {\rm Pf} \int  \frac{d^3{\bf x'}}{-4\pi}
\frac{1}{|{\bf x}-{\bf x'}|}~ {}_l f^{21}_{\rm match}
~\partial^{'}_{il} 
\frac{1}{r'_1}+ \frac{(n_{12}v_1)^2}{8r_{12}^2 r_2}
\right\}+1\leftrightarrow 2+{\cal O}(1) \; , \label{eq:KUint}
\end{eqnarray}
where $D$ denotes the operator $\partial_{1i} \partial_{2i}$.

\subsection{Elementary integrals}

\subsubsection{Nomenclature}

The inspection of the formula (\ref{eq:KUint}) for interaction terms
issued from $K_{ij} \partial_{ij} U$ suggests that we should
re-express this potential, as well as all the other ones, by means of
a restricted number of elementary integrals; basically one for each
source type. Hence the proposal for a useful systematic nomenclature,
which reflects their structure.  We shall introduce the following
notations (and ditto $1 \leftrightarrow 2$):
 
\begin{eqnarray} \label{eq:nomenclature}
& & \stackrel{(n,p)}{{\cal Y}}=
{\rm Pf} \int  \frac{d^3{\bf x'}}{-4\pi}
\frac{1}{|{\bf x}-{\bf x'}|}~r_1^{'n} r_2^{'p} \qquad \qquad
\sous{1}{{\cal F}}_{(P,Q)}^{12}=
\int \frac{d^3{\bf x}'}{-4 \pi} \frac{1}{|{\bf x}-{\bf x}'|}~
{}_{iQ} f^{12}_{\rm match}~ \partial^{'}_{iP} 
{\rm Pf} \frac{1}{r'_1} \nonumber \\ 
& & \soussur{1}{(n,p)}{{\cal N}}=
{\rm Pf} \int  \frac{d^3{\bf x'}}{-4\pi}
\frac{1}{|{\bf x}-{\bf x'}|}~
r_1^{'n} r^{'p}_2 \ln r'_1 \qquad 
\sous{1}{{\cal F}}_{(P,Q)}^{21}=
\int \frac{d^3{\bf x}'}{-4 \pi} \frac{1}{|{\bf x}-{\bf x}'|}~
{}_{iQ} f^{21}_{\rm match}~ \partial^{'}_{iP} {\rm Pf} \frac{1}{r'_1} 
\nonumber \\ & & \sous{1}{{\cal U}}_P=
\int \frac{d^3{\bf x}'}{-4 \pi} \frac{1}{|{\bf x}-{\bf x}'|}~ {}_k
g_{\rm match~k} ~\partial^{'}_P {\rm Pf} \frac{1}{r'_1} \quad \quad 
\sous{1}{{\cal F}}_{(P,Q)}=
\int \frac{d^3{\bf x}'}{-4 \pi} \frac{1}{|{\bf x}-{\bf x}'|}~
{}_{iQ} f_{\rm match}~ \partial^{'}_{iP} {\rm Pf} \frac{1}{r'_1}
\nonumber \\ & & \sous{1}{{\cal K}}_P= \int \frac{d^3{\bf x}'}{-4
\pi} \frac{1}{|{\bf x}-{\bf x}'|}~ g_{\rm match}~ \partial_P 
{\rm Pf} \frac{1}{r'_1} \qquad \quad \stackrel{(n,p)}{{\cal S}}
= {\rm Pf} \int
\frac{d^3{\bf x}'}{-4 \pi} |{\bf x}-{\bf x}'|~  r_1^{'n}
r_2^{'p} \nonumber \\ & & \sous{1}{{\cal G}}_{(P,Q)}=
\int \frac{d^3{\bf x}'}{-4 \pi} \frac{1}{|{\bf x}-{\bf x}'|}~
{}_{iQ} g_{\rm match}~ \partial^{'}_{iP} {\rm Pf} \frac{1}{r'_1} \quad 
 \soussur{1}{(n,p)}{{\cal M}}= {\rm Pf} \int \frac{d^3{\bf x}'}{-4 \pi}
|{\bf x}-{\bf x}'|~ r_1^{'n} r^{'p}_2 \ln r'_1  
\nonumber \\ & & {}_i \sous{1}{{\cal G}}_{(P,Q)}=
\int \frac{d^3{\bf x}'}{-4 \pi} \frac{1}{|{\bf x}-{\bf x}'|}~
{}_{iQ} g_{\rm match}~ \partial^{'}_P {\rm Pf} \frac{1}{r'_1} \quad 
\sous{1}{{\cal Q}}_{(P,Q)}= \int \frac{d^3{\bf x}'}{-4 \pi} |{\bf
x}-{\bf x}'|~ {}_{iQ} g_{\rm match}~ \partial^{'}_{iP} 
{\rm Pf} \frac{1}{r'_1} \nonumber \\  & & \sous{1}{{\cal H}}_{(P,Q)}=
\int \frac{d^3{\bf x}'}{-4 \pi} \frac{1}{|{\bf x}-{\bf x}'|}~
{}_{iQ} g_{\rm match}~ \partial^{'}_{iP} {\rm Pf} r'_1\;.   
\end{eqnarray}
The value of the previous integrals is not generally known at any
space location, except in some special cases. The reason is that their
sources involve three points, in addition to the integration variable
${\bf x}'$: the point ${\bf x}$ where the field is calculated, and the
two source points ${\bf y}_1$, ${\bf y}_2$.  The few ones that are
computable for any ${\bf x}$ include notably some self integrals like
${\cal Y}^{(n,0)}$ and ${\cal S}^{(n,0)}$, and the two integrals $D^2
{\cal N}^{(0,-1)}_1$ and $\partial_{2i} {\cal G}_{1 \; (i,0)}$
entering the interaction part of the $\hat{X}^{\rm (CNC)}$ potential
at the Newtonian order. Actually, there are no other cubic
contributions up to the 2.5PN order, and that is why we were able in
\cite{BFP98} to get the complete expression of the metric in the near
zone at this order. This property of the integrals $D^2 {\cal
N}^{(0,-1)}_1$ and $\partial_{2i} {\cal G}_{1 \; (i,0)}$ is linked to
the specific form of the integrands, which are made of products of two
second derivatives applied on appropriate functions, such that the
indices of the first derivative are contracted with those of the
second one: $ \partial_{ij} {\rm Pf} \ln r_1 \times \partial_{ij} {\rm
Pf} (1/r_2)$ and ${}_{i} g_j \times \partial_{ij} {\rm Pf}
(1/r_1)$. In both cases, particular solutions in the sense of
distributions of the corresponding Poisson equations

\begin{equation} \label{eq:KH}
\Delta K_1 =2 \partial_{ij} {\rm Pf} \ln r_1 ~ \partial_{ij}
{\rm Pf} \frac{1}{r_2} \;
,\qquad \qquad \Delta H_1 =2{}_{i} g_j ~\partial_{ij} {\rm Pf}
\frac{1}{r_1} \; . 
\end{equation}
can be exhibited \cite{BDI95,BFP98}. The solutions $K_1$ and $H_1$ of
(\ref{eq:KH}) that go to zero as $r \to \infty$ read:

\begin{mathletters}\label{eq:KH2}\begin{eqnarray}
K_1 &=& \left(\frac{1}{2}\Delta-\Delta_1\right)
\left[\frac{\ln r_1}{ r_2}\right]+
\frac{1}{2}\Delta_2\left[\frac{\ln r_{12}}{r_2}\right]
+\frac{r_2}{2r_{12}^2 r_1^2}+\frac{1}{r_{12}^2 r_2} \ ,  \\
H_1 &=& \frac{1}{2}\Delta_1\left[\frac{g}{r_1}+\frac{\ln r_1}{r_{12}}-
\Delta_1\left(\frac{r_1+r_{12}}{2} g \right)\right] \nonumber\\
&+&\partial_i\partial_{2i}\left[\frac{\ln r_{12}}{r_1}+
\frac{\ln r_1}{ 2r_{12}}\right] 
- \frac{1}{ r_1}\partial_{2i}[(\partial_i g)_{\textstyle
{}_1}]-\frac{r_2}{2r_1^2r_{12}^2} \; ,
\end{eqnarray}\end{mathletters}$\!\!$
with $\Delta_1= \partial_{1ii}$, $\Delta_2=\partial_{2ii}$; see also
(\ref{KHdev}) for the expanded forms of these solutions. Thus, we have

$$D^2 {\cal
N}^{(0,-1)}_1=\frac{1}{2}K_1\quad\hbox{and}\quad\partial_{2i} {\cal
G}_{1 \; (i,0)}=\frac{1}{2}H_1\;.$$ In the list
(\ref{eq:nomenclature}) above, we note the appearance of iterated
Poisson integrals such as ${\cal S}^{(-5,0)}$, ${\cal M}^{(0,-1)}$ and
$\partial_{2i} {\cal Q}_{1 \; (i,0)}$ which come from the 1PN
contribution to the retardation expansion of $\hat{X}^{\rm (CNC)}$,
and thus enter this potential through their second time
derivative. What we shall have to compute for our purpose is their
Hadamard regularized value. However, $\partial_{2i} {\cal Q}_{1 \;
(i,0)}$ is not available on the whole space (for any ${\bf x}$), so
that we cannot deduce 
$\partial_t^2 \partial_{2i} {\cal Q}_{1\; (i,0)}$ from it directly. 
We shall adopt then a different approach. In a first
stage, we express the operator $\partial_t^2$ with the help of the
partial derivatives $\partial_{1i}$ and $\partial_{2i}$: if $F({\bf
r}_1,{\bf r}_2)=\partial_{2i} {\cal Q}_{1 \; (i,0)}$, then

$$
\partial_t^2 F =
\partial_t \left[ v_1^i \partial_{1i}F\right] =
v_1^i v_1^j \partial_{1ij}F+ a_1^i \partial_{1i}F +1\leftrightarrow
2\; .  $$ Next, we commute the partial derivatives $\partial_{1L}$
with the integration sign, so that the derivatives act on the source
of the Poisson integral. This operation is legitimate only if the
derivatives $\partial_{1L}$ when acting on the integrands are viewed
as distributional (in the sense of Section IX in \cite{BFreg}). The
new integrands write then as a partie finie derivative of a product,
e.g. $\partial_{1L} ({}_{iQ} g~
\partial^{'}_{iP} \frac{1}{r'_1})$; but remember that we are not \emph{a
priori} allowed to develop them according to the Leibniz rule in our
formalism. In fact, these specific non Leibniz corrections happen to
give zero contribution to the 3PN potentials. This can be seen by
applying successively the formula (7.23) in \cite{BFreg} (which is
indeed sufficient since the ``test'' function $|{\bf x}-{\bf x}'|$ is
smooth at the points 1 and 2) to all the sources we are dealing
with. Therefore, we can employ the usual rule for derivatives of
products to perform our final transformation. In summary, we will have
for instance

\begin{eqnarray*}  & & \partial_{1kl}  \int \frac{d^3{\bf x}'}{-4
\pi} |{\bf x}-{\bf x}'|~ {}_{i} g_j ~\partial_{ij} {\rm Pf}
 \frac{1}{r'_1} =
\int \frac{d^3{\bf x}'}{-4 \pi} |{\bf x}-{\bf x}'|~ {}_{ikl}  g_j
~\partial_{ij} {\rm Pf} \frac{1}{r'_1} \mbox{} \qquad \qquad \qquad
\qquad \qquad \mbox{} \\ & & \qquad \qquad \qquad \quad+ 2 \int
\frac{d^3{\bf x}'}{-4 \pi} |{\bf x}-{\bf x}'|~ {}_{i(k} g_{\underline
j} ~\partial_{l)ij} {\rm Pf} \frac{1}{r'_1}+
\int \frac{d^3{\bf x}'}{-4 \pi} |{\bf x}-{\bf x}'|~ {}_{i}  g_j
~\partial_{ijkl} {\rm Pf} \frac{1}{r'_1} \; ,  
\end{eqnarray*}
(the index $\underline j$ means that $j$ is excluded from the
symmetrization operation).  On this form we can apply the partie finie
at 1 while staying in the same class of elementary integrals
(\ref{eq:nomenclature}).  We conclude by going back to the example of
the cubic term generated by $K_{ij} \partial_{ij} U$; we have, after
appropriate reshaping,

\begin{eqnarray*}
& & \int \frac{d^3{\bf x}'}{-4 \pi} \frac{1}{|{\bf x}-{\bf x}'|}~
K_{ij} \partial_{ij} U = 
G^3 {m}_1^3 \left\{
-\frac{a_1^i}{6}  \partial_{1i} \! \! \!  \!  \!
\stackrel{(-3,0)}{{\cal Y}} \right. +
\frac{v_1^i v_1^j}{30}  \partial_{1ij} \! \! \!  \!  \!
\stackrel{(-3,0)}{{\cal Y}} -
\frac{2v_1^2}{5}  \stackrel{(-5,0)}{{\cal Y}}  \nonumber \\
 & &\qquad \qquad \qquad \qquad \qquad - \left.
\frac{(n_1a_1)}{12r_1^2}+\frac{7(n_1v_1)^2}{150r_1^3}-
\frac{7 v_1^2}{450r_1^3} \right\} \nonumber \\ 
& & \qquad \qquad \qquad +G^3 {m}_1^2 {m}_2 \left\{
\frac{a_1^i}{4}   \partial_{2i} D 
\soussur{1}{(0,-1)}{{\cal N}} +
\frac{a_1^i}{48}\partial_{1i} D^2   
\soussur{1}{(2,-1)}{{\cal N}}  +
a_1^j  \partial_{2i} 
\sous{1}{{\cal F}}_{(i,j)}^{12} 
+  a_2^j  \partial_{2ij} 
\sous{1}{{\cal F}}_{(i,0)}^{21}
\right. \nonumber \\ & & \qquad
\qquad \qquad \qquad \qquad
  - \frac{(n_{12} a_1)}{8r_{12} r_2} +
\frac{v_1^i v_1^j}{8} \partial_{2ij} \! \! \!  \!  \!
\stackrel{(-2,-1)}{{\cal Y}} +
\frac{v_1^2}{16} D^2 
\soussur{1}{(0,-1)}{{\cal N}}  +
\frac{v_1^i v_1^j}{96}  \partial_{1ij} D^2  
\soussur{1}{(2,-1)}{{\cal N}}  
\nonumber \\ & & \qquad \qquad \qquad \qquad \qquad +
v_1^j v_1^k \partial_{2i} 
\sous{1}{{\cal F}}_{(i,jk)}^{12}  \left. + 
v_2^j v_2^k  \partial_{2ijk} 
\sous{1}{{\cal F}}_{(i,0)}^{21}+ \frac{(n_{12}v_1)^2}{8r_{12}^2 r_2}
\right\} +1\leftrightarrow 2+{\cal O}(1)  \; . 
\end{eqnarray*}
We refer to \cite{Fthese} for the expressions of the other non-linear
potentials expressed in this manner by means of the same nomenclature.
The problem is now to evaluate all the elementary integrals from which
the 3PN cubic potentials have been built.

\subsubsection{Parties finies of the elementary integrals}

As mentioned before, in most cases, we do not have at our disposal the
explicit values of the elementary integrals in all space.  This does
not matter since all we need is their Hadamard partie finie at point 1
(or 2). Notice that the partial derivative with respect to ${\bf y}_2$
is the only one that commutes with the partie finie operation at 1; to
be more explicit:

\begin{eqnarray*} & &\mbox{} \qquad [\partial_{2i}
F({\bf x},{\bf y}_1,{\bf y}_2)]_{\textstyle {}_1}= 
\partial_{2i} [F({\bf x},{\bf y}_1,{\bf y}_2)]_{\textstyle {}_1} \;, \\
&{\rm but}&  \mbox{} \qquad 
 [\partial_{1i} F({\bf x},{\bf y}_1,{\bf y}_2)]_{\textstyle {}_1} \neq
\partial_{1i} [F({\bf x},{\bf y}_1,{\bf y}_2)]_{\textstyle {}_1}   \qquad
({\rm with}~F, ~G \in {\cal F}).  \end{eqnarray*} Thus, for each
elementary integral, we shall determine first the partie finie of the
quantity figuring under the derivation symbol $\partial_2$ and, only
then, apply the latter operator. On the contrary, we cannot bring the
derivatives with respect to ${\bf y}_1$ out of the partie finie at 1,
so we are led to incorporate them to the sources, in the sense of
pseudo-functions, by permutation with the integration sign. As a
consequence, the integrals we are interested in are of the type:
\begin{eqnarray*}
&&\int \frac{d^3{\bf x}}{-4 \pi} 
|{\bf x}-{\bf x}'|^p ~ \partial_1 [{\rm Pf} F({\bf r}_1)] ~
\partial_1 [{\rm Pf} G({\bf r}_1,{\bf r}_2)]\\
&&\qquad\qquad\qquad\quad = -\frac{1}{4 \pi}<\partial_1 [{\rm Pf}
F({\bf r}_1)] ~\partial_1 [{\rm Pf} G({\bf r}_1, {\bf r}_2)] ~,~|{\bf
x}-{\bf x}'|^p>\; .
\end{eqnarray*}
They involve both a compact part (C) and a non-compact part (NC).  The
compact part is produced by the purely distributional contributions of
derivatives in the integrands: $$ -\frac{1}{4\pi} <{\sc D}_1[ F({\bf
r}_1)] ~{\rm Pf} \partial_1 G({\bf r}_1, {\bf r}_2) ~,~|{\bf x}-{\bf
x}'|^p>-\frac{1}{4\pi} <{\sc D}_1[G({\bf r}_1, {\bf r}_2)]~{\rm Pf}
\partial_1 F({\bf r}_1)~,~|{\bf x}-{\bf x}'|^p> \; . $$ As mentioned
at the beginning of the paragraph \ref{par:interaction}, the partie
finie derivatives reduce here, in our case, to those of the Schwartz
theory. This is obvious when $G=G({\bf r}_2)$, because then the source
$F({\bf r}_1)$, regarded as a linear functional ${\rm Pf}F({\bf
r}_1)$, acts on a function which is smooth in a neighbourhood of ${\bf
x}={\bf y}_1$; in the other cases, the result follows from explicit
calculations.  Note that the number $l$ of derivatives in front of
$G({\bf r}_1,{\bf r}_2)=g_{\rm match}$, $f^{12}_{\rm match}$,
$f^{21}_{\rm match}$, or $f_{\rm match}$ is always small enough so
that $\partial_{1\,L-1} G({\bf r}_1,{\bf r}_2)$ is bounded, hence
$\partial_{1L} {\rm Pf} G({\bf r}_1,{\bf r}_2)= {\rm Pf}
\partial_{1L}G({\bf r}_1,{\bf r}_2)$. Once we have in hands the
compact part, it remains to obtain its regularized value at point 1.
As a matter of fact, if the source is of the type ${\rm Pf} F
\delta_2$, where $F \in {\cal F}$, then

\begin{equation} \label{eq:Fdelta2}
 <{\rm Pf} (F \delta_2)({\bf x}')~ , ~\frac{1}{|{\bf x} - {\bf x}'|}> =
-\frac{1}{4\pi} \sum_{l \ge 0} \frac{(-)^l}{l!} \left(r_2^{'l} n_2^{'L} F
\right)_2 \partial_L \left(\frac{1}{r_2}\right)
\end{equation}
is smooth at point 1 and we need not call for the Hadamard
regularization. Therefore, we are allowed to replace directly $|{\bf
x}-{\bf x}'|$ by $r'_1$ in the left-hand-side of (\ref{eq:Fdelta2}).
When the source is of the generic type ${\rm Pf} F
\delta_1$, the same identity as (\ref{eq:Fdelta2}) holds, 
but with $2\rightarrow 1$. This shows that the integral $<{\rm Pf} (F
\delta_1)~ , ~|{\bf x} - {\bf x}'|^{-1}>$ is purely singular as $r_1
\to 0$, which means that it has no partie finie at ${\bf x}={\bf
y}_1$: the $\delta_1$ type sources do not contribute to the potentials
computed at body 1. Summarizing, we have

\begin{mathletters} \label{eq:Fdelta2prime}\begin{eqnarray}
 \left(<{\rm Pf} (F \delta_2)~ , ~\frac{1}{|{\bf x} - {\bf x}'|}>\right)_1 
&=& <{\rm Pf} (F \delta_2)~ , ~\frac{1}{r_1}>\;,\\
\left(<{\rm Pf} (F \delta_1)~ , ~\frac{1}{|{\bf x} - {\bf x}'|}>\right)_1 
&=& 0\;.\label{eq:Fdelta2primeb}
\end{eqnarray}\end{mathletters}$\!\!$
We refer the reader to equations (6.17)-(6.20) in \cite{BFreg} for
more details.

Let us now focus our attention on the non-compact parts of the
elementary integrals, whose integrands are made, by definition, of
ordinary functions.  The problem is to get the Hadamard partie finie
of the Poisson integral $P$ of $F \in {\cal F}$:

$$P({\bf x}')={\rm Pf} \int \frac{d^3{\bf x}}{-4 \pi} 
\frac{1}{|{\bf x}-{\bf x}'|}~
F({\bf x}) \; , $$ 
as well as the one of the iterated Poisson integral $Q$, given by:

$$Q({\bf x}')={\rm Pf} \int \frac{d^3{\bf x}}{-4 \pi} |{\bf x}-{\bf
x}'|~ F({\bf x}) \; . $$ [Each of these functions depends also on the
source points ${\bf y}_{1,2}$ and velocities ${\bf v}_{1,2}$.] Now,
the partie finie prescription applies only to functions admitting
power-like expansions near their singularities [see (\ref{II1})],
whereas $P$ or $Q$ may contain logarithmic coefficients in their
development: if we take for instance $F=1/r_1^3$, we shall have
$P=-[1+\ln (r'_1/s_1)]/r'_1$, where $s_1$ is the constant appearing in
(\ref{II3}). Following \cite{BFreg}, we shall simply include the
possible logarithms (i.e. $\ln r'_1$) appearing in the zeroth power
coefficient of the expansion of $P$ or $Q$ in the definition of the
partie finie: see the equation (5.4) in \cite{BFreg}. With this
generalized notion of ``partie finie'', we can give a sense to
$(P)_{\textstyle {}_1}$ and $(Q)_{\textstyle {}_1}$, as well as their
gradients $(\partial_i P)_{\textstyle {}_1}$ and $(\partial_i
Q)_{\textstyle {}_1}$. We make then the following statements (see
Section V of \cite{BFreg} for proofs and discussion):

\begin{mathletters} \label{eq:poissonth}
\begin{eqnarray} \left( P \right)_{\textstyle {}_1} 
&=& {\rm Pf}_{s_1,s_2} \int
\frac{d^3{\bf x}}{-4\pi} \frac{1}{r_1} F + \left[\ln
\left(\frac{r'_1}{s_1} \right)-1 \right] (r_1^2 F)_{\textstyle {}_1} \; ,
\label{eq:poissontha} \\ 
\left( Q \right)_{\textstyle {}_1} &=& {\rm Pf}_{s_1,s_2} \int
\frac{d^3{\bf x}}{-4\pi} r_1 F + \left[\ln
\left(\frac{r'_1}{s_1}\right)+\frac{1}{2} \right] 
(r_1^4 F)_{\textstyle {}_1} \; ,
\label{eq:poissonthb} \\
 \left(\partial_i P\right)_{\textstyle {}_1} &=& {\rm Pf}_{s_1,s_2}
 \int
\frac{d^3{\bf x}}{-4\pi} 
\frac{n_1^i}{r_1^2} F + \ln \left(\frac{r'_1}{s_1}\right) 
(n_1^i r_1 F)_{\textstyle {}_1} \; ,
\label{eq:poissonthc} \\
 \left(\partial_i Q \right)_{\textstyle {}_1} &=& {\rm Pf}_{s_1,s_2} \int
\frac{d^3{\bf x}}{4\pi} n_1^i F + \left[-\ln
\left(\frac{r'_1}{s_1}\right)+\frac{1}{2} \right] 
(n_1^i r_1^3 F)_{\textstyle {}_1} \; ,
\label{eq:poissonthd} 
\end{eqnarray}
\end{mathletters}$\!\!$
where the first terms in the right sides are made of some partie-finie
integrals in the sense of the definition (\ref{II3}). The ``constant''
$r'_1$ is the variable which tends toward zero when evaluating the
partie finie.  It is easy to show that the constant $s_1$ cancels out
between the two terms in each of the second members of the identities
(\ref{eq:poissonth}). Indeed, using the general dependence of the
partie-finie integral on the constants $s_1$, $s_2$ as given by (4.20)
in \cite{BFreg}, we easily see that $(P)_{\textstyle {}_1}$ given by
(\ref{eq:poissontha}), for instance, depends on the constants $r'_1$
and $s_2$ through the formula

\begin{equation}\label{const}
\left( P \right)_{\textstyle {}_1} = -\ln
\left(\frac{r_{12}}{r'_1} \right)(r_1^2 F)_{\textstyle {}_1} -\ln
\left(\frac{r_{12}}{s_2} \right)\left(\frac{r_2^3}{r_1} F\right)_{2}
+\dots
\end{equation}
where the dots indicate the terms that are independent of any
constants. As we see, the constant $s_1$ has been so to speak
``replaced'' by $r'_1$. This makes clearer why it is convenient to
keep the $\ln r'_1$ in the definition of the Hadamard partie finie; if
we had decided to exclude this logarithm from it, we would have found
some bare $\ln r_{12}$ in the first term of (\ref{const}) instead of a
nicer logarithm of a dimensionless quantity: $\ln ({r_{12}}/{r'_1})$;
but this is simply a matter of convenience, because we shall see that
in fact the ``constants'' $\ln r'_{1}$ and $\ln r'_{2}$ can be gauged
away from the 3PN equations of motion. The same argument is valid for
all cases in (\ref{eq:poissonth}). As a consequence, the acceleration
of the first body will depend only on two unspecified constants: $\ln
r_1'$, and of course $\ln s_2$ (and ditto for the acceleration of the
second body). See Section \ref{VII} for further discussion of these
constants.

The relations (\ref{eq:poissonth}) answer the problem of evaluating
the elementary integrals at the location of particle 1 without knowing
their values at an arbitrary field point. The subsequent task consists
of calculating the partie-finie integrals in the right-hand-sides,
which will turn out to be always possible.

\subsubsection{Integration methods}

The non compact parts of the regularized elementary integrals consist
essentially of some integrals ${\rm Pf} \int \!  d^3{\bf x}~F({\bf
r}_1,{\bf r}_2)$, with $F \in {\cal F}$. It is worth noting that the
sources depend on ${\bf r}_1$, ${\bf r}_2$ exclusively, and not on the
separate variables ${\bf x}$, ${\bf y}_1$, and ${\bf y}_2$, because
scalar products like $(xy_1)$, $(xy_2)$ or $(y_1 y_2)$ occurring in
$f^{12}_{\rm match}$, $f^{21}_{\rm match}$, and $f_{\rm match}$ are
killed by the derivatives that precede them in the integrand. In this
work, we make use of two different integration methods: (1) the
angular method, in which we determine successively angular and radial
integrals in spherical coordinates, (2) the analytic continuation
method, based on the so-called Riesz formula.

\medskip\paragraph{Angular method.}

Let $F=F({\bf r}_1,{\bf r}_2)$ be a function in the class ${\cal F}$.
We assume for a moment that $F$ is locally integrable at point 2; so
we are allowed to compute the integral over the whole space, deprived
from a small ball ${\cal B}_1(s)$ of center ${\bf y}_1$ and radius
$s>0$. Thus, we start with the well-defined quantity $\int_{{\mathbb
R}^3
\setminus {\cal B}_1(s)} \! d^3{\bf x} ~ F({\bf r}_1, {\bf r}_2)$,
but for convenience we write it in spherical coordinates
$(r_1,\theta_1,\phi_1)$, such that the azimuthal angle $\theta_1$
coincides with the separation angle between ${\bf y}_{12}$ and ${\bf
r}_1$,

$$\int_{{\mathbb R}^3 \setminus {\cal B}_1(s)} d^3{\bf x}~
F({\bf r}_1,{\bf r}_2)=
\int_{r_1}^{+\infty} dr_1 ~r_1^2 \int d\Omega_1~
F({\bf r}_1,{\bf r}_2) \; .  $$ Actually the function $F$ may be a
tensor with many indices, but the only unit vectors in the problem are
${\bf n}_1$, ${\bf n}_2$, ${\bf n}_{12}$, and only two of them, say
${\bf n}_1$ and ${\bf n}_{12}$, are independent, by virtue of the
relation $r_1 {\bf n}_1+r_{12} {\bf n}_{12} = r_2 {\bf n}_2$; and
therefore, $F$ can be expressed under the form of a finite sum of
tensorial products of type $n_1^L n_{12}^K$ ($l$, $k\in {\mathbb N}$);
moreover, each factor admits a symmetric trace-free (STF)
decomposition on the basis $\hat{n}_1^L={\rm STF} n_1^L$. Hence we
have:

$$F=\sum_{l=0}^{l_0} \sum_{k=0}^{k_0} \hat{n}_1^L n_{12}^K
G_{(lk)}(r_1,r_2)\;, \qquad l_0, ~k_0 \in {\mathbb N} \; . $$ Here,
the $G_{(lk)}$'s are some scalar functions of $r_1$ and $r_2$. Now,
$r_2$ is related to $r_1$ and the scalar product $(n_1 n_{12})$ by
$r_2=\sqrt{r_1^2+r_{12}^2+2 r_1 r_{12} (n_1 n_{12})}$. So that the
angular integral of $F$ can be obtained by means of the ``mean
formula''; see, e.g., the formula (A26) in \cite{BD86}. We get

\begin{equation} \label{eq:mean} \int d\Omega_1~F =
 2\pi \sum_{l=0}^{l_0} \sum_{k=0}^{k_0} \hat{n}_{12}^L n_{12}^K
\int_{-1}^1 dz~ G_{(lk)}\left(r_1,\sqrt{r_1^2+r_{12}^2+2
r_1 r_{12} z}\right) P_l(z)  \; , 
\end{equation}
where $P_l$ is the Legendre polynomial of order $l$. In the cases of
interest here, $G_{(lk)}$ is always a sum of rational fractions with
general structure $r_1^p r_2^q /(r_1+r_2+r_{12})^n$, with $p$, $q$
positive or negative integers and $n \in {\mathbb N}~\!$. It is easy
to check that the result of the angular integration depends on the
relative positions of $r_1$ and $r_{12}$. Therefore we must split the
radial integral into two contributions according to the integration
domains $r_1 \in ~\!]0,r_{12}]$ or $r_1 \in
~\!]r_{12},+\infty[$. Typical terms coming from the angular
integration are $r_1^p/(r_1+r_{12})^n$, as well as some more
complicated logarithmic terms such as $r_1^p \ln (1+r_{12}/r_1)$. Most
of the corresponding radial integrals are obtained straightforwardly
using some integrations by part, applying the partie finie at the
bound $r_1=s$ [i.e., removing the poles $1/s^k$, with $k\geq 1$, and
replacing $\ln s$ by $\ln s_1$]. In the case of the latter logarithmic
terms with $p=-1$, integrating by parts does not lead to anything, but
the radial integrals can be found in standard mathematical textbooks:

\begin{eqnarray}
& & \int_{r_{12}}^{+\infty} \frac{dr_1}{r_1}~ \ln \left(1 +
\frac{r_{12}}{r_1}\right)=
\int_0^{r_{12}} \frac{dr_1}{r_1}~ \ln \left(1 +
\frac{r_1}{r_{12}}\right) =\frac{\pi^2}{12} \;,
\nonumber \\
& & \int_{r_{12}}^{+\infty} \frac{dr_1}{r_1}~ \ln
\left(\frac{\displaystyle 1 +
\frac{r_{12}}{r_1}}{\displaystyle 1 - \frac{r_{12}}{r_1}}\right)
=\int_0^{r_{12}} \frac{dr_1}{r_1}~ \ln \left(\frac{\displaystyle 1 +
\frac{r_1}{r_{12}}}{\displaystyle 1 -
\frac{r_1}{r_{12}}}\right)=\frac{\pi^2}{4}\;.
\end{eqnarray}
It can be shown that the integrals diverging at ${\bf y}_1$ involve in
general a logarithm $\ln (r_{12}/s_1)$ but never any $\pi^2$
terms. The procedure we have just described indeed permits calculating
most of the elementary integrals.  Consider for instance the integral

\begin{equation}
{}_i \sous{1}{{\cal G}}_{(j,0)}=
\int \frac{d^3{\bf x}'}{-4 \pi} \frac{1}{|{\bf x}-{\bf x}'|}~
{}_{i} g~ \partial^{'}_j {\rm Pf} \frac{1}{r'_1} 
\end{equation}
(in which we replaced $g_{\rm match}$ by $g=\ln S$ since they merely
differ by a constant). We are interested in the value of this integral
at point 1, following the regularization. From
(\ref{eq:Fdelta2primeb}) we know that the distributional part of the
derivative will not contribute. Then, using (\ref{eq:poissontha}), we
readily find

$$\left({}_i \sous{1}{{\cal G}}_{(j,0)}\right)_1= {\rm Pf}\int
\frac{d^3{\bf x}}{-4 \pi} \frac{1}{r_1}~ {}_{i} g~ \partial_j
\frac{1}{r_1} +\frac{\delta^{ij}}{6 r_{12}} \left[\ln
\left(\frac{r'_1}{s_1}\right) - 1\right]\;.$$ The non-compact integral
has a sole divergence at point 1, so that we can apply the previous
method without any change, and get

\begin{equation}\label{int}
{\rm Pf}\int \frac{d^3{\bf x}}{-4 \pi} \frac{1}{r_1}~ {}_{i} g~
\partial_j \frac{1}{r_1} =-\frac{\delta^{ij}}{18
r_{12}}-\frac{n_{12}^i n_{12}^j}{12 r_{12}}-
\frac{\delta^{ij}}{6r_{12}} \ln \left(\frac{r_{12}}{s_1}\right) \; .
\end{equation}
As expected the constant $\ln s_1$ cancels out and we arrive at:

\begin{equation}
\left(_{i}  \sous{1}{{\cal G}}_{(j,0)} \right)_{\textstyle {}_1}=
-\frac{2 \delta^{ij}}{9 r_{12}}-\frac{n_{12}^i n_{12}^j}{12 r_{12}}-
\frac{\delta^{ij}}{6 r_{12}} \ln \left(\frac{r_{12}}{r'_1} \right)\; . 
\end{equation}

A few elementary integrals diverge at the locations ${\bf y}_1$ and
${\bf y}_2$ of both particles. In this occurrence, the integral
$\int_{{\mathbb R}^3 \setminus {\cal B}_1(s)} \! d^3{\bf x}~\!F$ has
no meaning, and the previous integration process is not adequate
anymore. However, Proposition 2 in Section IV B of
\cite{BFreg} allows us to extend it to this case. 
We introduce the auxiliary source

$$ \widetilde{F}_2=F-\sum_{b+3 \le 0} r_2^b \sous{2 \phantom{xx}}{f_b}
\; ,$$ which is locally integrable near ${\bf y}_2$ but does not
converge at infinity. As before, the angular integral of $\tilde{F}_2$
around ${\bf y}_1$ takes a different expression depending on whether
$r_1 \le r_{12}$ or $r_1 > r_{12}$, so we must split the radial
integration into the two domains $]0,r_{12}[$ and
$]r_{12},+\infty[$. Then, with full generality, the partie-finie
integral of the source $F$ is given by

\begin{eqnarray}\label{prop2}
{\rm Pf}_{s_1, s_2} \int d^3{\bf x}~F&=& {\rm Pf}_{s_1}
\int_0^{r_{12}} d r_1 ~r_1^2\int d\Omega_1~{\widetilde
F}_2~\!+\int_{r_{12}}^{+\infty} d r_1~r_1^2\int
d\Omega_1~\biggl[{\widetilde F}_2+\frac{1}{r_1^3}\left(r_2^3
F\right)_{\textstyle {}_2}\biggr]\nonumber\\ &+&4\pi \left(r_2^3
F\right)_{\textstyle {}_2} \ln\left(\frac{r_{12}}{s_2}\right)\;.
\end{eqnarray}
If this integral comes from a Poisson integral evaluated at 1, the
constant $s_1$ will be cancelled out as we have seen previously and
replaced by $r'_1$; but there will remain in general a constant $s_2$
coming from the singularity at the other point. With the angular
method we were able to obtain all the elementary integrals (and their
gradients) at point 1. See an appendix of \cite{Fthese} for the
complete list of those results.

\medskip\paragraph{Analytical continuation method.} 

The equivalence between the Hadamard partie finie prescription for
integrals and the analytic continuation regularization has long been
known (see e.g. \cite{Schwartz}), and we have recovered it in the
three-dimensional case by the Theorem 2 of \cite{BFreg}.  More
precisely, for any $F\in {\cal F}$ that behaves like $o(1/r^3)$ when
$r \to +\infty$, the integral $\int \! d^3{\bf x}~(r_1/s_1)^\alpha
(r_2/s_2)^\beta F$ of two complex variables $\alpha$ and $\beta$
admits an analytic continuation in the neighbourhood of
$\alpha=\beta=0$, and we have

\begin{equation}
{\rm Pf}_{s_1,s_2} \int \! d^3{\bf x}~F = {\rm
FP}_{\genfrac{}{}{0cm}{}{\alpha \to 0}{\beta \to 0}} \int \!  d^3{\bf
x} \left(\frac{r_1}{s_1}
\right)^\alpha \left(\frac{r_2}{s_2} \right)^\beta F =
{\rm FP}_{\genfrac{}{}{0cm}{}{\beta \to 0}{\alpha \to 0}} \int \!
d^3{\bf x} \left(\frac{r_1}{s_1} 
\right)^\alpha \left(\frac{r_2}{s_2} \right)^\beta F \; , 
\end{equation}
where ${\rm FP}_{\genfrac{}{}{0cm}{}{\alpha \to 0}{\beta \to 0}}$
means taking the finite part of the Laurent expansion of the (analytic
continuation of the) integral when $\alpha \to 0$ and $\beta \to 0$
successively.  This result is particularly useful in the case where
$F$ is of the type $r_1^p r_2^q$, with $p$ and $q$ relative integers,
since the integral is directly computable thanks to the Riesz formula
\cite{Riesz}:

\begin{equation} \label{eq:Riesz}
\int d^3{\bf x}~r_1^{\alpha+p} r_2^{\beta+q}=
\pi^{\frac{3}{2}} \frac{\Gamma\left(\frac{\alpha+p+3}{2}\right)
\Gamma\left(\frac{\beta+q+3}{2} \right)
\Gamma\left(-\frac{\alpha+\beta+p+q+3}{2}\right)}
{\Gamma\left(-\frac{\alpha+p}{2} 
\right) \Gamma\left(-\frac{\beta+q}{2}\right)
\Gamma\left(\frac{\alpha+\beta+p+q+6}{2}\right)} 
r_{12}^{\alpha+\beta+p+q+3}\;. 
\end{equation}
One may consult \cite{BFreg} for an example of practical computation.
Most of the time, the structure of the sources of elementary integrals
is more complicated than a simple $r_1^p r_2^q$, notably it involves
many ``free'' tensorial indices which imply that generally the sources
of elementary integrals, when fully developped, involve numerous
inverse powers of $S=r_1+r_2+r_{12}$. However, by considering all the
possible contractions of these free indices with vectors $n_{12}^i$
and Kronecker symbols $\delta^{ij}$, it happens that we reduce the
computation to that of several scalar integrals we can obtain thanks
to the Riesz formula (\ref{eq:Riesz}) (i.e., when performing the
contractions and after simplification of the result, we are always led
to the simple structure $r_1^p r_2^q$ without $1/S$ powers). It should
be noted that the set of scalar functions that we compute contains the
complete information about the complicated tensorial integral, i.e. it
permits to reconstitute it exactly.

Let us illustrate the method with the computation of the integral
(\ref{int}). It involves two free indices $ij$ and is necessarily of
the type

\begin{equation}\label{phipsi}
{\rm Pf} \int \frac{d^3{\bf x}}{-4\pi}~ \frac{1}{r_1} 
~\! {}_i g~\! \partial_j \frac{1}{r_1} =
\phi (r_{12}) ~\!n_{12}^i n_{12}^j+\psi(r_{12}) ~\!\delta^{ij}
\; ,
\end{equation}
where $\phi$ and $\psi$ are some unknown ``scalars'' depending on
$r_{12}$.  By contracting successively the integrand with $n_{12}^i
n_{12}^j$ and $\delta^{ij}$, and simplifying, we find:

\begin{eqnarray*}
n_{12}^i n_{12}^j ~\!{}_i g \partial_j \frac{1}{r_1} &=&
\frac{1}{4 r_1^3} - \frac{r_2}{4 r_1^4} +
\frac{1}{4 r_1 r_{12}^2}  - \frac{r_2}{4 r_1^2 r_{12}^2} 
- \frac{r_2^2}{4 r_1^3 
r_{12}^2}\\ &+& \frac{r_2^3}{4 r_1^4 r_{12}^2} + \frac{1}{4 r_1^2
r_{12}} -
\frac{r_2^2}{4 r_1^4 r_{12}} + \frac{r_{12}}{4 r_1^4} \;,\\
\delta^{ij}{}_i g \partial_j \frac{1}{r_1} &=&
\frac{1}{2 r_1^3} + \frac{1}{2 r_1^2 r_{12}} -
\frac{r_2}{2 r_1^3 r_{12}}\;,
\end{eqnarray*}
thus obtaining a sum of terms of the type $r_1^p r_2^q$ which can all
be integrated with the help of the formula (\ref{eq:Riesz}). This
computation yields a system of equations for the scalars $\phi$ and
$\psi$:

\begin{eqnarray*}
&&\phi+\psi = -\frac{5}{36r_{12}} -\frac{1}{6 r_{12}} 
\ln \left(\frac{r_{12}}{s_1}
\right) \;,\\
&&\phi+3\psi = -\frac{1}{4 r_{12}} -
\frac{1}{2 r_{12}} \ln \left( \frac{r_{12}}{s_1} \right)\;. 
\end{eqnarray*}
By solving the previous system, and inserting the results into
(\ref{phipsi}), we recover exactly the value given by (\ref{int}). We
have checked in this manner all the elementary integrals previously
computed with the angular method.

\subsubsection{Finite part of integrals diverging at infinity}

For the moment, we have left aside the case where our integrals are
defined by means of a finite part dealing with a divergence occuring
at infinity.  Such a study is needed to compute some 0.5PN integrals
encountered in Section
\ref{IV}. From now on, we suppose that the source $F$ 
admits an expansion when $r \to +\infty$ which is made of simple
powers $r^{n-3}$, with $n \leq n_{\rm max}$ (so the integral {\it a
priori} diverges at infinity when $n_{\rm max}\geq 0$), and we
consider the quantity:

\begin{equation} \label{eq:PFintF}
{\rm Pf} \bigg\{ {\rm FP}_{B \to 0} \int \! d^3{\bf x}~
\left(\frac{r}{r_0}\right)^B F \bigg\} \; .
\end{equation}
We split the integral above into two integrals $I_{\rm int}$ and
$I_{\rm ext}$ extending respectively over a domain $D_{\rm int}$
including the two local singularities ${\bf y}_{1,2}$, and the
complementary domain $D_{\rm ext}$ comprising the regions at
infinity. Using the integration variable ${\bf r}_1={\bf x}-{\bf
y}_1$, the external integral (on which the regularization ${\rm Pf}$
can be removed) reads as

$$I_{\rm ext} = {\rm FP}_{B \to 0} 
\int_{D_{\rm ext}} d^3{\bf r_1}
\left(\frac{|{\bf r}_1+{\bf y}_1|}{r_0}\right)^B 
F({\bf r_1}+{\bf y}_1) \;.$$ If we assume that the original integral
can generate only simple poles $\sim 1/B$ at infinity (which will be
always the case here), we can replace it by

\begin{eqnarray} I_{\rm ext} &=& {\rm FP}_{B \to 0} 
\int_{D_{\rm ext}} d^3{\bf r_1}
\left(\frac{r_1}{r_0}\right)^B F 
\nonumber \\ &+&
{\rm FP}_{B \to 0} ~\frac{B}{2} \int_{D_{\rm ext}} d^3{\bf r_1}
\left(\frac{r_1}{r_0}\right)^B  
\ln \left(1+2 (n_1y_1) \frac{1}{r_1}+\frac{y_1^2}{r_1^2} \right)F
\label{eq:Pfinf}
\end{eqnarray}
Indeed, in the case of simple poles, the other terms, involving at
least a factor $B^2$, will always give zero.  The second term is
non-zero only if the corresponding integral does admit a pole, and can
be calculated in a simple way by picking up the term of order
$1/r_1^3$ in the expansion of the integrand when $r\to
+\infty$. Notably, in the important case where the function $F$ goes
to zero like $1/r^3$ at infinity, its product with the log-term
behaves at least like $1/r^4$ and therefore the second term in
(\ref{eq:Pfinf}) gives no contribution. For instance, an integral
divergent at infinity that we encounter in the problem is

$${\rm FP}_{B \to 0} \int \! d^3{\bf x}~\left(\frac{r}{r_0}\right)^B
\partial_{ij}\left({\rm Pf}\frac{1}{r_1}\right)=-
\frac{4\pi}{3}\delta_{ij}\;.$$

\subsection{Lorentzian regularization of potentials}

All the potentials and their gradients (compact-support, quadratic and
non-compact potentials) which have been computed in this section and
the previous one were obtained at the point 1 using the standard
Hadamard regularization $(F)_{\textstyle {}_1}$. However, this
regularization, being defined within the hypersurface $t=$const of the
harmonic coordinates, must break at some point the Lorentz-invariance
properties of the potentials. That is, if a potential defined for a
smooth ``fluid'' behaves in a certain way under a Lorentz
transformation, we expect that its regularized value at the point 1 in
the sense of $(F)_{\textstyle {}_1}$ will generically not behave in
the same way. Nevertheless, the equations of motion in harmonic
coordinates, as computed with the regularization $(F)_{\textstyle
{}_1}$, are known to be Lorentz invariant up to the 2.5PN order
\cite{BFP98}. Perhaps not surprisingly because of this fact, it turns
out that the Lorentzian regularization $[F]_{\textstyle {}_1}$
(defined in \cite{BFregM}) yields no difference with respect to the
old regularization $(F)_{\textstyle {}_1}$ for all the 3PN potentials
but for one, namely the cubic-non-compact potential ${\hat X}^{\rm
(CNC)}$ defined by (\ref{eq:DXcNCCgen}) and which had to be computed
at the relative 1PN order. (Evidently we have $[F]_{\textstyle
{}_1}=(F)_{\textstyle {}_1}$ for all the potentials which are to be
computed with Newtonian accuracy.) Thus all the results obtained so
far but the one for $(\partial_i{\hat X}^{\rm (CNC)})_{\textstyle
{}_1}$ [the relevant quantity for the equations of motion] are valid
in the case of the new regularization.  In this subsection we compute
the remaining part $[\partial_i{\hat X}^{\rm (CNC)}]_{\textstyle
{}_1}-(\partial_i{\hat X}^{\rm (CNC)})_{\textstyle {}_1}$.

Since the regularization $[F]_{\textstyle {}_1}$ brings some new terms
with respect to the old one $(F)_{\textstyle {}_1}$ starting at the
relative 1PN order, and since ${\hat X}^{\rm (CNC)}$ is to be computed
at the 1PN order only, it is sufficient for this calculation to use
the lowest-order, Newtonian value of ${\hat X}^{\rm (CNC)}$.  From the
computation in \cite{BDI95,BFP98} we know the analytic closed-form
expression of ${\hat X}^{\rm (CNC)}$ at Newtonian order for any field
point $(t,{\bf x})$,

\begin{equation}\label{XCNC}
{\hat X}^{\rm (CNC)}=\frac{G^3 m_1^3}{ 12r_1^3}-G^3m_1^2m_2
\left\{\frac{1}{8r_2r_{12}^2}
+\frac{1}{ 16}K_1+H_1\right\}+{\cal O}(1)+1\leftrightarrow 2 \;,
\end{equation}
where the functions $K_1$ and $H_1$, which are solutions of certain
Poisson equations, are given explicitly by (\ref{eq:KH2}). The
complete developed forms of these functions are

\begin{mathletters}\label{KHdev}\begin{eqnarray}
K_1 &=& -\frac{1 }{ r_2^3} + \frac{1}{r_2r_{12}^2}-\frac{1 }
{ r_1^2r_2}+\frac{r_2}{
2r_1^2r_{12}^2}+\frac{r_{12}^2}{ 2r_1^2r_2^3} 
+\frac{r_1^2}{ 2r_2^3r_{12}^2} \;,\\
H_1 &=& -\frac{1 }{ 2r_1^3}-\frac{1}{ 4r_{12}^3} -
\frac{1}{4r_1^2r_{12}}-\frac{r_2 
}{ 2 r_1^2r_{12}^2}+\frac{r_2}{ 2r_1^3r_{12}} 
+\frac{3r_2^2}{ 4r_1^2r_{12}^3}+\frac{r_2^2}{ 2r_1^3r_{12}^2}
-\frac{r_2^3}{
2r_1^3r_{12}^3}  \;.
\end{eqnarray}\end{mathletters}$\!\!$
We replace these expressions into (\ref{XCNC}), and we implement all
the rules for the new regularization $[F]_{\textstyle {}_1}$ defined
in Section III of \cite{BFregM}. Equivalently, since the order of the
computation is limited to 1PN, we can use the closed form formula

\begin{equation}
[F]_{\textstyle {}_1}-(F)_{\textstyle {}_1}=\frac{1}{c^2}\left(({\bf
r}_1.{\bf v}_1)\left[\partial_tF+\frac{1}{2}v^i_1\partial_i
F\right]\right)_{\textstyle {}_1}+{\cal O}\left(4\right)\;,
\end{equation}
derived in Section IV of \cite{BFregM}. As a result, we obtain, for
the potential itself,

\begin{eqnarray}
[{\hat X}^{\rm (CNC)}]_{\textstyle {}_1}&-&({\hat X}^{\rm
(CNC)})_{\textstyle {}_1}\nonumber\\ &=&\frac{G^3m_1^2m_2}{
c^2r_{12}^3}\left\{\frac{43}{
40}(n_{12}v_1)^2-(n_{12}v_1)(n_{12}v_2)-\frac{43}{ 120}v_1^2
+\frac{1}{
3}(v_1v_2)\right\} +{\cal O}(4)\;.
\end{eqnarray}
In the case of the gradient needed for the equations of motion, we get
 
\begin{eqnarray}\label{diXCNC}
[\partial_i{\hat X}^{\rm (CNC)}]_{\textstyle {}_1}&-&(\partial_i{\hat
X}^{\rm (CNC)})_{\textstyle {}_1}\nonumber\\ &=&\frac{G^3m_1^2m_2}{
c^2r_{12}^4}\left\{\left[\frac{27}{ 56}(n_{12}v_1)^2-\frac{3}{
4}(n_{12}v_1)(n_{12}v_2)-\frac{27}{ 280}v_1^2+\frac{3}{
20}(v_1v_2)\right]n_{12}^i\right.\nonumber\\ &-&\left.\frac{27}{
140}(n_{12}v_1)v_1^i+\frac{3}{ 20}(n_{12}v_2)v_1^i+\frac{3}{
20}(n_{12}v_1)v_2^i
\right\}+{\cal O}(4)\;.
\end{eqnarray}
As we see, the new regularization brings some definite non-zero
contributions at the 1PN order in the case of this potential, which
will constitute a crucial contribution to the 3PN equations of
motion. The right-hand side of (\ref{diXCNC}) is not invariant by
itself under a Lorentz transformation --- it cannot be ---, but will
ensure finally the Lorentz-invariance of the 3PN equations of motion.

\section{Leibniz terms and non-distributivity}\label{VI} 

\subsection{Effect of a gauge transformation}

In this subsection we study the effect of a gauge transformation on
the 3PN equations of motion as well as energy of the two
particles. Let $\{x^\mu\}$ denote the harmonic coordinate system and
$g_{\mu\nu}(x)$ be the harmonic-coordinate metric, generated by the
two particles, that we have iterated in previous sections up to the
3PN order. The metric depends on the position ${\bf x}$ of the
field point, and on the coordinate time $t=x^0/c$ through the
trajectories ${\bf y}_{1,2}(t)$ and velocities ${\bf v}_{1,2}(t)$ of
the particles, i.e.

\begin{equation}\label{VI22}
g_{\mu\nu}(x)=g_{\mu\nu}[{\bf x};{\bf y}_{1}(t),{\bf y}_{2}(t);
{\bf v}_{1}(t),{\bf v}_{2}(t)]\;.
\end{equation}
We know that the dependence of the metric over the velocities arises
at the 1PN order (see e.g. the equations (7.2) in \cite{BFP98}),
namely the order ${\cal O}(4,3,4)$, where this notation is a shorthand
for saying ${\cal O}(4)={\cal O}(1/c^4)$ in $g_{00}$, ${\cal O}(3)$ in
$g_{0i}$ and ${\cal O}(4)$ in $g_{ij}$.  Consider an infinitesimal
coordinate transformation of the type

\begin{mathletters}\label{VI23}\begin{eqnarray}
&&{x'}^\mu=x^\mu+\xi^\mu(x)\;,\\
&&\xi^\mu(x)=\xi^\mu[{\bf x};{\bf y}_{1}(t),{\bf y}_{2}(t)]\;,
\end{eqnarray}\end{mathletters}$\!\!$
where, in order to simplify the presentation, we assume that the gauge
vector $\xi^\mu$ depends on the positions ${\bf y}_{1,2}$ of the
particles, but not on their velocities. Furthermore, we suppose that
this gauge transformation is at the level of the 3PN order, which
means that $\xi^0={\cal O}(7)$ and $\xi^i={\cal O}(6)$, or
equivalently $\xi^\mu={\cal O}(7,6)$. In addition, in a first stage,
we suppose that the vector $\xi^\mu(x)$ is a smooth function of the
coordinates even at the positions of the particles. The new metric in
the new coordinate system $\{{x'}^\mu\}$ is

\begin{equation}\label{VI24}
g'_{\mu\nu}(x')=g'_{\mu\nu}[{\bf x}';{\bf y}'_{1}(t'),{\bf
y}'_{2}(t');{\bf v}'_{1}(t'),{\bf v}'_{2}(t')]\;,
\end{equation}
where the new trajectories and velocities ${\bf y}'_{1,2}$, ${\bf
v}'_{1,2}$ are parametrized by the new coordinate time
$t'={x'}^0/c$. The coordinate change (\ref{VI23}), when applied at the
location of each of the particles, yields the relations between the
new and old trajectories, which, when retaining only the terms up to
the order ${\cal O}(6)$, read as

\begin{mathletters}\label{VI25}\begin{eqnarray}
{y'}^i_1(t')&=&{y}^i_1(t)+\xi^i(y_1)+{\cal O}(8)\;,\\
{y'}^i_2(t')&=&{y}^i_2(t)+\xi^i(y_2)+{\cal O}(8)\;,
\end{eqnarray}\end{mathletters}$\!\!$
where the $\xi^\mu(y_{1,2})$'s denote the gauge vector at the position
of the particles, for instance $\xi^\mu(y_1)=\xi^\mu[{\bf y}_1(t);{\bf
y}_{1}(t),{\bf y}_{2}(t)]$.  The new metric (\ref{VI24}), when
expressed in terms of the old variables, follows from this as

\begin{equation}\label{VI26}
g'_{\mu\nu}(x')=g'_{\mu\nu}(x)+\xi^i(x)\partial_ig_{\mu\nu}+\xi^i(y_1)
{}_{1}{\partial}_ig_{\mu\nu}+\xi^i(y_2){}_{2}{\partial}_ig_{\mu\nu}
+{\cal O}(10,9,10)\;,
\end{equation}
where we have used the fact that the dependence of the metric on the
velocities starts at the 1PN order, so the terms due to the
modification of the velocities do not contribute to (\ref{VI26}).
Since the 3PN metric depends on space ${\bf x}$ only through the two
distances ${\bf x}-{\bf y}_{1}$ and ${\bf x}-{\bf y}_{2}$, we have
$\partial_ig_{\mu\nu}+{}_1\partial_ig_{\mu\nu}+{}_2\partial_ig_{\mu\nu}=0$,
and so an equivalent form of (\ref{VI26}) is

\begin{equation}\label{VI27}
g'_{\mu\nu}(x')=g'_{\mu\nu}(x)
+[\xi^i(y_1)-\xi^i(x)]{}_{1}{\partial}_ig_{\mu\nu}+[\xi^i(y_2)-\xi^i(x)]
{}_{2}{\partial}_ig_{\mu\nu}+{\cal O}(10,9,10)\;.
\end{equation}
The equation (\ref{VI27}), when combined with the law of
transformation of tensors, i.e. in the present case

\begin{equation}\label{VI28}
g_{\mu\nu}(x)=g'_{\mu\nu}(x')+\partial_\mu\xi_\nu+\partial_\nu\xi_\mu+{\cal
O}(10,9,8)\;,
\end{equation}
where $\xi_\mu=\eta_{\mu\nu}\xi^\nu$, gives the metric variation or
Lie derivative $\delta_\xi g_{\mu\nu}=g'_{\mu\nu}(x)-g_{\mu\nu}(x)$
(where the same variable $x$ is used for both the transformed and
original metrics) as

\begin{equation}\label{VI29}
\delta_\xi g_{\mu\nu}=-\partial_\mu\xi_\nu-\partial_\nu\xi_\mu
+[\xi^i(x)-\xi^i(y_1)]{}_{1}{\partial}_ig_{\mu\nu}+[\xi^i(x)
-\xi^i(y_2)]{}_{2}{\partial}_ig_{\mu\nu}+ {\cal O}(10,9,8)\;.
\end{equation}
In fact, up to this order, only the $00$ component of the metric
includes a ``non-linear'' correction term; and, within that non-linear
term, the metric can be approximated by its Newtonian part, so

\begin{mathletters}\label{VI30}\begin{eqnarray}
\delta_\xi g_{00}&=&-2\partial_0\xi_0+\frac{2}{c^2}\biggl([\xi^i(x)
-\xi^i(y_1)]{}_{1}{\partial}_iU+[\xi^i(x)-\xi^i(y_2)]{}_{2}
{\partial}_iU\biggr)+{\cal O}(10)\;,\\
\delta_\xi g_{0i}&=&-\partial_0\xi_i-\partial_i\xi_0+{\cal O}(9)\;,\\
\delta_\xi g_{ij}&=&-\partial_i\xi_j-\partial_j\xi_i+{\cal O}(8)\;,
\end{eqnarray}\end{mathletters}$\!\!$
where $U=\frac{Gm_1}{r_1}+\frac{Gm_2}{r_2}$ is the Newtonian potential
(with a small inconsistency of notation with respect to previous
sections). Now, it is easy to check that, in the sense of
distributions,

$$
\Delta\biggl([\xi^i(x)-\xi^i(y_1)]{}_{1}{\partial}_iU+[\xi^i(x)-\xi^i(y_2)]
{}_{2}{\partial}_iU\biggr)=-2\partial_i\xi_j\partial_{ij}U
-\Delta\xi_i\partial_iU\;.
$$ Indeed, the delta-functions at the points 1 and 2, which come from
the Laplacian of $U$, are killed respectively by the factors
$\xi^i(x)-\xi^i(y_1)$ and $\xi^i(x)-\xi^i(y_2)$, which vanish
respectively at these two points, in front of them.  So, we can write
for $\delta_\xi g_{00}$ the simpler but equivalent expression

\begin{equation}\label{VI32}
\delta_\xi g_{00}=-2\partial_0\xi_0-\frac{2}{c^2}\Delta^{-1}
[2\partial_i\xi_j
\partial_{ij}U+\Delta\xi_i \partial_i U]+{\cal O}(10)\;,
\end{equation}
where $\Delta^{-1}$ denotes the usual Poisson integral.

The latter result can be generalized to our framework of singular
metrics by allowing the gauge vector $\xi^\mu$ to become singular at
the positions of the particles (in the sense that $\xi^\mu\in {\cal
F}$), provided that the integral appearing in (\ref{VI32}) is treated
as the
Hadamard partie finie of a Poisson integral in the way which is
investigated in Section V of \cite{BFreg}. Let us consider, for
example, the 3PN gauge vector given by

\begin{equation}\label{VI33}
\xi_\mu=\frac{G^3m^3}{c^6}\partial_\mu 
\left(\frac{\epsilon_1}{r_1}+\frac{\epsilon_2}{r_2}\right)\;,
\end{equation}
where $\epsilon_1$ and $\epsilon_2$ denote two dimensionless constants
or possibly functions of time $t$ (and where $m=m_1+m_2$). Note that
with this choice of gauge vector the new coordinates satisfy the
condition of harmonic coordinates outside the singularities (i.e. in
the sense of functions) at the 3PN order: indeed $\Box \xi^\mu={\cal
O}(9,8)$. When inserting (\ref{VI33}) into (\ref{VI32}), we must be
careful about evaluating the last term of (\ref{VI32}) in the sense of
distributions, taking into account the fact that $\Delta\xi_i$ is
distributional. For this term we obtain

$$
\Delta^{-1}[\Delta\xi_i \partial_i U]=\frac{G^3m^3}{c^6}\left[\gamma_1^i
\partial_i\left(\frac{\epsilon_1}{r_1}\right)+\gamma_2^i\partial_i\left(
\frac{\epsilon_2}{r_2}\right)\right]\;,
$$ where $\gamma_1^i$ and $\gamma_2^i$ are the Newtonian accelerations
of 1 and 2. Therefore, we find

\begin{equation}\label{VI35}
\delta_\xi g_{00}=-\frac{2G^3m^3}{c^8}\Bigg\{\Big[\partial_t^2
+\gamma_1^i\partial_i\Big]
\left(\frac{\epsilon_1}{r_1}\right)+2\Delta^{-1}\left[\partial_{ij}
\left(
\frac{\epsilon_1}{r_1}\right)\partial_{ij}U\right]\Biggr\}
+{\cal O}(10)+1\leftrightarrow 2\;.
\end{equation}
In the case where $\epsilon_1$ and $\epsilon_2$ are some pure
constants (independent on time) we can somewhat simplify the latter
expression by using the fact that the accelerations cancel out in the
first term. In this case, we obtain the full metric transformations as

\begin{mathletters}\label{VI36}\begin{eqnarray}
\delta_\xi g_{00}&=&-\frac{2G^3m^3}{c^8}\Bigg\{v_1^{ij}\partial_{ij}\left(
\frac{\epsilon_1}{r_1}\right)
+2\Delta^{-1}\left[\partial_{ij}\left(\frac{\epsilon_1}{r_1}
\right)\partial_{ij}U\right]\Biggr\}+{\cal O}(10)
+1\leftrightarrow 2\;,\\
\delta_\xi g_{0i}&=&\frac{2G^3m^3}{c^7}v_1^j\partial_{ij}
\left(\frac{\epsilon_1}{r_1}\right)
+{\cal O}(9)+1\leftrightarrow 2\;,\\
\delta_\xi g_{ij}&=&-\frac{2G^3m^3}{c^6}\partial_{ij}
\left(\frac{\epsilon_1}{r_1}\right)
+{\cal O}(8)+1\leftrightarrow 2\;.
\end{eqnarray}\end{mathletters}$\!\!$
By comparing this with the 3PN metric (\ref{III21}), we see that the
gauge transformation induces the following changes in the 3PN
potentials ${\hat T}$, ${\hat Y}_i$ and ${\hat Z}_{ij}$:

\begin{mathletters}\label{VI37}\begin{eqnarray}
\delta_\xi {\hat T}&=&-\frac{G^3m^3}{16}\Bigg\{v_1^{ij}\partial_{ij}\left(
\frac{\epsilon_1}{r_1}\right)+2\Delta^{-1}\left[\partial_{ij}
\left(\frac{\epsilon_1}{r_1}
\right)\partial_{ij}U\right]\Biggr\}+1\leftrightarrow 2\;,
\label{VI37a}\\
\delta_\xi {\hat Y}_i&=&-\frac{G^3m^3}{8}v_1^j\partial_{ij}
\left(\frac{\epsilon_1}{r_1}
\right)+1\leftrightarrow 2\;,\\
\delta_\xi {\hat Z}_{ij}&=&-\frac{G^3m^3}{8}\partial_{ij}
\left(\frac{\epsilon_1}{r_1}
\right)+1\leftrightarrow 2\;.
\end{eqnarray}\end{mathletters}$\!\!$
The computation of the non-linearity term in (\ref{VI37a}) is
straightforward, and we get

$$
\Delta^{-1}\left[\partial_{ij}\left(\frac{1}{ r_1}\right)
\partial_{ij}U\right]
=\frac{Gm_1}{2r_1^4}+Gm_2~{}_{ij}g_{ij}\;, $$ where $g=\ln
(r_1+r_2+r_{12})$ is a kernel satisfying $\Delta g=\frac{1}{r_1r_2}$
[see also (\ref{gf12})], and where we denote
${}_{ij}g_{ij}=\partial_{1ij}\partial_{2ij}g=D^2g$.  At last, we
insert the latter changes of the 3PN potentials into the (regularized)
equations of motion (\ref{III29}) with (\ref{III32}) (there is no need
to include a correction due to the non-distributivity), and obtain the
corresponding change in the acceleration of the particle 1 as

\begin{eqnarray}\label{VI38}
\delta_\xi a_1^i&=&\frac{2G^4m^3}{c^6 r_{12}^5}\Big(\epsilon_1 m_2
-\epsilon_2 m_1\Big)n_{12}^i
\nonumber\\
&+&\frac{G^3m^3\epsilon_2}{c^6r_{12}^4}\Big[-15(n_{12}v_{12})^2n_{12}^i
+3v_{12}^2n_{12}^i+6(n_{12}v_{12})v_{12}^i\Big]\;.
\end{eqnarray}
In the case where $\epsilon_1$ and $\epsilon_2$ depend on time, there
are some extra contributions proportional to $\dot\epsilon_2$ and
$\ddot\epsilon_2$. A good check of (\ref{VI38}) is the
fact that to the change in the acceleration (\ref{VI38}) always
corresponds a change in the associated energy; that is, the gauge
transformation does not modify the existence of a conserved energy
(see Section \ref{VII} for the computation of the 3PN energy). Namely,
we find that the combination $m_1\delta_\xi a_1^i v_1^i+m_2\delta_\xi
a_2^i v_2^i$ is a total time-derivative, and from this we obtain 
the gauge transformation of the energy as

\begin{equation}\label{VI39}
\delta_\xi E=\epsilon_2\frac{G^3m^3m_1}{c^6r_{12}^3}
\Bigg[\frac{Gm_2}{r_{12}}
-3(n_{12}v_{1})(n_{12}v_{12})+(v_{1}v_{12})\Bigg]+1\leftrightarrow 2\;.
\end{equation}

\subsection{Leibniz contributions}

An important ingredient of the present computation is the novel
distributional derivative associated with the Hadamard regularization
which has been introduced in \cite{BFreg} (see also Section \ref{II}).
This derivative permits us to derive in a systematic and consistent
way all the integrals encountered in the problem; however it
represents merely a {\it mathematical} tool, which is maybe not
connected to any relevant Physics.  Therefore, it is important to know
exactly the role played by this derivative in the 3PN equations of
motion, with respect to say the Schwartz distributional derivative
\cite{Schwartz}.  We know that our distributional derivative affects
the computation of two types of terms: (i) the ``self'' terms entering
{\it a priori} in the non-linear potentials ${\hat X}$, ${\hat T}$ and
${\hat Y}_i$ and which are ill-defined in the case of the Schwartz
derivative (see Section \ref{V}), (ii) the ``Leibniz'' terms which
account for the violation of the Leibniz rule during the 3PN iteration
of the metric as discussed in Section \ref{III}. In the present
subsection, we compute the Leibniz terms and combine the result with
the one of Section
\ref{V} concerning the self terms. 
The conclusion is that the terms coming from the use of the
distributional derivative are necessary for keeping track of the
Lorentz invariance of the equations of motion, and that no other
Physics is involved with them in the present formalism (see also
Section \ref{VII}). We do the computation for both the ``particular''
derivative defined by (\ref{II8}) and the more correct one given by
(\ref{II9})-(\ref{II10}).

The Leibniz terms discussed in Section \ref{III} consist of those
contributions of the type (\ref{III20}) and alike which arise in the
process of simplification of the 3PN metric $h^{\mu\nu}$ by means of
the Leibniz rule. These terms depend only on the distributional part of
the derivative, ${\sc D}^{\rm part}_i[F]$ or ${\sc D}_i[F]$. The
formulas giving the complete Leibniz terms in $h^{\mu\nu}$, not being
very attractive, are relegated to Appendix \ref{D}. When reducing
explicitly these formulas we find that all the terms take the same
simple structure and, not surprisingly, arise only at the 3PN order. As
already announced in (\ref{III24}), the Leibniz terms imply {\it a
priori} a net contribution to the 3PN potentials ${\hat T}$, ${\hat
Y}_i$ and ${\hat Z}_{ij}$. Actually, in the case of the particular
derivative, their contributions to the vector and tensor potentials
${\hat Y}_i$ and ${\hat Z}_{ij}$ turn out to be zero,

\begin{mathletters}\label{VI1}\begin{eqnarray}
&&\delta_{\rm Leibniz}{\hat Y}_i=0\;,\\
&&\delta_{\rm Leibniz}{\hat Z}_{ij}=0\;,
\end{eqnarray}\end{mathletters}$\!\!$
while the contribution to the scalar potential ${\hat T}$, also in the
case of ${\sc D}^{\rm part}_i[F]$, is found to be

\begin{equation}\label{VI2}
\delta_{\rm Leibniz}{\hat T}
=-\frac{G^3m_1^3}{ 96}v_1^iv_1^j\partial_{ij}
\left(\frac{1}{ r_1}\right)+\frac{11}{
36} \frac{G^4m_1^3m_2}{r_{12}^2}~\!n_{12}^i\partial_i
\left(\frac{1}{ r_1}\right)
+1\leftrightarrow
2 \;.
\end{equation}
The modification of the acceleration of body 1 which is generated by
the latter Leibniz terms reads as

\begin{equation}\label{VI3}
\delta_{\rm Leibniz}a_1^i=\frac{G^3m_2^3}{ c^6r_{12}^4}
\left[\frac{5}{
2}(n_{12}v_2)^2n_{12}^i-\frac{1}{ 2}v_2^2n_{12}^i
-(n_{12}v_2)v_2^i\right] 
-\frac{88}{ 9}\frac{G^4m_1m_2^3}{ c^6r_{12}^5}n_{12}^i\;.
\end{equation}

On the other hand, we computed in Sections \ref{IV} and \ref{V} many
distributional terms associated with the derivative of the non-linear
potentials in the right-hand side of the field equations [see
(\ref{III13})].  Most of these terms are simply given by the Schwartz
distributional derivative. The only terms which require the new
distributional derivative of \cite{BFreg} come from the computation of
the ``self'' parts of the non-compact potentials (see Section
\ref{V}). In this case, the modifications of the potentials have been
found to be given by (\ref{self}), from which we obtain the following
modification of the acceleration:

\begin{equation}\label{VI4}
\delta_{\rm self}a_1^i=\frac{G^3m_2^3}{ c^6r_{12}^4}\left[-\frac{1}{
2}(n_{12}v_2)^2n_{12}^i+\frac{1}{ 10}v_2^2n_{12}^i+\frac{1}{
5}(n_{12}v_2)v_2^i\right] 
+\frac{151}{ 9}\frac{G^4m_1m_2^3}{ c^6r_{12}^5}n_{12}^i\;.
\end{equation}
Adding up (\ref{VI3}) and (\ref{VI4}) we therefore obtain the total
effect of the (particular) distributional derivative as

\begin{equation}\label{VI5}
\delta_{\rm distribution}a_1^i=\frac{G^3m_2^3}{
c^6r_{12}^4}\left[2(n_{12}v_2)^2 
n_{12}^i-\frac{2}{ 5}v_2^2n_{12}^i-\frac{4}{ 5}(n_{12}v_2)v_2^i\right]
+7\frac{G^4m_1m_2^3}{ c^6r_{12}^5}n_{12}^i \;.
\end{equation}
Interestingly, this quite simple piece of the acceleration of particle
1 involves the velocity of particle 2 alone, and therefore does not
stay by itself invariant under a Lorentz transformations, or, rather,
at this order, a Galilean transformation (indeed, for this to be true
the term should depend on the relative velocity ${\bf v}_{12}={\bf
v}_1-{\bf v}_2$). Therefore, if we are correct, since we are using
harmonic coordinates and have employed a Lorentzian
regularization, the result (\ref{VI5}) has to combine with other
pieces in the acceleration so as to maintain the Lorentz invariance of
the equations. We have found that this is exactly what happens: the
dependence of (\ref{VI5}) over the velocity ${\bf v}_2$ is mandatory
for the Lorentz invariance of the final 3PN equations to work. This
constitutes, in our opinion, an important check of the relevance of
the distributional derivative introduced in \cite{BFreg}. It shows
also that this derivative is merely a tool for preventing a breakdown
of the Lorentz invariance when performing integrations by parts of
complicated divergent integrals [the last term in (\ref{VI5}), which
is not checked by the Lorentz invariance, will turn out to be absorbed
into the adjustement of a certain constant, see Section \ref{VII}].

The previous check has been done with the ``particular''
distributional derivative (\ref{II8}), and it is interesting to redo
the computation in the case of the distributional derivative defined
by (\ref{II9})-(\ref{II10}), that we recall is more satisfying than
the particular one because it obeys the rule of commutation of
successive derivatives. (But note in passing that we have verified
that the particular derivative does not yield any ambiguity at the 3PN
order which would be due to the non-commutation of derivatives;
however, such ambiguities could arise at higher orders, in which case
the ``correct'' derivative would be more appropriate.) In particular,
while the particular derivative is entirely deterministic, the
derivative (\ref{II9})-(\ref{II10}) depends on a constant $K$, and it
is important to know the fate of this constant in the final equations
of motion, and how the test of the Lorentz invariance will manage to
be satisfied {\it in fine}. Like for the particular derivative we find
that the incidence of this derivative is through two distinct
contributions, Leibniz and self. Consider the Leibniz contribution: we
perform exactly the same computation as before, i.e. based on the
formulas in the appendix \ref{D}, and find that in the case of the new
derivative (\ref{II9})-(\ref{II10}) the terms in the potentials ${\hat
Y}_i$ and ${\hat Z}_{ij}$ are no longer zero, but are given by

\begin{mathletters}\label{VI51}\begin{eqnarray}
&&{\delta_{\rm Leibniz}{\hat
Y}_i}_{~\!\big|_K}=\left(-\frac{1}{15}+\frac{2}{15}K\right)G^3m_1^3
v_1^j\partial_{ij}\left(\frac{1}{ r_1}\right)+1\leftrightarrow 2\;,\\
&&{\delta_{\rm Leibniz}{\hat
Z}_{ij}}_{~\!\big|_K}=\left(-\frac{1}{15}+\frac{2}{15}K\right)G^3m_1^3
\partial_{ij}\left(\frac{1}{ r_1}\right)+1\leftrightarrow 2
\;.\label{VI51b}
\end{eqnarray}\end{mathletters}$\!\!$
Our convention is that the explicit indication in the left-hand side
of the dependence over $K$ means that the computation is performed
using the ``correct'' distributional derivative. In the case of the
modification of the potential ${\hat T}$ the things are a little more
complicated because we have to take into account, in addition to a
``linear'' contribution similar to those of (\ref{VI51}), the
``non-linear'' term which is generated by the modification of the
tensor potential ${\hat Z}_{ij}$ shown in (\ref{VI51b}); {\it cf} the
source term ${\hat Z}_{ij}\partial_{ij}V$ in the definition
(\ref{III24a}) of ${\hat T}$. We obtain

\begin{eqnarray}\label{VI52}
{\delta_{\rm Leibniz}{\hat
T}}_{~\!\big|_K}&=&\left(-\frac{53}{480}+\frac{2}{5}K\right)G^3m_1^3
v_1^iv_1^j\partial_{ij}\left(\frac{1}{ r_1}\right)\nonumber\\
&+&\left(\frac{19}{288}+\frac{47}{24}K\right)
\frac{G^4m_1^3m_2}{r_{12}^2}n_{12}^i
\partial_{i}\left(\frac{1}{ r_1}\right)\nonumber\\
&+&\Delta^{-1}\left[{\delta_{\rm Leibniz}
{\hat Z}_{ij}}_{~\!\big|_K}\partial_{ij}U\right]
+1\leftrightarrow 2\;,
\end{eqnarray}
where $U=\frac{Gm_1}{r_1}+\frac{Gm_2}{r_2}$. Now, using the results of
the previous subsection, we see that many of these terms are in the
form of a gauge transformation corresponding to a gauge vector
$\xi^\mu$ of the type (\ref{VI33}). Indeed, we pose

\begin{mathletters}\label{VI53}\begin{eqnarray}
{\epsilon_1}_{~\!\big|_K}&=&\frac{8}{15}\left(1-2K\right)
\left(\frac{m_1}{m}\right)^3\;,\\ 
{\epsilon_2}_{~\!\big|_K}&=&\frac{8}{15}\left(1-2K\right)
\left(\frac{m_2}{m}\right)^3\;.
\end{eqnarray}\end{mathletters}$\!\!$
With this choice the Leibniz corrections in ${\hat Y}_i$ and
${\hat Z}_{ij}$ become pure gauge,

\begin{mathletters}\label{VI54}\begin{eqnarray}
{\delta_{\rm Leibniz}{\hat Y}_i}_{~\!\big|_K}&=&{\delta_\xi 
{\hat Y}_{i}}_{~\!\big|_K}\;,\\
{\delta_{\rm Leibniz}{\hat Z}_{ij}}_{~\!\big|_K}&=&{\delta_\xi 
{\hat Z}_{ij}}_{~\!\big|_K}\;,
\end{eqnarray}\end{mathletters}$\!\!$
while we can re-write (\ref{VI52}) in the simplified form
 
\begin{eqnarray}\label{VI55}
{\delta_{\rm Leibniz}{\hat
T}}_{~\!\big|_K}&=&\left(-\frac{37}{480}+\frac{1}{3}K\right)G^3m_1^3
v_1^iv_1^j\partial_{ij}\left(\frac{1}{ r_1}\right)
\nonumber\\
&+&\left(\frac{19}{288}+\frac{47}{24}K\right)
\frac{G^4m_1^3m_2}{r_{12}^2}n_{12}^i
\partial_{i}\left(\frac{1}{ r_1}\right)+1\leftrightarrow 2
+{\delta_\xi {\hat T}}_{~\!\big|_K}\;.
\end{eqnarray}
The corresponding modification of the acceleration of particle 1 is
found to be

\begin{eqnarray}\label{VI56}
\delta_{\rm Leibniz}{a_1^i}_{~\!\big|_K}&=&
\left(\frac{37}{2}-80K\right)\frac{G^3m_2^3}{
c^6r_{12}^4}\left[(n_{12}v_2)^2
n_{12}^i-\frac{1}{5}v_2^2n_{12}^i-\frac{2}{
5}(n_{12}v_2)v_2^i\right]\nonumber\\
&+&\left(-\frac{19}{9}-\frac{188}{3}K\right)
\frac{G^4m_1m_2^3}{c^6r_{12}^5}n_{12}^i+{\delta_\xi
a_1^i}_{~\!\big|_K} \;,
\end{eqnarray}
where the last term represents the gauge term (\ref{VI38}) but
computed with (\ref{VI53}). Therefore, modulo a change of gauge, we
see that the Leibniz modification of the acceleration brought about by
the correct derivative has exactly the same form as that, given by
(\ref{VI3}), due to the particular one. However, we must also include
the contribution of the self terms. We have redone the computation of
the self terms as in Section \ref{V} but using the $K$-dependent
derivative and compared the corresponding acceleration with the
previous result (\ref{VI4}). We get

\begin{eqnarray}\label{VI57}
\delta_{\rm self}{a_1^i}_{~\!\big|_K}-\delta_{\rm
self}{a_1^i}&=&\frac{G^3m_2^3}{
c^6r_{12}^4}\left[\frac{9}{2}(n_{12}v_2)^2n_{12}^i
-\frac{9}{10}v_2^2n_{12}^i
-\frac{9}{5}(n_{12}v_2)v_2^i\right]\nonumber\\
&+&\left(-\frac{20}{3}+\frac{44}{3}K\right)\frac{G^4m_1m_2^3}{
c^6r_{12}^5}n_{12}^i \;.
\end{eqnarray}
Subtracting (\ref{VI3}) and (\ref{VI56}) for the Leibniz terms, and
adding up the difference of self terms given by (\ref{VI57}), we
thereby obtain the difference between the total effects of the two
distributional derivatives in the acceleration as

\begin{eqnarray}\label{VI58}
\delta_{\rm distribution}{a_1^i}_{~\!\big|_K}-\delta_{\rm
distribution}{a_1^i}&=&\left(\frac{41}{2}-80K\right)\frac{G^3m_2^3}{
c^6r_{12}^4}\left[(n_{12}v_2)^2
n_{12}^i-\frac{1}{5}v_2^2n_{12}^i-\frac{2}{
5}(n_{12}v_2)v_2^i\right]\nonumber\\
&+&\left(1-48K\right)\frac{G^4m_1m_2^3}{c^6r_{12}^5}n_{12}^i+{\delta_\xi
a_1^i}_{~\!\big|_K} \;.
\end{eqnarray}
As we see, there is a dependence on the individual velocity ${\bf
v}_2$ which is left out. Anticipating the result that $a_1^i$,
computed with the particular derivative, is invariant under Lorentz
transformations, this means that the $K$-dependent derivative breaks
down the Lorentz invariance for general values of $K$ (indeed the
gauge term cannot modify the behaviour under Lorentz
transformations). Fortunately, we are now able to {\it fine tune} the
constant $K$ so that the velocity-dependent terms in (\ref{VI58})
vanish. Therefore, we obtain a unique value,

\begin{equation}\label{VI59}
K=\frac{41}{160}\;,
\end{equation}
for which the equations of motion computed with the help of the
correct derivative (\ref{II9})-(\ref{II10}) are Lorentz invariant, as
they are with the particular derivative. 

Thus, in the case of the correct derivative, the equations of motion
are not in general Lorentz-invariant, despite the use of the
Lorentzian regularization. The likely reason is that the
distributional derivatives were not defined in a ``Lorentzian'' way
(their distributional terms involve the delta-pseudo-function ${\rm
Pf}\delta_1$ and not the Lorentzian one ${\rm Pf}\Delta_1$). Recall
that the Lorentzian regularization permitted to add some crucial
contributions, proportional to $m_1^2m_2$ in the acceleration of
particle 1 [see for instance (\ref{diXCNC})], which are mandatory for
satisfying the Lorentz invariance. In the case of the correct
derivative, we find that there is still a limited class of terms,
proportional to $m_2^3$, which do not obey the Lorentz invariance,
unless $K$ is adjusted to the unique value (\ref{VI59}). Finally we
obtain

\begin{equation}\label{VI60}
\delta_{\rm distribution}{a_1^i}_{~\!\big|_{{\textstyle\frac{41}{160}}}}
-\delta_{\rm distribution}{a_1^i} =-\frac{113}{10}\frac{G^4m_1m_2^3}{
c^6r_{12}^5}n_{12}^i +{\delta_\xi
a_1^i}_{~\!\big|_{{\textstyle\frac{41}{160}}}}\;.
\end{equation}
We shall see in Section \ref{VII} that the effect of the first term is
simply to modify a logarithmic constant $\ln (r'_2/s_2)$ that we shall
adjust when we look for a conserved 3PN energy. After adjustement of
this constant we find that the 3PN equations of motion computed with
the two derivatives are physically the same since they merely differ
by the gauge transformation appearing in (\ref{VI60}).

\subsection{Non-distributivity contributions}

The distributive parts of the linear momentum $P_1^i$ and force
$F_1^i$ densities have been written down in (\ref{III32}). They were
obtained under the uncorrect hypothesis of distributivity, that is
$[FG]_{\textstyle {}_1}=[F]_{\textstyle {}_1}[G]_{\textstyle {}_1}$,
and we must now correct for this. (As explained in Section \ref{III},
our strategy has been to delineate as much as possible the problems,
by concentrating our attention first on the computation of the
regularized values of the potentials when taken individually, second
on the corrections due to the non-distributivity,
i.e. $[FG]_{\textstyle {}_1}\not=[F]_{\textstyle {}_1}[G]_{\textstyle
{}_1}$.) Again, we find that such a subtlety as the non-distributivity
makes a difference starting precisely at the 3PN order. We get for
the required corrections in $P_1^i$ and $F_1^i$:

\begin{mathletters}\label{VI9}\begin{eqnarray}
P_1^i-(P_1^i)_{\textstyle {}_{\rm distr}}&=&\frac{1}{
c^4}\biggl(v_1^i[V^2]_{\textstyle {}_1}-v_1^i[V]_{\textstyle
{}_1}^2\biggr)\nonumber\\ &+&\frac{1}{
c^6}\biggl(12v_1^i[V_jV_j]_{\textstyle {}_1}-12v_1^i[V_j]_{\textstyle
{}_1}[V_j]_{\textstyle {}_1}+2v_1^i[V^3]_{\textstyle {}_1}\nonumber\\
&&\quad -3v_1^i[V]_{\textstyle {}_1}[V^2]_{\textstyle
{}_1}+v_1^i[V]_{\textstyle {}_1}^3\nonumber\\ &&\quad -8[V_j{\hat
W}_{ij}]_{\textstyle {}_1}+8[V_j]_{\textstyle {}_1}[{\hat
W}_{ij}]_{\textstyle {}_1}+8v_1^j[V{\hat W}_{ij}]_{\textstyle
{}_1}-8v_1^j[V]_{\textstyle {}_1}[{\hat W}_{ij}]_{\textstyle
{}_1}\nonumber\\ &&\quad -8[V^2V_i]_{\textstyle
{}_1}+4[V^2]_{\textstyle {}_1}[V_i]_{\textstyle {}_1}+4[V]_{\textstyle
{}_1}^2[V_i]_{\textstyle {}_1}\nonumber\\ &&\quad +\frac{1}{
2}v_1^2v_1^i[V^2]_{\textstyle {}_1}-\frac{1}{
2}v_1^2v_1^i[V]_{\textstyle {}_1}^2-16v_1^j[V_iV_j]_{\textstyle
{}_1}+16v_1^j[V_i]_{\textstyle {}_1}[V_j]_{\textstyle
{}_1}\biggr)\;,\\ F_1^i-(F_1^i)_{\textstyle {}_{\rm
distr}}&=&\frac{1}{ c^2}\biggl(-2[V\partial_iV]_{\textstyle
{}_1}+2[V]_{\textstyle {}_1}[\partial_iV]_{\textstyle
{}_1}\biggr)\nonumber\\ &+&\frac{1}{ c^4}\biggl(-v_1^2[V]_{\textstyle
{}_1}[\partial_iV]_{\textstyle {}_1}+v_1^2[V\partial_iV]_{\textstyle
{}_1} -8[V_j]_{\textstyle {}_1}[\partial_iV_j]_{\textstyle
{}_1}+8[V_j\partial_iV_j]_{\textstyle {}_1}\nonumber\\ &&\quad
+[V]_{\textstyle {}_1}^2[\partial_iV]_{\textstyle
{}_1}-[V^2]_{\textstyle {}_1}[\partial_iV]_{\textstyle {}_1} +
2[V^2\partial_iV]_{\textstyle {}_1} - 2[V]_{\textstyle
{}_1}[V\partial_iV]_{\textstyle {}_1}\biggr)\nonumber\\ &+&\frac{1}{
c^6}\biggl(-\frac{1}{ 4}v_1^4[V]_{\textstyle
{}_1}[\partial_iV]_{\textstyle {}_1}+\frac{1}{
4}v_1^4[V\partial_iV]_{\textstyle {}_1}\nonumber\\ &&\quad +\frac{3}{
2}v_1^2[V]_{\textstyle {}_1}^2[\partial_iV]_{\textstyle
{}_1}-\frac{3}{ 2}v_1^2[V^2]_{\textstyle
{}_1}[\partial_iV]_{\textstyle
{}_1}+3v_1^2[V^2\partial_iV]_{\textstyle {}_1} - 3v_1^2[V]_{\textstyle
{}_1}[V\partial_iV]_{\textstyle {}_1}\nonumber\\ &&\quad +\frac{8}{
3}[V]_{\textstyle {}_1}^3[\partial_iV]_{\textstyle
{}_1}-3[V]_{\textstyle {}_1}[V^2]_{\textstyle
{}_1}[\partial_iV]_{\textstyle {}_1}+\frac{2}{ 3}[V^3]_{\textstyle
{}_1}[\partial_iV]_{\textstyle {}_1}\nonumber\\ &&\quad
-3[V]_{\textstyle {}_1}^2[V\partial_iV]_{\textstyle
{}_1}+2[V^2]_{\textstyle {}_1}[V\partial_iV]_{\textstyle
{}_1}+2[V]_{\textstyle {}_1}[V^2\partial_iV]_{\textstyle {}_1} -
\frac{4}{ 3}[V^3\partial_iV]_{\textstyle {}_1}\nonumber\\ &&\quad 
-16[\partial_iV_j]_{\textstyle {}_1}[{\hat R}_{j}]_{\textstyle
{}_1}+16[\partial_iV_j{\hat R}_{j}]_{\textstyle
{}_1}-16[V_j]_{\textstyle {}_1}[\partial_i{\hat R}_{j}]_{\textstyle
{}_1}+16[V_j\partial_i{\hat R}_{j}]_{\textstyle {}_1}\nonumber\\
&&\quad +8[V]_{\textstyle {}_1}[\partial_iV_j]_{\textstyle
{}_1}[V_j]_{\textstyle {}_1}+8[V]_{\textstyle
{}_1}[\partial_iV_jV_j]_{\textstyle {}_1}-
16[V\partial_iV_jV_j]_{\textstyle {}_1}\nonumber\\ &&\quad
+4[\partial_iV]_{\textstyle {}_1}
[V_j]_{\textstyle {}_1}[V_j]_{\textstyle {}_1}
+4[\partial_iV]_{\textstyle {}_1}[V_jV_j]_{\textstyle {}_1}
-8[\partial_iVV_jV_j]_{\textstyle {}_1}\nonumber\\ &&\quad
-12v_1^2[\partial_iV_j]_{\textstyle {}_1}[V_j]_{\textstyle
{}_1}+12v_1^2[\partial_iV_jV_j]_{\textstyle {}_1} \nonumber\\ &&\quad
+4v_1^j[V]_{\textstyle {}_1}^2[\partial_iV_j]_{\textstyle
{}_1}+4v_1^j[V^2]_{\textstyle {}_1}[\partial_iV_j]_{\textstyle {}_1} -
8v_1^j[V^2\partial_iV_j]_{\textstyle {}_1}\nonumber\\ &&\quad
+8v_1^j[V]_{\textstyle {}_1}[\partial_iV]_{\textstyle
{}_1}[V_j]_{\textstyle {}_1}+8v_1^j[V\partial_iV]_{\textstyle
{}_1}[V_j]_{\textstyle {}_1}- 16v_1^j[V\partial_iV V_j]_{\textstyle
{}_1}\nonumber\\ &&\quad +8v_1^j[V_k]_{\textstyle
{}_1}[\partial_i{\hat W}_{jk}]_{\textstyle
{}_1}-8v_1^j[V_k\partial_i{\hat W}_{jk}]_{\textstyle
{}_1}+8v_1^j[\partial_iV_k]_{\textstyle {}_1}[{\hat
W}_{jk}]_{\textstyle {}_1}-8v_1^j[\partial_iV_k{\hat
W}_{jk}]_{\textstyle {}_1}\nonumber\\ &&\quad
-4v_1^jv_1^k[V]_{\textstyle {}_1}[\partial_i{\hat W}_{jk}]_{\textstyle
{}_1}+4v_1^jv_1^k[V\partial_i{\hat W}_{jk}]_{\textstyle
{}_1}\nonumber\\ &&\quad 
-4v_1^jv_1^k[\partial_iV]_{\textstyle {}_1}[{\hat W}_{jk}]_{\textstyle
{}_1}+4v_1^jv_1^k[\partial_iV{\hat W}_{jk}]_{\textstyle
{}_1}\nonumber\\ &&\quad +16v_1^jv_1^k[\partial_iV_k]_{\textstyle
{}_1}[V_j]_{\textstyle
{}_1}-16v_1^jv_1^k[\partial_iV_kV_j]_{\textstyle {}_1}
\nonumber\\     
&&\quad +8[{\hat X}]_{\textstyle {}_1}[\partial_iV]_{\textstyle
{}_1}-8[{\hat X}\partial_iV]_{\textstyle {}_1}+8[V]_{\textstyle
{}_1}[\partial_i{\hat X}]_{\textstyle {}_1}-8[V\partial_i{\hat
X}]_{\textstyle {}_1}\biggr)\;.
\end{eqnarray}\end{mathletters}$\!\!$
These formulas look complicated but are in fact rather simple to
evaluate because they require only some lower-order post-Newtonian
precision in the potentials, with notably all the difficult
non-compact potentials needed at the Newtonian order only (hence the
interest of separating out the problems as we did). Note that it is
crucial here to employ the Lorentzian regularization $[F]_{\textstyle
{}_1}$. The net result of this computation is

\begin{mathletters}\label{VI10}\begin{eqnarray}
P_1^i-(P_1^i)_{\textstyle {}_{\rm distr}}&=&\frac{G^3m_1^2m_2}{
c^6r_{12}^3}\biggl(\frac{2}{ 5}(n_{12}v_{12})n_{12}^i -\frac{2}{
15}v_{12}^i\biggr)\;,\\ F_1^i-(F_1^i)_{\textstyle {}_{\rm
distr}}&=&\frac{G^3m_1^2m_2}{ c^6r_{12}^4}\biggl(\left[\frac{241}{
70}\frac{Gm_1}{ r_{12}}-\frac{51}{ 70}\frac{Gm_2}{
r_{12}}\right]n_{12}^i\nonumber\\ &+&\frac{723}{
28}(n_{12}v_{12})^2n_{12}^i -\frac{723}{
140}v_{12}^2n_{12}^i-\frac{723}{ 70}(n_{12}v_{12})v_{12}^i\biggr)\;.
\end{eqnarray}\end{mathletters}$\!\!$
Therefore the supplement of acceleration linked to the
non-distributivity is

\begin{eqnarray}\label{VI11}
a_1^i-(a_1^i)_{\textstyle {}_{\rm distr}}&=&\frac{G^3m_1^2m_2}{
c^6r_{12}^4}\biggl(\left[\frac{779}{ 210}\frac{Gm_1}{
r_{12}}-\frac{97}{ 210}\frac{Gm_2}{ r_{12}}\right]n_{12}^i\nonumber\\
&+&\frac{779}{ 28}(n_{12}v_{12})^2n_{12}^i -\frac{779}{
140}v_{12}^2n_{12}^i-\frac{779}{ 70}(n_{12}v_{12})v_{12}^i\biggr)\;.
\end{eqnarray}
Since only the relative velocity ${\bf v}_{12}$ is involved this part
of the acceleration is Galilean invariant. Furthermore, it can be
expressed in a simpler way by introducing an infinitesimal gauge
transformation of the type (\ref{VI33}). We pose

\begin{mathletters}\label{VI11'}\begin{eqnarray}
(\epsilon_1)_{\textstyle {}_{\rm
distr}}&=&-\frac{779}{420}\frac{m_1m_2^2}{m^3}\;,\\
(\epsilon_1)_{\textstyle {}_{\rm
distr}}&=&-\frac{779}{420}\frac{m_1^2m_2}{m^3}\;,
\end{eqnarray}\end{mathletters}$\!\!$
and easily obtain

\begin{equation}\label{VI11''}
a_1^i-(a_1^i)_{\textstyle {}_{\rm distr}}=(\delta_\xi
a_1^i)_{\textstyle {}_{\rm
distr}}+\frac{G^4m_1m_2^2}{c^6r_{12}^5}\left[-\frac{97}{210}m_1
+\frac{779}{210}m_2\right]n_{12}^i\;.
\end{equation}
Thus, the only Physics brought about by the non-distributivity
(i.e. which is not affected by a gauge transformation) is constituted
by the quartic ($G^4$) term displayed in the right-hand side of
({\ref{VI11''}).

\section{The 3PN equations of motion}\label{VII}

\subsection{Existence of the conserved energy}

At present, the equations of motion are complete. We want now to look
for the conserved energy associated with these equations at the 3PN
order (considering of course only the conservative part of the
equations, i.e. excluding the radiation reaction acceleration at the
2.5PN order). We shall see that the existence of an energy is not
immediate, but requires the adjustment of a certain constant.

We proved in Section \ref{V} that the equations of motion of body 1
depend on two arbitrary constants, which are the constant $r'_1$,
tending to zero as we approach the particle 1 (but considered here as
taking some finite non-zero value), and the constant $s_2$ associated
with the Hadamard regularization near the other particle 2 [see
(\ref{II3})]. Similarly, the equations of body 2 depend on the
constants $r'_2$ and $s_1$. All these constants appear inside the
logarithms entering the equations of motion in harmonic
coordinates. Gathering the results for the ``logarithmic'' part of the
equations of body 1, we obtain

\begin{eqnarray}\label{VI12}
a_1^i&=&\frac{44}{3}\frac{G^4m_1^3m_2}{c^6r_{12}^5}n_{12}^i\ln
\left(\frac{r_{12}}{r'_1}\right)-\frac{44}{3}\frac{G^4m_1m_2^3}
{c^6r_{12}^5}n_{12}^i\ln\left(\frac{r_{12}}{s_2}\right)\nonumber\\
&+&\frac{G^3m_1^2m_2}{c^6r_{12}^4}\Big[110(n_{12}v_{12})^2n_{12}^i
-22v_{12}^2n_{12}^i-44(n_{12}v_{12})v_{12}^i\Big]
\ln\left(\frac{r_{12}}{r'_1}
\right)+\dots\;,
\end{eqnarray}
where the dots indicate the terms which do not contain any logarithms.
The terms shown in (\ref{VI12}) contain the whole dependence of the
acceleration of 1 over $r'_1$ and $s_2$; there are no other constants
elsewhere. Notice that $s_2$ enters a single quartic-order term
proportional to $G^4m_1m_2^3$. Now, most of the terms in (\ref{VI12})
can in fact be gauged away. To see this, we apply the formula
(\ref{VI38}) with the particular choice

\begin{mathletters}\label{VI40}\begin{eqnarray}
(\epsilon_1)_{\ln}
&=&-\frac{22}{3}\frac{m_1m_2^2}{m^3}
\ln\left(\frac{r_{12}}{r'_2}\right)\;,\\
(\epsilon_2)_{\ln}
&=&-\frac{22}{3}\frac{m_1^2m_2}{m^3}
\ln\left(\frac{r_{12}}{r'_1}\right)\;.
\end{eqnarray}\end{mathletters}$\!\!$
The corresponding transformation of the acceleration is
(\ref{VI38}), except that $(\epsilon_1)_{\ln}$ and
$(\epsilon_2)_{\ln}$ depend on time through the orbital separation
$r_{12}$, so in fact this formula should contain also some
terms proportional to the time-derivatives of $(\epsilon_1)_{\ln}$ and
$(\epsilon_2)_{\ln}$; but the point for us is that these extra terms
are free of any logarithms. Therefore, modulo the dots indicating the
logarithmic-free terms, we can write

\begin{equation}\label{VI13}
a_1^i=(\delta_\xi a_1^i)_{\ln}
-\frac{44}{3}\frac{G^4m_1m_2^3}{c^6r_{12}^5}n_{12}^i
\ln\left(\frac{r'_2}{s_2}\right)+\dots\;,
\end{equation}
where $(\delta_\xi a_1^i)_{\ln}$ denotes the coordinate change of the
acceleration.

The term in (\ref{VI13}) which is left out after this coordinate
change depends only on the ratio between $r'_2$ and $s_2$ (similarly,
in the equations of motion of body 2, we would find the ratio of
$r'_1$ and $s_1$). This term is of the same type as the one in
(\ref{VI60}) giving the difference of accelerations, modulo a change
of gauge, when different distributional derivatives are used. Notice
that the constant $r'_2$ was originally absent from the equations of
motion of 1, but has to be introduced in order to ``remove'' these
logarithms by the coordinate transformation. Therefore, the only
physical freedom remaining in the equations of motion is the yet
unspecified constant $\ln\left(\frac{r'_2}{s_2}\right)$. Now we use
this freedom to find a conserved energy associated with the equations
of motion, which means a local-in-time functional $E$ of the
trajectories and velocities of the two particles, i.e.

\begin{equation}\label{VI14}
E=E[{\bf y}_{1}(t),{\bf y}_{2}(t);{\bf v}_{1}(t),{\bf v}_{2}(t)]\;,
\end{equation}
which is constant as a consequence of the 3PN equations of motion,
i.e.

\begin{equation}\label{VI16}
\frac{dE}{dt}\equiv v_1^i\frac{\partial E}{\partial y_1^i}
+v_2^i\frac{\partial E}{\partial y_2^i}
+a_1^i\frac{\partial E}{\partial v_1^i}+a_2^i\frac{\partial
E}{\partial v_2^i}={\overline {\cal O}}(7)\;.
\end{equation} 
The accelerations ${\bf a}_1$ and ${\bf a}_1$ are to be replaced by
the functionals of the positions and velocities given by the 3PN
equations of motion. Our special notation for the remainder means a
radiation-reaction term which is purely of order 2.5PN plus the
neglected terms at 3.5PN; schematically ${\overline {\cal
O}}(7)=\frac{1}{c^5}F_5+{\cal O}(7)$. See (\ref{VII2'}) below for the
expression of the term $\frac{1}{c^5}F_5$. If an energy exists, the
quantity $m_1 a_1^iv_1^i+m_2 a_2^iv_2^i$ must be a total time
derivative. In practive, we look for a local-in-time functional
$D[{\bf y}_{1},{\bf y}_{2};{\bf v}_{1},{\bf v}_{2}]$ such that

\begin{equation}\label{VI15}
m_1 a_1^iv_1^i+m_2 a_2^iv_2^i+\frac{d D}{dt}={\overline {\cal O}}(7)\;,
\end{equation} 
and we obtain the energy as $E=\frac{1}{2}m_1 {\bf
v}_1^2+\frac{1}{2}m_2 {\bf v}_2^2+D$.  Now, the computation with our
3PN equations of motion (obtained by means of,
say, the particular derivative) shows that the quantity $D$ does not
exist for any values of the constants $\ln\left(r'_2/s_2\right)$ and
$\ln\left(r'_1/s_1\right)$. However, we find that this ``nearly''
works, because we can determine some ${\hat D}$ such that

\begin{equation}\label{VI17}
m_1 a_1^iv_1^i+m_2 a_2^iv_2^i+\frac{d{\hat D}}{dt}=
-\frac{44}{3}\frac{G^4m_1^2m_2^2}{c^6r_{12}^5}\left\{m_2(n_{12}v_1)
\left[\ln\left(\frac{r'_2}{s_2}\right)-\frac{159}{308}\right]
+1\leftrightarrow 2 \right\}+{\overline {\cal O}}(7)\;.
\end{equation}
From the computation we obtain ${\hat D}$ as a well-defined local
functional of the positions and velocities of the particules
[containing in particular some logarithms
$\ln\left(r_{12}/r'_1\right)$ and $\ln\left(r_{12}/r'_2\right)$].  The
right-hand side of (\ref{VI17}) cannot be written, for generic values
of $\ln\left(r'_2/s_2\right)$ and $\ln\left(r'_1/s_1\right)$, in the
form of a total time-derivative. It would be possible, for this to be
the case, to adopt the simplest choice that both these constants are
numerically equal to $\frac{159}{308}$. However, this choice does not
represent the most general solution for obtaining a total
time-derivative. Indeed, nothing prevents $\ln\left(r'_1/s_1\right)$
and $\ln\left(r'_2/s_2\right)$ to depend also on the masses $m_1$ and
$m_2$, and therefore such a dependence on the masses should in fact be
mandatory (totalitarian principle). Since the regularization procedure
followed in this paper is more mathematical than physical, we can be
confident that no Physics will be overlooked only if at each step we
obtain the most general solution allowed by the
process. Unfortunately, the most general solution in this case
contains an arbitrary parameter.

The necessary and sufficient condition for the right side of
(\ref{VI17}) to be a total time-derivative is that the factor of
$(n_{12}v_1)$ in (\ref{VI17}) be invariant by exchanging the
particle's labels 2 and 1, i.e.

\begin{equation}\label{VI18'}
m_2\left[\ln\left(\frac{r'_2}{s_2}\right)-\frac{159}{308}\right]
=m_1\left[\ln\left(\frac{r'_1}{s_1}\right)-\frac{159}{308}\right]\;.
\end{equation} 
Denoting by $\lambda m$ the common value of both sides of
(\ref{VI18'}), where $\lambda$ is a constant and $m=m_1+m_2$, we
obtain the most general solution as

\begin{mathletters}\label{VI18}\begin{eqnarray}
\ln\left(\frac{r'_2}{s_2}\right)&=&\frac{159}{308}
+\lambda\frac{m}{m_2}\;,\label{VI18a}\\
\ln\left(\frac{r'_1}{s_1}\right)&=&\frac{159}{308}
+\lambda\frac{m}{m_1}\;.
\end{eqnarray}\end{mathletters}$\!\!$
This $\lambda$ is a dimensionless quantity which is the same for the
two particles 1 and 2. We now prove that $\lambda$ is necessarily a
pure {\it numerical} constant, independent of the masses. Notice that
the $\lambda$-term in (\ref{VI18a}) will yield a contribution to the
acceleration of 1 which is, as concerns the dependence over the
masses, of the type $m_1m_2^2~\!m~\!\lambda$ [see (\ref{VI13})]. If
$\lambda$ depends on the masses, it must be a symmetric function of
$m_1$ and $m_2$, and therefore it can be expressed solely in terms of
the symmetric mass ratio $\nu=\frac{m_1m_2}{m^2}$. Suppose that
$\lambda=\sum_{-\infty}^{+\infty} \lambda_i\nu^i$, where the
$\lambda_i$'s are numerical constants, so the $\lambda$-term in the
acceleration of 1 is of the type $m_1m_2^2~\!m\sum
\lambda_i\nu^i$. First, we see that all the cases $i\leq -1$ are
excluded because the equations of motion would not have the correct
perturbative limit when $\nu\to 0$; for instance, in the case $i=-1$,
we get a term of the type $m^3 m_2$ which tends to $m_2^4$ in this
limit, and therefore modifies the geodesic motion of a test particle
around a Schwarzschild black hole, which is of course
excluded. Second, the cases $i\geq 1$, though they pass the simplest
physical requirements, imply that the individual particle
accelerations are no longer polynomials in the two individual masses
$m_1$ and $m_2$, because of the appearance of inverse powers of the
total mass $m=m_1+m_2$. For instance, the case $i=1$ leads to a term
of the type $\frac{m_1^2~\!m_2^3}{m}$. But we know that when doing a
diagrammatic expansion of the $N$ body problem based on the
post-Minkowskian expansion (see \cite{DEF96} for the details of the
method) that each successive diagram is a polynomial of the $N$
masses. Therefore, we exclude the possibility that some inverse powers
of the total mass appear, and find, in conclusion, that $\lambda$ is a
pure constant ($\lambda=\lambda_0$).

At last, we have succeeded in finding a conserved energy at the 3PN
order by specifying an unknown logarithmic ratio, but at the price of
having introduced an arbitrary purely numerical constant
$\lambda$. The constant $\lambda$ will be left undetermined in the
present work. So the final 3PN equations of motion we obtain in this
paper, as well as the final 3PN energy, depend on the unknown
parameter $\lambda$. The appearance of $\lambda$ suggests that the
present formalism, based on a point-mass regularization, is physically
incomplete. The resulting ambiguity is equivalent to the ``static''
ambiguity found by Jaranowski and Sch\"afer \cite{JaraS99}. It is
probably linked to the fact that one can write the Einstein field
equations into many different forms, which are all equivalent in the
case of regular sources, but which are in general not equivalent in
the case of point-particles because the distributional derivative does
not obey the Leibniz rule. If we had chosen initially a different form
of the field equations, the Leibniz terms we computed in Section
\ref{VI}, could have been different.  More precisely, only that part
of the Leibniz terms which is Galilean-invariant and consequently is
not required by the Lorentz-invariance symmetry could change. But we
have seen in (\ref{VI3}) and (\ref{VI60}) that the Galilean-invariant
part of the Leibniz terms is precisely made of only one term, which is
of the same type (proportional to $G^4m_1m_2^3$) as the term
containing the constant $\ln (r'_2/s_2)$ [see (\ref{VI13})] that
we have adjusted in the equations (\ref{VI18a}), resulting in the
appearance of the constant $\lambda$. Thus, in agreement with
Jaranowski and Sch\"afer \cite{JaraS99}, we might say that $\lambda$
encodes an ambiguity associated with the violation of the Leibniz rule
by the distributional derivative. At a deeper level, this would mean
that the ambiguity is a consequence of a theorem of Schwartz
\cite{Schwartz54} according to which it is impossible to define at
once a multiplication of distributions which agrees with the ordinary
product for continuous functions, and a derivation of distributions
which satisfies the Leibniz rule and reduces to the ordinary
derivative in the case of $C^1$ functions. If this explanation is
correct, it is unlikely that the constant $\lambda$ could be
determined within the present formalism.

We find by combining (\ref{VI18}) and (\ref{VI17}) that the dependence
of the 3PN energy $E$ on $\lambda$ is

\begin{equation}\label{VI20}
E={\hat E}-\frac{11}{3}\lambda \frac{G^4m_1^2m_2^2m}{c^6r_{12}^4}\;,
\end{equation}
where ${\hat E}$ does not depend on $\lambda$, while we obtain, using
(\ref{VI13}), that the acceleration writes

\begin{equation}\label{VI21}
a_1^i={\hat a}_1^i-\frac{44}{3}\lambda \frac{G^4m_1m_2^2m}
{c^6r_{12}^5}n_{12}^i\;,
\end{equation}
where similarly ${\hat a}_1^i$ is independent of $\lambda$.  On the
other hand, the acceleration and energy depend also on the two
constants $r'_1$ and $r'_2$, but from the previous discussion this is
not a problem because $r'_1$ and $r'_2$ are associated with an
arbitrariness in the choice of coordinates: the 3PN equations of
motion contain the logarithms $\ln\left(r_{12}/r'_1\right)$ and
$\ln\left(r_{12}/r'_2\right)$ which have been shown in (\ref{VI12}) to
be in the form of the gauge transformation associated with
(\ref{VI40}). In particular, the ``constants'' $\ln r'_1$ and $\ln
r'_2$, which might be said to be formally infinite because $r'_1$ and
$r'_2$ were tending to zero [recall the discussion after
(\ref{eq:poissonth})], will never appear in any physical
result. Similarly, the dependence of the energy on the logarithms
$\ln\left(r_{12}/r'_1\right)$ and $\ln\left(r_{12}/r'_2\right)$ is
pure gauge. From (\ref{VI39}) we get

\begin{equation}\label{VI40'}
E=\frac{22}{3}\frac{G^3m_1^3m_2}{c^6r_{12}^3}\Bigg[-\frac{Gm_2}{r_{12}}
+3(n_{12}v_{1})(n_{12}v_{12})-(v_{1}v_{12})\Bigg]\ln\left(
\frac{r_{12}}{r'_1}\right)+1\leftrightarrow
2+\dots\;,
\end{equation}
where the dots denote the terms independent of logarithms [this result
can also be checked directly using (\ref{VI12})].

Finally, to be more specific about the influence of the distributional
derivative, notice that the solution we have obtained in (\ref{VI18})
corresponds to the ``particular'' distributional derivative defined by
(\ref{II8}). If one uses the ``correct'' derivative
(\ref{II9})-(\ref{II10}) instead, with the value $K=\frac{41}{160}$ we
have obtained in (\ref{VI59}) from the Lorentz invariance, we obtain
the same equation to be solved as (\ref{VI17}) but with the rational
fraction $+\frac{783}{3080}$ instead of $-\frac{159}{308}$. This is
easily seen thanks to (\ref{VI60}). So, the solution becomes in this
case

\begin{mathletters}\label{VI19}\begin{eqnarray}
\ln\left(\frac{r'_2}{s_2}\right)&=&-\frac{783}{3080}
+\lambda\frac{m}{m_2}\;,\\
\ln\left(\frac{r'_1}{s_1}\right)&=&-\frac{783}{3080}
+\lambda\frac{m}{m_1}\;.
\end{eqnarray}\end{mathletters}$\!\!$
Replacing this into the equations of motion (and associated energy),
it is then clear that they are physically the same as those computed
with the other derivative, because they differ by the mere change of
gauge,

\begin{equation}\label{VI19'}
{a_1^i}_{~\!\big|_{{\textstyle\frac{41}{160}}}}-{a_1^i}
={\delta_\xi a_1^i}_{~\!\big|_{{\textstyle\frac{41}{160}}}}\;,
\end{equation}
that we obtained in (\ref{VI60}). We give it here thoroughtly for
completeness:

\begin{eqnarray}\label{VI19''}
{\delta_\xi
a_1^i}_{~\!\big|_{{\textstyle\frac{41}{160}}}}&=&\frac{13}{25}
\frac{G^4m_1m_2}{c^6
r_{12}^5}\Big(m_1^2-m_2^2\Big)n_{12}^i
\nonumber\\
&+&\frac{13}{50}\frac{G^3m_2^3}{c^6r_{12}^4}
\Big[-15(n_{12}v_{12})^2n_{12}^i
+3v_{12}^2n_{12}^i+6(n_{12}v_{12})v_{12}^i\Big]\;.
\end{eqnarray}

\subsection{End results}

We present the 3PN equations of motion of the particle 1 in harmonic
coordinates, which are obtained by summing up all the contributions of
the potentials computed in Sections \ref{IV} and \ref{V}, as well as
the pieces due to the non-distributivity and the Leibniz terms (see
Section \ref{VI}). The equations depend on two gauge constants $r'_1$
and $r'_2$ through some logarithms, and on one unknown purely
numerical coefficient $\lambda$. The equations of the particle 2 are
obtained by exchanging all the labels $1\leftrightarrow 2$.

\begin{eqnarray}\label{VII1}
a_1^i &=& -\frac{G m_2 n_{12}^i}{r_{12}^2} \nonumber\\ &+& 
\frac{1}{c^2}
\Bigg\{\bigg[\frac{5 G^2 m_1 m_2}{r_{12}^3} 
+ \frac{4 G^2 m_2^2}{r_{12}^3 }+
\frac{G m_2}{r_{12}^2} \bigg(\frac{3}{2} 
(n_{12}v_2)^2 - v_1^2 + 4 (v_1v_2) - 2 v_2^2\bigg)\bigg] n_{12}^i
\nonumber\\ & &
\qquad + \frac{G m_2}{r_{12}^2} \bigg(4 (n_{12}v_1)  - 3
(n_{12}v_2)\bigg) (v_1^i - v_2^i) \Bigg\} \nonumber\\ &+&
 \frac{1}{c^4}
\Bigg\{\bigg[-\frac{57 G^3 m_1^2 m_2}{4 r_{12}^4}-  \frac{69 G^3 
m_1 m_2^2}{2 r_{12}^4}- \frac{9 G^3 m_2^3}{r_{12}^4} + \frac{G
m_2}{r_{12}^2} \bigg(-\frac{15}{8} (n_{12}v_2)^4 + 
\frac{3}{2} (n_{12}v_2)^2 v_1^2 \nonumber\\ & & \qquad 
- 6 (n_{12}v_2)^2 (v_1v_2) - 2 (v_1v_2)^2 +
\frac{9}{2} (n_{12}v_2)^2 v_2^2 + 4 (v_1v_2) v_2^2 - 2 v_2^4\bigg) 
\nonumber\\
& & 
\qquad + \frac{G^2 m_1 m_2}{ r_{12}^3} 
\bigg(\frac{39}{2} (n_{12}v_1)^2 - 39
(n_{12}v_1) (n_{12}v_2) +
\frac{17}{2} (n_{12}v_2)^2 - \frac{15}{4} v_1^2 - \frac{5}{2} (v_1v_2) 
\nonumber\\ & & \qquad +
\frac{5}{4} v_2^2\bigg) + \frac{G^2 m_2^2}{r_{12}^3} \bigg(2 (n_{12}v_1)^2 -
4  (n_{12}v_1) (n_{12}v_2) - 
6 (n_{12}v_2)^2 - 8 (v_1v_2) + 4 v_2^2\bigg)\bigg] n_{12}^i \nonumber\\ & &
\quad 
+ \bigg[\frac{G^2 m_2^2}{r_{12}^3} \bigg(-2 (n_{12}v_1) 
- 2 (n_{12}v_2)\bigg)
+ \frac{G^2 m_1 
m_2}{r_{12}^3} \bigg(-\frac{63}{4} (n_{12}v_1) +
\frac{55}{4} (n_{12}v_2)\bigg) \nonumber\\ & & \qquad 
+ \frac{G m_2}{r_{12}^2} \bigg(-6 (n_{12}v_1) (n_{12}v_2)^2 + 
\frac{9}{2} (n_{12}v_2)^3 + (n_{12}v_2) v_1^2 - 4 (n_{12}v_1) (v_1v_2) 
\nonumber\\ & & \qquad + 4 (n_{12}v_2) (v_1v_2) 
+ 4 (n_{12}v_1) v_2^2 - 5
(n_{12}v_2) 
 v_2^2\bigg)\bigg] (v_1^i - v_2^i) \Bigg\} \nonumber\\ &+& 
\frac{1}{c^5}\Bigg\{\bigg[\frac{208 G^3 m_1 m_2^2}{15 r_{12}^4}
\bigg((n_{12}v_1) -  (n_{12}v_2)\bigg) 
- \frac{24 G^3 m_1^2 m_2}{5 r_{12}^4}
\bigg((n_{12}v_1) - (n_{12}v_2)\bigg) \nonumber\\ & & \qquad
+ \frac{12 G^2 m_1 m_2}{5 r_{12}^3}
\bigg((n_{12}v_1)- (n_{12}v_2) \bigg) [v_1^2 - 2(v_1v_2)+v_2^2]  
\bigg] n_{12}^i + \bigg[\frac{8 G^3 m_1^2
m_2}{5 r_{12}^4}\nonumber\\ & & \qquad 
- \frac{32 G^3 m_1 m_2^2}{5 r_{12}^4} 
- \frac{4 G^2 m_1 m_2}{5 r_{12}^3}  [v_1^2 - 2(v_1v_2)+v_2^2] \bigg] 
(v_1^i - v_2^i)\Bigg\}
\nonumber\\ 
&+& \frac{1}{c^6} \Bigg\{\bigg[\frac{G m_2}{r_{12}^2}
\bigg(\frac{35}{16} 
(n_{12}v_2)^6 
- \frac{15}{8} (n_{12}v_2)^4 v_1^2 
+ \frac{15}{2} (n_{12}v_2)^4 (v_1v_2) + 3
(n_{12}v_2)^2 (v_1v_2)^2 \nonumber\\ & & \qquad
- \frac{15}{2} (n_{12}v_2)^4 v_2^2 + \frac{3}{2}
(n_{12}v_2)^2 v_1^2 v_2^2 - 12 (n_{12}v_2)^2 (v_1v_2) v_2^2 
- 2 (v_1v_2)^2
v_2^2 \nonumber\\ & & \qquad 
+ \frac{15}{2} (n_{12}v_2)^2 v_2^4 + 4 (v_1v_2) v_2^4 - 2
v_2^6\bigg) + \frac{G^2 m_1 m_2}{ r_{12}^3 } \bigg(-\frac{171}{8}
(n_{12}v_1)^4 \nonumber\\ & & \qquad 
+ \frac{171}{2} (n_{12}v_1)^3 (n_{12}v_2) 
- \frac{723}{4} (n_{12}v_1)^2
(n_{12}v_2)^2 + 
\frac{383}{2} (n_{12}v_1) (n_{12}v_2)^3 \nonumber\\ & & \qquad
- \frac{455}{8} (n_{12}v_2)^4 +
\frac{229}{4} (n_{12}v_1)^2 v_1^2 - \frac{205}{2} (n_{12}v_1) 
(n_{12}v_2)
v_1^2 + \frac{191}{4} (n_{12}v_2)^2 v_1^2 - \frac{91}{8} v_1^4 
\nonumber\\ & &
\qquad 
-  \frac{229}{2} (n_{12}v_1)^2 (v_1v_2) + 244 (n_{12}v_1) (n_{12}v_2)
(v_1v_2) - 
\frac{225}{2} (n_{12}v_2)^2 (v_1v_2)\nonumber\\ & & \qquad
+ \frac{91}{2} v_1^2 (v_1v_2) -
\frac{177}{4} (v_1v_2)^2 + \frac{229}{4} (n_{12}v_1)^2 v_2^2 
- \frac{283}{2}
(n_{12}v_1) (n_{12}v_2) v_2^2 \nonumber\\ & & \qquad
+ \frac{259}{4} (n_{12}v_2)^2 v_2^2 -
\frac{91}{4} v_1^2 v_2^2 + 43 (v_1v_2) v_2^2 - \frac{81}{8} v_2^4\bigg) +
\frac{G^2 
m_2^2}{ r_{12}^3 } \bigg(-6 (n_{12}v_1)^2 (n_{12}v_2)^2 
\nonumber\\ & & \qquad
+ 12 (n_{12}v_1)  (n_{12}v_2)^3 + 6 
(n_{12}v_2)^4 + 4 (n_{12}v_1) (n_{12}v_2) (v_1v_2) + 12 (n_{12}v_2)^2
(v_1v_2)\nonumber\\ & & \qquad
+ 4 (v_1v_2)^2 - 4 (n_{12}v_1) (n_{12}v_2) v_2^2 - 12 (n_{12}v_2)^2
v_2^2 - 8 (v_1v_2) v_2^2 + 4 v_2^4\bigg)  \nonumber\\ & & \qquad  
+ \frac{G^3 m_2^3}{r_{12}^4}  \bigg(-(n_{12}v_1)^2 + 
2 (n_{12}v_1) (n_{12}v_2) + \frac{43}{2} (n_{12}v_2)^2 + 18 (v_1v_2) - 9
v_2^2\bigg) \nonumber\\ & & \qquad
+ \frac{G^3 m_1 m_2^2}{r_{12}^4} \bigg(\frac{415}{8}  (n_{12}v_1)^2 -
\frac{375}{4}  
(n_{12}v_1) (n_{12}v_2) + \frac{1113}{8} (n_{12}v_2)^2 
\nonumber\\ & & \qquad
- \frac{615}{64} \bigg((n_{12}v_1)-(n_{12}v_2)\bigg)^2 \pi^2 + 18 v_1^2 +
\frac{123}{64} \pi^2 (v_1-v_2)^2 + 
33 (v_1v_2) - \frac{33}{2} v_2^2 \bigg) \nonumber\\ & & \qquad
+ \frac{G^3 m_1^2 m_2}{r_{12}^4}
\bigg(-\frac{45887}{168} (n_{12}v_1)^2 + 
\frac{24025}{42} (n_{12}v_1) (n_{12}v_2) - \frac{10469}{42} 
(n_{12}v_2)^2
\nonumber\\ & & \qquad +
\frac{48197}{840} v_1^2 - \frac{36227}{420} (v_1v_2) 
+ \frac{36227}{840}
v_2^2 + 110 \bigg((n_{12}v_1) - (n_{12}v_2)\bigg)^2 \ln
\left(\frac{r_{12}}{r'_1} \right)  \nonumber\\ & & \qquad
- 22 (v_1 - v_2)^2 \ln \left(\frac{r_{12}}{r'_1} \right) \bigg) 
+ \frac{16 G^4
m_2^4}{r_{12}^5} + \frac{G^4 m_1^2 m_2^2}{r_{12}^5} 
\bigg(\frac{34763}{210} -
\frac{44 \lambda}{3} - \frac{41}{16} \pi^2\bigg) \nonumber\\ & & \qquad + 
\frac{G^4 m_1^3 m_2}{r_{12}^5} 
\bigg(-\frac{3187}{1260} + \frac{44}{3} \ln
\left(\frac{r_{12}}{r'_1} \right)\bigg) + 
\frac{G^4 m_1 m_2^3}{r_{12}^5} \bigg(\frac{10478}{63} 
- \frac{44 \lambda}{3} -
\frac{41}{16}  \pi^2  \nonumber\\ & & \qquad-
\frac{44}{3} \ln \left(\frac{r_{12}}{r'_2} \right)\bigg)\bigg] 
n_{12}^i +
\bigg[\frac{G m_2}{r_{12}^2} \bigg(\frac{15}{2} (n_{12}v_1) 
(n_{12}v_2)^4 -
\frac{45}{8} (n_{12}v_2)^5 - 
\frac{3}{2} (n_{12}v_2)^3 v_1^2 \nonumber\\ & & \qquad
+ 6 (n_{12}v_1) (n_{12}v_2)^2 (v_1v_2) - 6
(n_{12}v_2)^3 (v_1v_2) - 2 (n_{12}v_2) (v_1v_2)^2 \nonumber\\ & & \qquad 
- 12 (n_{12}v_1)(n_{12}v_2)^2 v_2^2 + 12 (n_{12}v_2)^3 v_2^2 +
(n_{12}v_2) v_1^2 
v_2^2 - 4 (n_{12}v_1) (v_1v_2) v_2^2 \nonumber\\ & & \qquad 
+ 8 (n_{12}v_2) (v_1v_2) v_2^2 
+ 4 (n_{12}v_1)  v_2^4 - 7 (n_{12}v_2) v_2^4\bigg) \nonumber\\ & & \qquad 
+\frac{G^2 
m_2^2}{r_{12}^3} \bigg(-2 (n_{12}v_1)^2 
(n_{12}v_2) + 8 (n_{12}v_1) (n_{12}v_2)^2 \nonumber\\ & & \qquad
+ 2 (n_{12}v_2)^3 + 2 (n_{12}v_1)
(v_1v_2) + 4 (n_{12}v_2) (v_1v_2) - 2 (n_{12}v_1) v_2^2  - 4 (n_{12}v_2)
v_2^2\bigg) \nonumber\\ & & \qquad
+ \frac{G^2 m_1 m_2}{r_{12}^3} \bigg(-\frac{243}{4} (n_{12}v_1)^3
+ \frac{565}{4} (n_{12}v_1)^2 
(n_{12}v_2) - \frac{269}{4} (n_{12}v_1) (n_{12}v_2)^2 
\nonumber\\ & & \qquad
- \frac{95}{12} (n_{12}v_2)^3 + \frac{207}{8} (n_{12}v_1) v_1^2 -
\frac{137}{8} (n_{12}v_2) 
v_1^2 - 36 (n_{12}v_1) (v_1v_2) \nonumber\\ & & \qquad
+ \frac{27}{4} (n_{12}v_2) (v_1v_2)+
\frac{81}{8} (n_{12}v_1) v_2^2 + \frac{83}{8} (n_{12}v_2) v_2^2\bigg) 
\nonumber\\ & & \qquad +
\frac{G^3 m_2^3}{r_{12}^4} \bigg(4 (n_{12}v_1) 
+ 5 (n_{12}v_2)\bigg) +
\frac{G^3  m_1 m_2^2}{r_{12}^4} \bigg(-\frac{307}{8} 
(n_{12}v_1) + \frac{479}{8} (n_{12}v_2) \nonumber\\ & & \qquad
+ \frac{123}{32} \bigg((n_{12}v_1)-
(n_{12}v_2)\bigg) \pi^2 \bigg) + \frac{G^3 m_1^2 m_2}{r_{12}^4}
\bigg(\frac{31397}{420} 
(n_{12}v_1) - \frac{36227}{420} (n_{12}v_2) \nonumber\\ & & \qquad
- 44 \bigg((n_{12}v_1) -  (n_{12}v_2)\bigg) \ln
\left(\frac{r_{12}}{r'_1} \right) \bigg)\bigg] (v_1^i - v_2^i) 
\Bigg\}+{\cal O}(7)\;.
\end{eqnarray}
These equations are in full agreement with the known results valid up
to the 2.5PN order \cite{DD81a,Dthese,D82,BFP98}. They have the
correct perturbative limit given by the geodesics of the Schwarzschild
metric at the 3PN order. Most importantly, the equations are invariant
under Lorentz transformations (developed to 3PN order); this can be
checked using for instance the formulas developed in
\cite{BFregM}. Finally, as we have seen previously, the equations of
motion admit a conserved energy at the 3PN order. The study of the
Lagrangian (and Hamiltonian) formulation of these equations is
reported in a separate work \cite{ABF00}. The energy is given by

\begin{eqnarray}\label{VII2}
E &=& \frac{ m_1 v_1^2}{2 } -\frac{G m_1 m_2}{2 r_{12}} \nonumber\\ &+&  
\frac{1}{c^2} \Bigg\{\frac{G^2 m_1^2 m_2}{2 r_{12}^2 } + \frac{3  
m_1 v_1^4}{8 }+ \frac{G m_1 m_2}{r_{12}}
\bigg(-\frac{1}{4}(n_{12}v_1) (n_{12}v_2) + \frac{3}{2} v_1^2  -
\frac{7}{4} (v_1v_2) \bigg)\Bigg\} \nonumber\\ &+&
\frac{1}{c^4}\Bigg\{-\frac{G^3 m_1^3 m_2}{2 r_{12}^3}- \frac{19 G^3
m_1^2 m_2^2}{8 r_{12}^3} + \frac{5 m_1 v_1^6}{16} +
\frac{G m_1 m_2}{r_{12}} \bigg(\frac{3}{8} (n_{12}v_1)^3 
(n_{12}v_2)\nonumber\\ & & \quad + \frac{3}{16} (n_{12}v_1)^2 (n_{12}v_2)^2
- \frac{9}{8} 
(n_{12}v_1) (n_{12}v_2) v_1^2 - 
 \frac{13}{8} (n_{12}v_2)^2 v_1^2 + \frac{21}{8} v_1^4  \nonumber\\ & & \quad
+ 
\frac{13}{8}   (n_{12}v_1)^2 (v_1v_2) + \frac{3}{4} (n_{12}v_1) (n_{12}v_2)
(v_1v_2)  - \frac{55}{8} v_1^2 (v_1v_2) + 
\frac{17}{8} (v_1v_2)^2  + \frac{31}{16} v_1^2 v_2^2 \bigg) \nonumber\\ & &
\quad + 
\frac{G^2 m_1^2 m_2}{r_{12}^2} 
\bigg(\frac{29}{4} (n_{12}v_1)^2 - \frac{13}{4} (n_{12}v_1) (n_{12}v_2) +
\frac{1}{2}(n_{12}v_2)^2 - \frac{3}{2} v_1^2 + \frac{7}{4} v_2^2\bigg)\Bigg\}
\nonumber\\ &+& \frac{1}{c^6} \Bigg\{\frac{35  m_1 v_1^8}{128 }+ \frac{G m_1
m_2}{ 
r_{12}}  \bigg(-\frac{5}{16} (n_{12}v_1)^5 (n_{12}v_2) 
- \frac{5}{16}   (n_{12}v_1)^4 (n_{12}v_2)^2 \nonumber\\ & & \quad - 
\frac{5}{32} (n_{12}v_1)^3 (n_{12}v_2)^3 +
\frac{19}{16} (n_{12}v_1)^3 (n_{12}v_2) v_1^2 + \frac{15}{16} (n_{12}v_1)^2
(n_{12}v_2)^2 v_1^2 \nonumber\\ & & \quad
+ \frac{3}{4} (n_{12}v_1) (n_{12}v_2)^3 v_1^2 +
\frac{19}{16} (n_{12}v_2)^4 v_1^2 - \frac{21}{16} (n_{12}v_1) (n_{12}v_2)
v_1^4 - 2 (n_{12}v_2)^2 v_1^4 + \frac{55}{16} v_1^6 \nonumber\\ & & \quad
- \frac{19}{16} (n_{12}v_1)^4 (v_1v_2) - (n_{12}v_1)^3 (n_{12}v_2) (v_1v_2) -
\frac{15}{32} (n_{12}v_1)^2 (n_{12}v_2)^2 (v_1v_2) \nonumber\\ & & \quad
+ \frac{45}{16}
(n_{12}v_1)^2 v_1^2 (v_1v_2) + \frac{5}{4} (n_{12}v_1) (n_{12}v_2) v_1^2
(v_1v_2) + \frac{11}{4}(n_{12}v_2)^2 v_1^2 (v_1v_2)  \nonumber\\ & & \quad -
\frac{139}{16} v_1^4 (v_1v_2) - \frac{3}{4} 
(n_{12}v_1)^2 (v_1v_2)^2 + \frac{5}{16} (n_{12}v_1) (n_{12}v_2)
(v_1v_2)^2 +
\frac{41}{8} v_1^2 (v_1v_2)^2  \nonumber\\ & & \quad 
+\frac{1}{16} (v_1v_2)^3  - \frac{45}{16}
(n_{12}v_1)^2 v_1^2 v_2^2 - \frac{23}{32} (n_{12}v_1) (n_{12}v_2) v_1^2 v_2^2+
 \frac{79}{16} v_1^4 v_2^2 - \frac{161}{32} v_1^2 (v_1v_2) v_2^2   \bigg) 
\nonumber\\ & & \quad + \frac{G^2 m_1^2 m_2}{r_{12}^2}
\bigg(-\frac{49}{8} (n_{12}v_1)^4 + 
\frac{75}{8} (n_{12}v_1)^3 (n_{12}v_2) - \frac{187}{8} (n_{12}v_1)^2
(n_{12}v_2)^2 \nonumber\\ & & \quad + \frac{247}{24} (n_{12}v_1)
(n_{12}v_2)^3 + \frac{49}{8} (n_{12}v_1)^2 v_1^2 + \frac{81}{8}
(n_{12}v_1) (n_{12}v_2) v_1^2 -
\frac{21}{4} 
(n_{12}v_2)^2 v_1^2 + \frac{11}{2} v_1^4 \nonumber\\ & & \quad
- \frac{15}{2} (n_{12}v_1)^2
(v_1v_2) - \frac{3}{2} (n_{12}v_1) (n_{12}v_2) (v_1v_2) + \frac{21}{4}
(n_{12}v_2)^2 (v_1v_2) - 27 v_1^2 (v_1v_2) \nonumber\\ & & \quad
+ \frac{55}{2} (v_1v_2)^2 +
\frac{49}{4} (n_{12}v_1)^2 v_2^2 - \frac{27}{2} (n_{12}v_1) (n_{12}v_2) v_2^2
+ \frac{3}{4} (n_{12}v_2)^2 v_2^2 + \frac{55}{4} v_1^2 v_2^2 \nonumber\\ & &
\quad 
- 28 (v_1v_2) v_2^2 + \frac{135}{16} v_2^4\bigg) + \frac{3 G^4 m_1^4 m_2}{8
r_{12}^4}+ \frac{G^4 m_1^3
m_2^2}{r_{12}^4} 
\bigg(\frac{5809}{280} - \frac{11}{3} \lambda - 
\frac{22}{3} \ln \left(\frac{r_{12}}{r'_1} \right)\bigg) \nonumber\\ & & \quad
+ \frac{G^3 m_1^2 m_2^2}{r_{12}^3} \bigg(\frac{547}{12} (n_{12}v_1)^2 - 
\frac{3115}{48} (n_{12}v_1) (n_{12}v_2)  -
\frac{123}{64} (n_{12}v_1)^2 \pi^2  \nonumber\\ & & \quad + \frac{123}{64}
(n_{12}v_1) 
(n_{12}v_2) \pi^2  - \frac{575}{18} v_1^2 +
\frac{41}{64} \pi^2 v_1^2 + \frac{4429}{144} (v_1v_2) - \frac{41}{64} \pi^2
(v_1v_2) \bigg) \nonumber\\ & & \quad +
\frac{G^3  m_1^3  m_2}{r_{12}^3} 
\bigg(-\frac{44627}{840} (n_{12}v_1)^2 + \frac{32027}{840} (n_{12}v_1)
(n_{12}v_2)  
+  \frac{3}{2} (n_{12}v_2)^2 + \frac{24187}{2520} v_1^2 \nonumber\\ & & \quad
- 
\frac{27967}{2520} (v_1v_2) + \frac{5}{4} v_2^2 + 22 (n_{12}v_1)^2 \ln
\left(\frac{r_{12}}{r'_1}  \right) - 22 (n_{12}v_1) (n_{12}v_2) \ln
\left(\frac{r_{12}}{r'_1} \right) \nonumber\\ & &\quad -
\frac{22}{3} v_1^2 \ln \left(\frac{r_{12}}{r'_1} \right) + \frac{22}{3}
(v_1v_2) \ln \left(\frac{r_{12}}{r'_1} \right)\bigg)
\Bigg\}+1\leftrightarrow 2 +{\cal O}(7)\;. \end{eqnarray} This energy
is conserved in the sense that its time-derivative computed with the
3PN equations of motion equals the radiation reaction effect
at the 2.5PN order, namely

\begin{eqnarray}\label{VII2'}
\frac{dE}{dt}&=&\frac{4}{5}\frac{G^2m_1^2m_2}{c^5 r_{12}^3}\bigg[(v_1v_{12})
\left(-v_{12}^2+2\frac{Gm_1}{r_{12}}-8\frac{Gm_2}{r_{12}}\right)\nonumber\\
& &\quad
+(n_{12}v_1)(n_{12}v_{12})\left(3v_{12}^2-6\frac{Gm_1}{r_{12}}+\frac{52}{3}
\frac{Gm_2}{r_{12}}\right)\bigg]+1\leftrightarrow
2 +{\cal O}(7) \;.
\end{eqnarray}

The rather complicated expressions (\ref{VII1})-(\ref{VII2}) simplify
drastically in the case where the orbit is circular [apart from the
gradual inspiral associated with the balance equation (\ref{VII2'})]
and where we place ourselves in the center-of-mass frame. The circular
orbit corresponds to the physical situation of the inspiralling
compact binaries which motivate our work
\cite{3mn,CFPS93,FCh93,CF94,TNaka94,P95,DIS98}. Here, we give the
result concerning circular orbits without proof (see also
\cite{BF00}). The relative acceleration reads

\begin{equation}\label{VII3}
\frac{d {\bf v}_{12}}{ dt} = - \omega^2 {\bf y}_{12}+ {\bf
F}_{\rm reac} + {\cal O}(7)\;, 
\end{equation}
where ${\bf F}_{\rm reac}$ is the standard radiation-reaction force in
harmonic coordinates,

\begin{equation}\label{VII3'}
{\bf F}_{\rm reac}=-\frac{32}{5}\frac{G^3 m^3\nu}{c^5r_{12}^4}{\bf v}_{12}
\end{equation}
($\nu=\frac{m_1m_2}{m^2}$ being the symmetric mass ratio), and where
the orbital frequency $\omega$ of the relative circular motion is
given to the 3PN order by

\begin{eqnarray}\label{VII4}
 \omega^2 &=& \frac{G m}{ r_{12}^3}\Biggl\{ 1+\left(-3+\nu\right)
 \gamma + \left(6+\frac{41}{4}\nu +\nu^2\right) \gamma^2 \nonumber\\
 &+&\left(-10+\left[-\frac{67759}{840}+\frac{41}{64}\pi^2+22
 \ln\left(\frac{r_{12}}{r'_0}\right)+\frac{44}{3}\lambda \right]\nu
 +\frac{19}{2}\nu^2+\nu^3\right) \gamma^3 +{\cal O}(8)
\Biggr\}\;.
\end{eqnarray}
The post-Newtonian parameter is defined by $\gamma=\frac{G m}{r_{12}
c^2}$, and we recall that $r_{12}=|{\bf y}_1-{\bf y}_2|$ is the
orbital separation in harmonic coordinates. The constant $r'_0$
appearing in the logarithm is related to the two constants $r'_1$ and
$r'_2$ by

\begin{equation}
\ln r'_0=\frac{m_1}{ m}\ln r'_1+\frac{m_2}{ m}\ln r'_2\;.
\end{equation}
The 3PN energy $E$ in the center of mass of the particles, which is
such that $\frac{dE}{dt}=0$ as a consequence of the conservative
equations of motion, is obtained as
 
\begin{eqnarray}\label{VII5}
 E &=& -\frac{1}{2}\mu c^2 \gamma\Biggl\{1
 +\left(-\frac{7}{4}+\frac{1}{4}\nu\right) \gamma + 
 \left(-\frac{7}{8}+\frac{49}{8}\nu +\frac{1}{8}\nu^2\right) \gamma^2
 \nonumber\\ 
&+&\left(-\frac{235}{64}+\left[\frac{106301}{6720}-\frac{123}{64}\pi^2+
 \frac{22}{3}\ln\left(\frac{r_{12}}{r'_0}\right)-
 \frac{22}{3}\lambda\right]\nu   
 +\frac{27}{32}\nu^2+\frac{5}{64}\nu^3\right) \gamma^3 +{\cal O}(8) 
\Biggr\}\;.\nonumber\\
\end{eqnarray}
The invariant 3PN energy follows from the replacement of the
post-Newtonian parameter $\gamma$ by its expression in terms of the
frequency $\omega$ as deduced from computing the inverse of
(\ref{VII4}). We find

\begin{eqnarray}\label{VII6}
 E &=& -\frac{1}{2}\mu c^2 x
 \Biggl\{1+\left(-\frac{3}{4}-\frac{1}{12}\nu\right) x + 
 \left(-\frac{27}{8}+\frac{19}{8}\nu -\frac{1}{24}\nu^2\right) x^2 
\nonumber\\
 &+&\left(-\frac{675}{64}+\left[\frac{209323}{4032}-\frac{205}{96}\pi^2-
\frac{110}{9}\lambda\right]\nu
 -\frac{155}{96}\nu^2-\frac{35}{5184}\nu^3\right) x^3 +{\cal O}(8) 
\Biggr\}\;,
\end{eqnarray}
where the parameter $x$ is defined by

\begin{equation}\label{VII7}
x=\left(\frac{G m \omega}{c^3}\right)^{2/3}\;.
\end{equation}
Note that the logarithm disappeared from the invariant expression of
the energy (\ref{VII6}), in agreement with the fact that it is pure
gauge. However, the constant $\lambda$ stays in the final formula; the
static ambiguity constant $\omega_s$ of Jaranowski and Sch\"afer
\cite{JaraS99} is related to it by
$\omega_s=-\frac{11}{3}\lambda-\frac{1987}{840}$ (see \cite{BF00}).

\acknowledgments

Most of the algebraic manipulations reported in this article have been
done with the help of the softwares Mathematica and MathTensor. One of
us (G.F.) would like to acknowledge Jean-Marc Conan and Vincent
Michau for letting him complete this work in the context of his
military service.

\appendix

\section{Sum of Leibniz terms}\label{D}

In this appendix we give the sum of all the terms of the type
$\delta_{\rm Leibniz}T$ introduced by (\ref{III20}) that we have
encountered during the process of simplification of the 3PN
potentials. The reduction of these terms using the distributional
derivative is done in Section \ref{VI}.

\begin{eqnarray}\label{D1}
&&\delta_{\rm Leibniz}\left(\frac{h^{00}+h^{ii}}{
2}\right)=\Box_{\cal R}^{-1}\Bigg\{-\frac{8}{ c^4}\bigg(\partial_i~\!{\rm
Pf}~\!V\partial_i~\!{\rm Pf}~\!V+V\Box ~\!{\rm Pf}~\!V-\frac{1}{2}\Box
(~\!{\rm Pf}~\!V^2)-\frac{1}{ c^2}(\partial_t~\!{\rm Pf}~\!V)^2\bigg) 
\nonumber\\
&&\quad -\frac{8}{ c^6}\bigg(\partial_i~\!{\rm
Pf}~\!V\partial_i~\!{\rm Pf}~\!{\hat W}+\frac{1}{2}V\Box ~\!{\rm
Pf}~\!{\hat W}+\frac{1}{2}{\hat W}\Box ~\!{\rm Pf}~\!V-\frac{1}{2}\Box
(~\!{\rm Pf}~\!V {\hat W})-\frac{1}{ c^2}\partial_t~\!{\rm
Pf}~\!V\partial_t~\!{\rm Pf}~\!{\hat W}\bigg)
\nonumber\\
&&\quad -\frac{32}{ c^6}\bigg(V\partial_i~\!{\rm
Pf}~\!V\partial_i~\!{\rm Pf}~\!V+\frac{1}{2}V^2\Box ~\!{\rm
Pf}~\!V-\frac{1}{ c^2}V(\partial_t~\!{\rm Pf}~\!V)^2-\frac{1}{6}\Box
(~\!{\rm Pf}~\!V^3)\bigg)
\nonumber\\
&&\quad -\frac{32}{ c^6}\bigg(\partial_i~\!{\rm Pf}~\!V\partial_i(~\!{\rm
Pf}~\!V^2)-2V\partial_i~\!{\rm Pf}~\!V\partial_i~\!{\rm
Pf}~\!V\bigg)\nonumber\\
&&\quad -\frac{16}{ c^8}\bigg(V\partial_t^2(~\!{\rm Pf}~\!V^2)
-2V(\partial_t~\!{\rm Pf}~\!V)^2-2V^2\partial_t^2~\!{\rm Pf}~\!V\bigg)
\nonumber\\
&&\quad -\frac{16}{ c^8}\bigg(\partial_t~\!{\rm
Pf}~\!V\partial_t(~\!{\rm Pf}~\!V^2) -2V(\partial_t~\!{\rm
Pf}~\!V)^2\bigg)\nonumber\\ &&\quad-\frac{32}{
c^8}\bigg(V_i\partial_t\partial_i(~\!{\rm Pf}~\!V^2)
-2VV_i\partial_t\partial_i~\!{\rm Pf}~\!V-2V_i\partial_t~\!{\rm
Pf}~\!V\partial_i~\!{\rm Pf}~\!V\bigg)
\nonumber\\
&&\quad -\frac{64}{ c^8}\bigg(V^2\partial_i~\!{\rm
Pf}~\!V\partial_i~\!{\rm Pf}~\!V+\frac{1}{3}V^3\Box ~\!{\rm
Pf}~\!V-\frac{1}{12}\Box (~\!{\rm Pf}~\!V^4)\bigg)\nonumber\\
&&\quad-\frac{32}{ c^8}\bigg(\partial_i(~\!{\rm
Pf}~\!V^2)\partial_i(~\!{\rm Pf}~\!V^2)-4V^2\partial_i~\!{\rm
Pf}~\!V\partial_i~\!{\rm Pf}~\!V \bigg)
\nonumber\\
&&\quad +\frac{144}{ c^8}\bigg(V\partial_i~\!{\rm
Pf}~\!V\partial_i(~\!{\rm Pf}~\!V^2)-2V^2\partial_i~\!{\rm
Pf}~\!V\partial_i~\!{\rm Pf}~\!V
\bigg)\nonumber\\ 
&&\quad -\frac{128}{ 3c^8}\bigg(\partial_i~\!{\rm
Pf}~\!V\partial_i(~\!{\rm Pf}~\!V^3)-3V^2\partial_i~\!{\rm
Pf}~\!V\partial_i~\!{\rm Pf}~\!V \bigg)
\nonumber\\
&&\quad +\frac{64}{ c^8}\bigg(\partial_i~\!{\rm
Pf}~\!V_j\partial_j(~\!{\rm Pf}~\!VV_i)-V\partial_i~\!{\rm
Pf}~\!V_j\partial_j~\!{\rm Pf}~\!V_i-V_i\partial_i~\!{\rm
Pf}~\!V_j\partial_j~\!{\rm Pf}~\!V \bigg)
\nonumber\\
&&\quad -\frac{16}{ c^8}\bigg({\hat W}_{ij}\partial_{ij}(~\!{\rm
Pf}~\!V^2)-2V{\hat W}_{ij}\partial_{ij}~\!{\rm Pf}~\!V-2{\hat
W}_{ij}\partial_i~\!{\rm Pf}~\!V\partial_j~\!{\rm Pf}~\!V
\bigg)\nonumber\\ &&\quad -\frac{4}{ c^8}\bigg(\partial_i~\!{\rm
Pf}~\!{\hat W}\partial_i~\!{\rm Pf}~\!{\hat W}+{\hat W}\Box ~\!{\rm
Pf}~\!{\hat W}-\frac{1}{2}\Box (~\!{\rm Pf}~\!{\hat W}^2)\bigg)
\nonumber\\
&&\quad -\frac{16}{ c^8}\bigg({\hat W}\partial_i~\!{\rm
Pf}~\!V\partial_i~\!{\rm Pf}~\!V+2V\partial_i~\!{\rm
Pf}~\!V\partial_i~\!{\rm Pf}~\!{\hat W}+V{\hat W}\Box ~\!{\rm Pf}~\!V
+\frac{1}{2}V^2\Box ~\!{\rm Pf}~\!{\hat W}-\frac{1}{2}\Box (~\!{\rm
Pf}~\!V^2{\hat W})\bigg)
\nonumber\\
&&\quad +\frac{8}{ c^8}\bigg({\hat W}\Box(~\!{\rm Pf}~\!V^2)-2V{\hat
W}\Box ~\!{\rm Pf}~\!V -2{\hat W}\partial_i~\!{\rm
Pf}~\!V\partial_i~\!{\rm Pf}~\!V\bigg)\nonumber\\ &&\quad -\frac{16}{
c^8}\bigg(\partial_i~\!{\rm Pf}~\!{\hat W}\partial_i(~\!{\rm
Pf}~\!V^2)-2V\partial_i~\!{\rm Pf}~\!{\hat W}\partial_i~\!{\rm
Pf}~\!V\bigg)
\nonumber\\
&&\quad -\frac{32}{ c^8}\bigg(\partial_i~\!{\rm
Pf}~\!V\partial_i(~\!{\rm Pf}~\!V{\hat W})-V\partial_i~\!{\rm
Pf}~\!V\partial_i~\!{\rm Pf}~\!{\hat W}-{\hat W}\partial_i~\!{\rm
Pf}~\!V\partial_i~\!{\rm Pf}~\!V\bigg)
\nonumber\\ 
&&\quad +\frac{8}{ c^8}\bigg(\partial_k~\!{\rm Pf}~\!{\hat
W}_{ij}\partial_k~\!{\rm Pf}~\!{\hat W}_{ij}+{\hat W}_{ij}\Box ~\!{\rm
Pf}~\!{\hat W}_{ij}-\frac{1}{2}\Box (~\!{\rm Pf}~\!{\hat W}_{ij}{\hat
W}_{ij})\bigg)
\nonumber\\
&&\quad -\frac{64}{ c^8}\bigg(\partial_i~\!{\rm
Pf}~\!V\partial_i~\!{\rm Pf}~\!({\hat X}+\frac{1}{2}{\hat
Z})+\frac{1}{2}V\Box ~\!{\rm Pf}~\!({\hat X}+\frac{1}{2}{\hat
Z})+\frac{1}{2}({\hat X}+\frac{1}{2}{\hat Z})\Box ~\!{\rm
Pf}~\!V\nonumber\\ &&\quad-\frac{1}{2}\Box(~\!{\rm Pf}~\!V({\hat
X}+\frac{1}{2}{\hat Z}))\bigg)\Bigg\} +{\cal O}(10)\;,\\ &&
\delta_{\rm Leibniz}h^{0i}=\nonumber\\ &&\quad\Box_{\cal
R}^{-1}\Bigg\{-\frac{16}{ c^5}\bigg(\partial_j~\!{\rm
Pf}~\!V\partial_j~\!{\rm Pf}~\!V_i+\frac{1}{2}V\Box ~\!{\rm Pf}~\!V_i
+\frac{1}{2}V_i\Box ~\!{\rm Pf}~\!V-\frac{1}{2}\Box (~\!{\rm
Pf}~\!VV_i)-\frac{1}{ c^2}\partial_t~\!{\rm Pf}~\!V\partial_t~\!{\rm
Pf}~\!V_i\bigg)
\nonumber\\
&&\quad +\frac{24}{ c^7}\bigg(\partial_t~\!{\rm
Pf}~\!V\partial_i(~\!{\rm Pf}~\!V^2)-2V\partial_t~\!{\rm
Pf}~\!V\partial_i~\!{\rm Pf}~\!V\bigg)\nonumber\\ &&\quad +\frac{24}{
c^7}\bigg(\partial_i~\!{\rm Pf}~\!V\partial_t(~\!{\rm
Pf}~\!V^2)-2V\partial_i~\!{\rm Pf}~\!V\partial_t~\!{\rm Pf}~\!V\bigg)
\nonumber\\
&&\quad -\frac{32}{ c^7}\bigg(\partial_j~\!{\rm
Pf}~\!V\partial_j~\!{\rm Pf}~\!{\hat R}_i+\frac{1}{2}V\Box ~\!{\rm
Pf}~\!{\hat R}_i+\frac{1}{2}{\hat R}_i\Box ~\!{\rm
Pf}~\!V-\frac{1}{2}\Box(~\!{\rm Pf}~\!V{\hat R}_i)\bigg)\nonumber\\
&&\quad -\frac{8}{ c^7}\bigg(\partial_j~\!{\rm
Pf}~\!V_i\partial_j~\!{\rm Pf}~\!{\hat W}+\frac{1}{2}V_i\Box ~\!{\rm
Pf}~\!{\hat W}+\frac{1}{2}{\hat W}\Box ~\!{\rm
Pf}~\!V_i-\frac{1}{2}\Box(~\!{\rm Pf}~\!V_i{\hat W})\bigg)\nonumber\\
&&\quad +\frac{32}{ c^8}\bigg(\partial_j~\!{\rm
Pf}~\!V\partial_i(~\!{\rm Pf}~\!VV_j)-V\partial_j~\!{\rm
Pf}~\!V\partial_i~\!{\rm Pf}~\!V_j-V_j\partial_j~\!{\rm
Pf}~\!V\partial_i~\!{\rm Pf}~\!V\bigg)\nonumber\\ &&\quad -\frac{32}{
c^8}\bigg(\partial_j~\!{\rm Pf}~\!V\partial_j(~\!{\rm
Pf}~\!VV_i)-V\partial_j~\!{\rm Pf}~\!V\partial_j~\!{\rm
Pf}~\!V_i-V_i\partial_j~\!{\rm Pf}~\!V\partial_j~\!{\rm
Pf}~\!V\bigg)\nonumber\\ &&\quad +\frac{32}{
c^7}\bigg(\partial_i~\!{\rm Pf}~\!V_j\partial_j(~\!{\rm
Pf}~\!V^2)-2V\partial_i~\!{\rm Pf}~\!V_j\partial_j~\!{\rm
Pf}~\!V\bigg)\nonumber\\ &&\quad -\frac{32}{
c^7}\bigg(\partial_j~\!{\rm Pf}~\!V_i\partial_j(~\!{\rm
Pf}~\!V^2)-2V\partial_j~\!{\rm Pf}~\!V_i\partial_j~\!{\rm
Pf}~\!V\bigg)
\nonumber\\
&&\quad +\frac{16}{ c^7}\bigg(\partial_k~\!{\rm
Pf}~\!V_j\partial_k~\!{\rm Pf}~\!{\hat W}_{ij}+\frac{1}{2}V_j\Box
~\!{\rm Pf}~\!{\hat W}_{ij}+\frac{1}{2}{\hat W}_{ij}\Box ~\!{\rm
Pf}~\!V_j-\frac{1}{2}\Box(~\!{\rm Pf}~\!V_j{\hat
W}_{ij})\bigg)\nonumber\\ &&\quad -\frac{32}{
c^7}\bigg(V_i\partial_j~\!{\rm Pf}~\!V\partial_j~\!{\rm
Pf}~\!V+2V\partial_j~\!{\rm Pf}~\!V\partial_j~\!{\rm
Pf}~\!V_i+\frac{1}{2}V^2\Box ~\!{\rm Pf}~\!V_i+V V_i\Box ~\!{\rm
Pf}~\!V-\frac{1}{2}\Box(~\!{\rm Pf}~\!V^2V_i)\bigg)\Bigg\}
\nonumber\\
&&\quad +{\cal O}(9)\;,\\ && \delta_{\rm Leibniz}h^{ij}=\nonumber\\
&&\quad\Box_{\cal R}^{-1}\Bigg\{\frac{16}{
c^6}\bigg(\partial_{(i}~\!{\rm Pf}~\!V\partial_{j)}(~\!{\rm
Pf}~\!V^2)-2V\partial_{(i}~\!{\rm Pf}~\!V\partial_{j)}~\!{\rm
Pf}~\!V\bigg)\nonumber\\ &&\quad-\frac{8}{
c^6}\delta_{ij}\bigg(\partial_{k}~\!{\rm Pf}~\!V\partial_{k}(~\!{\rm
Pf}~\!V^2)-2V\partial_{k}~\!{\rm Pf}~\!V\partial_{k}~\!{\rm
Pf}~\!V\bigg)\Bigg\}+{\cal O}(8) \;.
\end{eqnarray}

\end{document}